\documentclass[showpacs, preprintnumbers, nofootinbib, aps, prd, superscriptaddress,10pt, showkeys, notitlepage, twocolumn]{revtex4-1}

\usepackage{graphicx,amssymb,amsmath,amsthm,amsfonts,mathrsfs,epsfig}

\usepackage[linktocpage]{hyperref}
\usepackage[usenames,dvipsnames]{color}
\usepackage{epstopdf}
\usepackage{aas_macros}
\usepackage{pifont}
\definecolor{darkred}{rgb}{0.5,0,0}
\definecolor{darkgreen}{rgb}{0,0.5,0}
\definecolor{darkblue}{rgb}{0,0,0.5}
\definecolor{prussian}{rgb}{0.0, 0.19, 0.33}
\definecolor{richelectricblue}{rgb}{0.03, 0.57, 0.82}
\definecolor{teal}{rgb}{0.0, 0.5, 0.5}
\definecolor{mediumseagreen}{rgb}{0.24, 0.7, 0.44}
\definecolor{lust}{rgb}{0.9, 0.13, 0.13}
\definecolor{ballblue}{rgb}{0.13, 0.67, 0.8}
\definecolor{darkcyan}{rgb}{0.0, 0.55, 0.55}
\definecolor{mountainmeadow}{rgb}{0.19, 0.73, 0.56}
\definecolor{palecarmine}{rgb}{0.69, 0.25, 0.21}
\definecolor{richcarmine}{rgb}{0.84, 0.0, 0.25}
\definecolor{tangelo}{rgb}{0.98, 0.3, 0.0}
\definecolor{venetian}{rgb}{0.784,0.031,0.082}
\definecolor{bdfrance}{rgb}{0.192,0.549,0.906}

\hypersetup{colorlinks=true,
            citecolor=venetian,
            linkcolor=bdfrance,
            urlcolor=lust}

\usepackage{amsmath,amssymb}
\usepackage{tensor}
\usepackage{mathtools}
\usepackage{amsbsy}
\usepackage{bm}
\usepackage{float}

\newcommand{\be}{\begin{equation}}
\newcommand{\ee}{\end{equation}}
\newcommand{\ba}{\begin{aligned}}
\newcommand{\ea}{\end{aligned}}
\newcommand{\bear}{\begin{eqnarray}}
\newcommand{\eear}{\end{eqnarray}}

\newcommand{\p}{\prime}

\newcommand{\rp}{{\rm p}}

\newcommand{\re}{{\rm e}}

\newcommand{\rc}{{\rm c}}
\newcommand{\bnabla}{\boldsymbol{\nabla}}
\newcommand{\bv}{\mathbf{v}}
\newcommand{\bB}{\mathbf{B}}
\newcommand{\bphi}{\boldsymbol{\hat{\varphi}}}

\newcommand{\bJ}{\mathbf{J}}

\newcommand{\cC}{{\cal C}}
\newcommand{\tcC}{{\tilde{\mathcal{C}}}}

\newcommand{\cF}{{\cal F}}

\newcommand{\Rs}{R_{\star}}
\newcommand{\Ms}{M_{\star}}

\newcommand{\rN}{{\rm N}}
\newcommand{\nn}{\nonumber}

\begin{document}

\title{Magnetic equilibria of relativistic axisymmetric stars: The impact of flow constants}

\author{Arthur G. Suvorov}
\email{arthur.suvorov@manlyastrophysics.org}
\affiliation{Manly Astrophysics, 15/41-42 East Esplanade, Manly, New South Wales 2095, Australia}
\affiliation{Theoretical Astrophysics, Eberhard Karls University of T{\"u}bingen, T{\"u}bingen, D-72076, Germany}

\author{Kostas Glampedakis}
\email{kostas@um.es}
\affiliation{Departamento de F\'isica, Universidad de Murcia, Murcia, E-30100, Spain}
\affiliation{Theoretical Astrophysics, Eberhard Karls University of T{\"u}bingen, T{\"u}bingen, D-72076, Germany}

\date{\today}

\begin{abstract}
\noindent Symmetries and conservation laws associated with the ideal Einstein-Euler system, for stationary and axisymmetric stars, can be utilized to define a set of \emph{flow constants}. 
These quantities are conserved along flow lines in the sense that their gradients are orthogonal to the four-velocity. They are also conserved along surfaces of constant magnetic flux, making 
them powerful tools to identify general features of neutron star equilibria. One important corollary of their existence is that mixed poloidal-toroidal fields are inconsistent with the absence of 	
meridional flows except in some singular sense, a surprising but powerful result first proven by Bekenstein and Oron. In this work, we revisit the flow constant formalism to rederive this result 
together with several new ones concerning both nonlinear and perturbative magnetic equilibria. Our investigation is supplemented by some numerical solutions for multipolar magnetic fields 
on top of a Tolman-VII background, where strict power-counting of the flow constants is used to ensure a self-consistent treatment.
\end{abstract} 

\maketitle


\section{Introduction}
\label{sec:intro}

The intricate magnetic fields of neutron stars govern most, if not all, of their persistent and transient activity. Outbursts from the magnetar subclass are likely powered by an ultra-strong internal field \cite{dt92,td93}, for instance, while it is the magnetospheric field that largely controls the morphology of radio pulsars \cite{besk93,mel16} and X-ray binaries \cite{pw21,gs21}. Magnetohydrodynamic (MHD) modelling of neutron stars, particularly within the realm of general relativity (GR), is thus crucial to interpret their emissions.

Though theorised earlier from stability arguments \cite{p56,wri73,tay73}, the pioneering simulations by Braithwaite and collaborators \cite{bs04,bn06,bs06} led to the popularity of the `twisted torus' model: an almost axisymmetric field with a toroidal component that is confined to an equatorial torus. These configurations are stable over many dynamical (Alfv{\'e}n) timescales (see also Refs.~\cite{bra09,akg13,mit15}), and thus are a leading candidate for the field structure of mature neutron stars (though it has been argued that non-axisymmetric end states are more likely \cite{bra08,bec22}). MHD evolutions of pulsar magnetospheres suggest that the inclination angle made between the magnetic and rotation axes tends to zero at late times \cite{phil14}, further promoting the degree of axisymmetry. There is thus reasonable motivation for exploring the set of stationary and axisymmetric (GR-)MHD equilibria, even if a neutron star never truly attains such a state (because of processes related to finite thermal \cite{pons09} and electrical \cite{gour22} conductivity, for example).

It is generally advantageous to exploit symmetries when approaching a theoretical problem. In cases of stationary and 
axisymmetric MHD equilibria, there exists a number of \emph{flow constants} (also known as \emph{streamline invariants}~\cite{mark17}) 
which reduce the complexity of the system \cite{c56,w59,mestelbook}. More precisely, these `flow constants' are functions which are conserved along flow lines in the sense that 
their gradient is orthogonal to the (four-)velocity. Equivalently, these are functions which are conserved along surfaces of constant magnetic flux. Their introduction allows for the 
basic (GR)MHD equations
to be rewritten in a form that is more conducive to analytic and numerical investigation, like the familiar example of the Grad-Shafranov (GS) equation which may contain some general toroidal function of the magnetic flux (i.e. a flow constant) \cite{love86,is03}.

In the GR context, these flow constants were derived and detailed by Bekenstein and Oron (hereafter BO; \cite{bek78,bek79}) and others since, including Ioka and Sasaki (IS; \cite{is03,is04}) and Gourgoulhon and collaborators \cite{gour06,gour11,mark17,uryu19,uryu22}. Using this formalism, BO were able to prove a surprising result about GRMHD equilibria: configurations with mixed poloidal-toroidal fields are singular in the limit of a vanishing meridional flow. That is, the limit of vanishing meridional flow also corresponds to the limit of either a purely poloidal or purely toroidal magnetic field and a circular spacetime \cite{oron02}. The former is at least intuitive on a dynamical level: inducing a toroidal field requires a poloidal current, and hence meridional flow \citep{fr12}. It is remarkable though that one cannot completely suppress the meridional flow if considering a sequence of mixed-field equilibria. 
As purely poloidal or toroidal fields are unstable \cite{lasky11,rezz11,kiu12,rezz12}, the BO theorem suggests that meridional flows are a necessary ingredient in realistic neutron star models (see also Ref.~\cite{gusakov18}). This result seems to have gone largely unnoticed in the literature.

The goal of this work is to revisit the flow constant formalism and its corollaries, both at the non-linear and perturbative levels. Indeed, owing to the complexity of non-linear GRMHD, most models of neutron star equilibria, such as those of Colaiuda et al. (C08; \cite{col08}) and Ciolfi et al. (C09; \cite{c09}), invoke a perturbative scheme using an expansion in the magnetic field. We discuss the various scenarios in the context of the BO results, aiming to provide a fresh look at the theory landscape and its implications for neutron star structure. The special roles of the induction and GS equations are discussed in detail. Our results are complemented by some numerical solutions for perturbative equilibria, which differ from existing ones in the literature: we argue in Sec.~\ref{sec:previous} that previous choices may not be physically admissible.

This work is split into two parts. A reader whose interests are more focussed towards non-linear and general corollaries of the flow constant formalism can contain their reading to Secs.~\ref{sec:flowconstants} and \ref{sec:corollaries}, while one who is primarily interested in perturbative equilibria may skip ahead to Sec.~\ref{sec:perturbative}. Numerical constructions of multipolar `twisted-torus' solutions are presented in Sec.~\ref{sec:workedex}, with discussion and conclusions given in Sec.~\ref{sec:conclusions}.

Natural units $c = G = 1$ are assumed throughout, unless explicitly stated otherwise. Greek indices run over spacetime coordinates, 
while Latin indices are reserved for purely \emph{meridional} components ($r$,$\theta$). As a quick guide, some previous literature that is frequently referred to is abbreviated as follows: BO78 = Ref.~\cite{bek78}, BO79 = Ref.~\cite{bek79}, IS03 = Ref.~\cite{is03}, IS04 = Ref.~\cite{is04}, C08 = Ref.~\cite{col08}, and C09 = Ref.~\cite{c09}.


\section{General-relativistic MHD equilibria with flow constants}
\label{sec:flowconstants}

This section presents certain key relationships between flow constants and `primitive' quantities, such as the poloidal magnetic field and the fluid velocity. Although the results presented 
in this section can largely be found in previous literature (most notably Refs.~\cite{bek78,bek79,is03,is04,gour06,gour11,mark17,uryu19,uryu22,oron02}), we provide a summary here, with the eventual 
aim of pointing out important corollaries and various limits in Section \ref{sec:corollaries}. 
A self-contained derivation of the flow constants (following the steps of the original derivation of BO78 and BO79) can be found in Appendix~\ref{sec:appa}. For completeness and to 
highlight the role of GR, an analogous derivation in the Newtonian limit is provided in Appendix \ref{sec:appNewton}.

\subsection{Metric tensor and conventions}

At the full non-linear level, we work with a general stationary and axisymmetric metric $g_{\mu \nu}$ in spherical-like coordinates $\{t,r,\theta,\varphi\}$ with signature $\{-,+,+,+\}$.
Later on it will be convenient to introduce a static and spherically symmetric `background', where the magnetic field is treated 
perturbatively (see Sec.~\ref{sec:perturbative}). Here and in Sec.~\ref{sec:corollaries}, MHD dynamics are treated non-linearly, though the fluid is assumed 
to be perfect, infinitely conducting, and barotropic. 

The stress-energy tensor appearing in the Einstein equations contains both fluid and electromagnetic components. The fluid piece is written as
\be
\label{eq:fluidstress}
T^{\mu \nu}_{\rm fluid} = \left( \epsilon+ p \right) u^{\mu} u^{\nu} + p g^{\mu \nu},
\ee
where $\epsilon, p, u^\mu$ denote the fluid energy density, pressure, and four-velocity, respectively. The former parameter can be expressed in terms
of the rest-mass density and specific internal energy  $\tilde{\epsilon}$ as
\be
 \epsilon =  \rho ( 1+  \tilde{\epsilon} ).
 \label{densities}
\ee
With regard to these densities, there is a certain degree of notational confusion in the literature.
In Refs.~\cite{c09,c10}, for instance, $\rho$ is used to denote the \emph{energy density}, which makes it appear at first glance that the resulting GS equation
disagrees with others in the literature \cite{is03,is04}, though this is not the case (cf. Sec. \ref{sec:previous}). The notation used here is the typical one
when working with the Tolman-Oppenheimer-Volkoff (TOV) equations, for example. For a barotropic fluid, the energy density and rest-mass density are related 
through the first law of thermodynamics, $d(\epsilon / \rho) = - p d(1  / \rho)$.

Expression \eqref{eq:fluidstress} corresponds to the stress-energy of a single fluid, where we have $\rho = m n$ for baryon mass, $m$, and number density, $n$. 
In reality, neutron star matter comprises multiple particle species ($e^{-}, p, n, \ldots$) 
and a multi-fluid description is needed (see, for instance, Refs.~\cite{prix05,g12}). The distinction between particle constituents is important when considering the dynamics 
of magnetised stars, as the four-current is entirely sourced by the flow of the electrons relative to the other charged particles in the limit of charge neutrality. 
The single fluid description is based on the additional assumptions of a negligible electron inertia and a common velocity for the rest of the particles 
(charged and uncharged).

The flow constants we introduce in the next section that arise from the induction equation should thus be strictly associated with the electron number density 
and not the global one. This point is discussed further in Sec. \ref{sec:induction}. For a single fluid in the ideal MHD limit, the electromagnetic stress-energy tensor is given by
\be \label{eq:emstress}
T^{\alpha\beta}_{\rm EM}  = g^{\alpha\beta} \frac{B^2}{8\pi}  + \frac{1}{4\pi} \left ( B^2 u^\alpha u^\beta - B^\alpha B^\beta \right ),
\ee
where $B^\mu$ is the magnetic field four-vector. The electric and magnetic components are generally given as $E_{\mu} = F_{\mu \nu} u^{\nu}$ and 
$B_{\alpha} = \tfrac{1}{2}\sqrt{-g} \epsilon_{\alpha \beta \gamma \delta} u^{\beta} F^{\gamma \delta}$ for Levi-Civita symbol $\epsilon_{\alpha \beta \gamma \delta}$ 
and Faraday tensor $F_{\mu \nu} = A_{\nu,\mu} - A_{\mu,\nu}$. Here $A_{\mu}$ is the vector potential, whose existence is implied by Maxwell's equations, 
$\nabla_{[\alpha} F_{\mu \nu]} = 0$, and $E_{\mu} = 0$ defines the ideal MHD condition of infinite conductivity.

A result of immediate interest concerns a theorem by Carter \cite{cart69,cart73}. While a strictly azimuthal flow $(u^{i} = 0)$ implies that one 
can consider a circular spacetime, $g_{ti} = g_{\varphi i } = 0$, inclusion of meridional flows or mixed poloidal-toroidal fields generally forces these extra 
off-diagonal terms to be non-zero and means one cannot work with the Papapetrou metric \cite{pap66}. Intuitively, if one considers a radial flow ($u^{r}\neq0$), for instance, 
it is clear that the physical dynamics are not symmetric under time reversal if the system is also rotating. 
Similarly, for a mixed magnetic field there will generally be a poloidal current and the same logic applies \citep{fr12}. Spacetimes with purely toroidal fields are still circular 
however, since one can then exploit the azimuthal coordinate/gauge freedom \cite{gour11}. Additionally, one can always set $g_{r \theta} = 0$ regardless.

The necessity of non-circular components is one key reason why constructing mixed poloidal-toroidal equilibria at a non-linear level is 
so difficult in GR \cite{boq95};  it is only recently that such equilibria have been numerically constructed \cite{uryu19} and evolved \cite{uryu22} 
self-consistently\footnote{It has been argued that even for mixed, virial-strength fields, the degree of non-circularity is of order $\lesssim 10^{-3}$ \cite{oron02,pili17}. 
Whether this remains true for all physical setups is unclear owing to the numerical nature of the statement, but for computational purposes 
a circular approximation may be reasonable.} (see also Refs.~\cite{love86,fuji13,gusakov18} for some Newtonian solutions). Aside from mathematical complexity, 
these extra pieces enrich the physical system as discussed in Sec.~\ref{sec:conclusions}.

\subsection{The BO flow constants}
\label{sec:BO}

In this section we list the formulae for the flow constants, as derived in the BO papers \cite{bek78,bek79} and the aforementioned references. A proper derivation 
is relegated to Appendix \ref{sec:appa} owing to the length of the calculations involved. Each of the constants here are derived from the conservation law 
$\nabla_{\mu} T^{\mu \nu} = \nabla_{\mu} \left( T^{\mu \nu}_{\rm fluid} + T^{\mu \nu}_{\rm EM}\right) = 0$ together with the Einstein equations 
$R_{\mu \nu} - \tfrac{1}{2} R g_{\mu \nu} = 8 \pi T_{\mu \nu}$ of a general stationary and axisymmetric spacetime.

Although we use the term `flow', Eq.~\eqref{levelsurf1} from Appendix \ref{sec:appa} shows that for any conserved $Q$ (i.e. $u^{\alpha} Q_{,\alpha} = 0$) 
one necessarily has that $B^{\alpha} Q_{,\alpha} = 0$. Introducing the magnetic potential $\psi = A_{\varphi}$, which can be shown to satisfy $B^{\alpha} \psi_{,\alpha} = 0$, 
we express the flow constants in a more familiar form using $\psi$. For circular spacetimes, it is always possible to introduce a zero-angular-momentum observer such that 
the poloidal field admits a Chandrasekhar-like description with $B^{r} \propto \psi_{,\theta}$ and $B^{\theta} \propto \psi_{,r}$ up to metric factors \cite{boq95}. 
Although the decomposition is not as simple in general, we still have that $B^{i} \propto A_{\varphi}$ by definition and thus $B^i = \mathcal{O}(\psi)$ to leading order since 
$u^{i} A_{t,i} = 0$ from the ideal MHD condition $u^{\nu} F_{t \nu} = 0$ (see also Sec. 2.1.2 in C08).

Adopting the notation of IS03 and IS04, we have the following relations amongst the five flow constants $\cC (\psi), \Omega (\psi), E(\psi), L(\psi),$ and $D(\psi)$:
\be
E  = - \left ( \mu + \frac{B^2}{4\pi n} \right ) u_t - \frac{\cC}{4\pi} ( u_t + \Omega u_\varphi ) B_t,
\label{eqE1}
\ee
\be
L  =  \left ( \mu + \frac{B^2}{4\pi n} \right ) u_\varphi + \frac{\cC}{4\pi} ( u_t + \Omega u_\varphi ) B_\varphi,
\label{eqL1}
\ee
\be
D  = E - \Omega L,
\label{eqD}
\ee
where $\mu$ is the specific enthalpy (this is defined in Appendix~\ref{sec:appa} and should not be confused with the chemical potential unless the temperature is everywhere zero).  
Some of these constants can be used to express the magnetic field $B^\mu$ and meridional flow $u^i$ ($i=\{r,\theta\}$) via
\be
B^\mu = - \cC n  \left [ \, ( u_t + \Omega u_\varphi ) u^\mu + \delta_t^\mu + \Omega \delta^\mu_\varphi  \, \right ],
\label{eqBmu}
\ee
and
\be
 B^2 = - \cC n  ( B_t + \Omega B_\varphi ).
 \label{eqB2}
\ee
Using~\eqref{eqB2} in Eqs.~\eqref{eqE1} and \eqref{eqL1} one finds the alternative expressions
\begin{align}
E &= - \mu  u_t + \frac{\cC \Omega }{4\pi} ( u_t B_\varphi   -u_\varphi B_t  ) ,
\label{eqE2}
\\ \nn
\\
L &=   \mu u_\varphi + \frac{\cC}{4\pi}  ( u_t B_\varphi   -u_\varphi B_t  ) .
\label{eqL2}
\end{align}
It is illuminating to decompose~\eqref{eqBmu} into components, viz.
\begin{align}
B^i &= - \cC n ( u_t + \Omega u_\varphi ) u^i,
\label{eqBi1}
\\
\nn \\
B^\varphi &= - \cC n \left [  \Omega +  ( u_t + \Omega u_\varphi ) u^\varphi   \right ],
\label{eqBphi1}
\\
\nn \\
B^t &= - \cC n \left [ 1+   ( u_t + \Omega u_\varphi ) u^t   \right ].
\label{eqBt1}
\end{align}
Following IS04 and BO79, these constants can be interpreted as follows. $E(\psi)$ and $L(\psi)$ represent the total enthalpy of the plasma and specific enthalpic 
angular momentum of the magnetic field, respectively; 
 $D(\psi)$ symbolises the specific plasma energy according to a corotating observer and $\Omega(\psi)$ relates to the angular 
 velocity\footnote{Note that what is called $\Omega$ here generally differs from the actual angular velocity $u^{\varphi}/u^{t}$, with agreement only in the 
 limit that the toroidal field vanishes; see Eq.~(4.42) of Ref.~\cite{gour11}, noting that what we call $\Omega$ is denoted there by $\omega$ and $-A$ by BO. 
 Because of its hybrid nature, $\Omega$ could also be considered `induction-like'.} of the fluid. 
These constants can be thought of as `Bernoulli-like', in the sense that there is a simple but non-trivial hydrodynamic limit; the constant 
$D$ in particular can be directly identified with the Bernoulli integral (see Appendices). 
The remaining constant $\cC(\psi)$ -- representing the magnetic  field strength relative to the magnitude of meridional flow -- is instead `induction-like'. 
The existence of this constant is a direct consequence of the ideal assumption $E_{\mu} = 0$ and becomes trivial in the non-magnetic limit. 
Moreover, expression \eqref{eqBi1} shows that in the limit of vanishing meridional flow, $u^{i} \rightarrow 0$, the poloidal field generally 
vanishes unless $\cC \rightarrow \infty$.


\subsection{The Newtonian flow constants}
\label{sec:BOnewton}

Flow constants analogous to the ones found by BO78 in the context of GRMHD equilibria are known to exist in Newtonian axisymmetric-stationary systems. 
These are discussed, for instance, in Mestel's textbook \cite{mestelbook} but their origins can be traced back to older work~\cite{p56, c56, w59}. 
Here we summarise the expressions relating these constants with the various hydromagnetic parameters (adopting standard vectorial notation); 
a detailed derivation of these expressions can be found in Appendix~\ref{sec:appNewton}.

The defining property of a Newtonian flow constant $Q$ is $\bv \cdot \bnabla Q=0$. It is easy to show (see Appendix~\ref{sec:appNewton}) that the same flow constant is 
a function $Q(\psi)$, where the magnetic scalar potential $\psi$ is defined via the poloidal field component
\be
\bB_{\rm p}= \bnabla \psi \times \bnabla \varphi.
\ee
As in GR, the Newtonian constants can be grouped according to their `equation of origin', namely the ideal induction [Eq.~\eqref{inductNewt}] or the
Euler equation [Eq.~\eqref{EulerNewt}]. In the first group we find the flow constants $\cC_\rN$ and $\Omega_\rN$ which, as their notation suggests, are analogous to the GR ones 
$\cC$ and $\Omega$. These appear in the following relation between the magnetic field and fluid velocity
\be
\bB =  \cC_\rN \rho  \left( \bv - \varpi \Omega_\rN \bphi \right ),
\ee
where $\varpi = r \sin\theta$. In particular, the poloidal/meridional component of this expression is
\be 
\bB_{\rm p} = \cC_\rN \rho \bv_{\rm p}.
\ee
In the second group we find the following three Bernoulli-like flow constants which result by projecting the Euler equation along the $\bv$ and $\bB$ vectors
\begin{align}
E_\rN &=  \frac{1}{2}  v^2  +  \Phi + h  -  \frac{\Omega_\rN \cC_\rN}{4\pi} \varpi  B_\varphi,
 \\
L_\rN &= \varpi  \left (v_\varphi - \frac{\cC_\rN}{4\pi} B_\varphi \right ),
\\
D_\rN &=\frac{1}{2} v^2 + \Phi + h -\varpi \Omega_\rN v_\varphi,
\end{align}
where $h$ is the fluid enthalpy and $\Phi$ is the gravitational potential. 


\section{Corollaries of the flow constant formalism}
\label{sec:corollaries}

In this section, we make use of the flow constant formalism to establish some corollaries 
regarding the possible set of magnetic equilibria.

\subsection{The `no toroidal field theorem' and necessity of meridional flow}
\label{sec:notor}

  One of the most surprising consequences of the flow constant MHD formalism is the theorem derived in BO79 under a circular spacetime 
assumption (see their Sec. VI for more context):  ``the absence of meridional circulation implies the vanishing of the toriodal [sic] magnetic field.''

The validity of this theorem hinges on the simultaneous presence of the meridional flow and the flow constant $\cC$ in the `induction-originated' equation~\eqref{eqBmu}. 
Indeed, setting $u^i=0$ in~\eqref{eqBi1}, and in combination with a non-vanishing poloidal field $B^i \neq 0 $, leads to the requirement $\cC \to \infty$.
This divergence would cause the hydrodynamical constants $E,L$ to diverge too, unless [see Eqs.~\eqref{eqE1}, \eqref{eqL1}]
\begin{align}
B_t &= g_{tt} B^t + g_{t\varphi} B^\varphi  + g_{ti} B^i =0,
\\
B_\varphi &= g_{\varphi\varphi} B^\varphi + g_{t\varphi} B^t + g_{\varphi i} B^i =0.
\end{align}
The determinant of this linear system is non-vanishing in the absence of horizons,
and as a consequence the unique solution is the trivial one, $B^t = B^\varphi = 0$.

As an additional intuitive argument supporting the theorem's conclusion, consider how the toroidal field appears to a locally inertial observer.
The relevant tetrad component reads
$ B_{(\varphi)} \propto L(\psi)/\cC(\psi)$
up to metric factors (e.g. \cite{is04,c09}; see Sec.~\ref{sec:workedex}). Since $\cC \to \infty$ in the $u^{i} \to 0$ limit, keeping the above non-zero requires 
that $L \to \infty$ at an equally fast rate 
if the observer is to witness a toroidal field. This is clearly incompatible with the hydrodynamic limit, and thus we are forced to conclude that the toroidal field must vanish to maintain consistency.

Therefore a stationary-axisymmetric MHD equilibrium without meridional flow must be purely poloidal as well as purely spacelike. 
As far as astrophysically relevant MHD equilibria are concerned this is not an acceptable situation as it is well known that purely poloidal fields are generically 
unstable (e.g. \cite{lasky11,rezz12}). This means that a stable mixed poloidal-toroidal configuration should be accompanied by some amount of meridional flow. 

As an aside we note that a vanishing meridional flow is still compatible with a purely toroidal field (which is also
generically unstable \cite{fr12,kiu12}). Under these circumstances, setting $u^i = B^i =0$ in~\eqref{eqBi1} does not imply a divergent flow constant $\cC$. 
Moreover, the vanishing poloidal field implies $F_{r\varphi} = F_{tr} =0$ and again $\Omega \neq u^\varphi / u^t$ (see Appendix~\ref{sec:appa}). 
The two non-vanishing field components then read $B^t = \cC n  ( u^\varphi  -\Omega u^t ) u_\varphi$ and $B^\varphi =  - \cC n  ( u^\varphi  - \Omega u^t )  u_t$; {these expressions can be compared with those found in Ref.~\cite{kiu08}, which studied nonlinear toroidal equilibria.}


\subsection{The Grad-Shafranov equation}
\label{sec:GS}

Without presenting details, which are carefully laid out by IS03, the Euler equation can be shown to lead to the GS 
equation\footnote{The following conversion rule should be applied when adapting formulae from IS to our notation:  
$\mu_{\rm IS} = \mu/m$,  $Q_{\rm IS} = Q/m$ where $Q = \{\cC,L,E,D\}$.}
\be
\ba
0 =& J^\varphi - \Omega J^t + \frac{1}{\cC \sqrt{-g}} \left [\,  ( \mu u_r )_{,\theta} - ( \mu u_\theta )_{,r} \, \right ] \\
&+ n u^\varphi \left  ( L^\p - \Lambda \cC^\p \right )  - n u^t \left [ \,  E^\p - \Lambda ( \cC \Omega )^\p \, \right ],  
 \label{eqGS1}
\ea
\ee
where
\be
\Lambda =  \frac{1}{4\pi}  ( u_t B_\varphi   -u_\varphi B_t  ) 
\ee
and $J^{\mu} = \nabla_\nu F^{\mu \nu} /4 \pi$ is the four-current. Many studies of equilibria concern themselves only with Eq.~\eqref{eqGS1}, discarding other information 
under the assumption that the other ideal MHD equations are trivial. This is not really the case, however, as the number of free functions in~\eqref{eqGS1} make conceptually 
obvious. As pointed out in Sec.~\ref{sec:BO}, the flow constant $\cC$ arises directly from the induction equation and thus key physical points are 
missed if one considers~\eqref{eqGS1} in isolation (see also Sec.~\ref{sec:induction}). One such point was described in Sec.~\ref{sec:notor}, namely 
that setting the meridional flow to zero but keeping a mixed poloidal-toroidal field is inconsistent, even though such a choice leaves \eqref{eqGS1} well behaved. 
Another concerns the nature of \emph{Alfv{\'e}n points} (i.e. where the magnetic energy density matches the kinetic energy density of the flow) and the stellar surface, which 
we discuss in Sec. \ref{sec:alfvenpoints}.


\subsection{The role of the induction equation}
\label{sec:induction}

Although we work within a relativistic framework in this paper, many of the issues we discuss carry over at a Newtonian level. 
This is exemplified by the presence of BO-like flow constants in Newtonian systems (Sec.~\ref{sec:BOnewton}).
Owing to the complexity of multifluid dynamics in GR, especially when convective motions and mixed magnetic fields are permitted, we opt to describe some physical features related 
to induction here in a Newtonian language. In particular, the fact that `induction-like' flow constants are only associated with the electron fluid while the others are associated 
with the bulk applies to both frameworks. This bares on the physical interpretation of various limits, as discussed throughout, some of which have been addressed in detail by Prix \cite{prix05}.

Our objective in this section, therefore, is to provide a careful rederivation of the induction-originated $\cC_\rN$ and $\Omega_\rN$ flow constants 
by accounting for the distinct proton and electron fluid velocities. These are denoted, respectively, as $\bv_\rc$ and $\bv_\re$.  The induction equation itself 
can be derived from Faraday's law, $\bnabla \times \mathbf{E} = - \partial_t \bB /c$, and the Euler equation for the electron fluid with the inertial terms omitted, i.e.
\be
\mathbf{E} = -\frac{1}{c} \bv_\re \times \bB + \bnabla ( \mu_\re + m_\re \Phi ).
\ee
The resulting equation is
\be
\partial_t \bB = \bnabla \times (\bv_\re \times \bB ).
\label{inductionN1}
\ee
The manipulations of the `single-fluid' induction equation described in Appendix~\ref{sec:appNewton} can be exactly repeated if we change 
$\bv \to \bv_\re$ and $\rho \to \rho_\re$. In terms of the poloidal/meridional and toroidal/azimuthal components of the vectors $\bB, \bv$,
\be
 \bB = \bB_{\rm p} + B_\varphi \bphi, \qquad \bv_\re = \bv_{\rm ep} + v_{ e \varphi} \bphi, 
\ee
we find
\be
\bB_\rp = \cC_\rN \rho_\re \bv_{\rm e p},
\label{inductionN2}
\ee
where $\cC_\rN$ is constant along poloidal field lines, i.e.
\be
\bB_\rp \cdot \bnabla \cC_\rN = 0.
\ee
In axisymmetry this is equivalent to
\be
\bv_\re \cdot \bnabla \cC_\rN = 0,
\ee
which shows that $\cC_\rN$ is also conserved with respect to the electron flow. 

The poloidal field can be written in a divergence-free form in terms of the scalar potential $\psi (r,\theta)$,
\be
\bB_\rp = \bnabla \psi \times \bnabla \varphi.
\label{psi_def}
\ee
In this parametrisation the poloidal field lines lie in constant $\psi$ surfaces and therefore the same must be true for 
the flow lines of $\bv_{\rm ep}$. As a consequence, $\cC_\rN = \cC_\rN (\psi)$.

Further manipulation of Eq.~\eqref{inductionN1} leads to the flow constant $\Omega_\rN (\psi)$ as part of 
an expression  for the toroidal field
\be
B_\varphi = \cC_\rN \rho_\re  \left( v_\varphi^\re - \varpi \Omega_\rN \right ).
\ee
In combination with~\eqref{inductionN2}, this allows us to write the Newtonian counterpart of Eq.~\eqref{eqBmu} as
\be
\bB =  \cC_\rN \rho_\re  \left( \bv_\re - \varpi \Omega_\rN \bphi \right ).
\label{inductionN3}
\ee
The final step consists in expressing $\bv_\re$ in terms of the total current $\bJ$ and the proton velocity (which in non-superfluid matter can 
be taken to coincide with the neutron fluid velocity). Making use of the assumed charge neutrality of the system, $n_\re = n_\rc$, 
we have
\be
\bv_\re = \bv_\rc - \bJ / n_\re,
\label{eqJ}
\ee 
which then implies that
\be \label{eq:bbcurrent}
\bB =  \cC_\rN \rho_\re  \left( \bv_\rc - \varpi \Omega_\rN \bphi \right ) - \cC_\rN m_\re \bJ .
\ee
Assuming $n_\re =0$ at the surface, this expression can easily accommodate a non-vanishing magnetic field as long as $\bJ \neq 0$, thus
alleviating any pathological behaviour at the surface (see related discussion in the following section). In practice, this may not be an issue 
since astrophysical neutron stars are endowed with a non-vacuum magnetosphere  in which $n_\re$ is non-vanishing~\cite{gj69}. Eq.~\eqref{eq:bbcurrent} may also 
invalidate the BO result applying in the limit $\bv_{\rm c p} \to 0$ (see the final paragraph of Appendix~\ref{sec:appNewton}), as $\cC_{\rN} \to \infty$ is no longer necessary 
to maintain $\bB_{p} \neq 0$ if the poloidal current $\bJ_{\rm p}$ does not vanish. 

As a final point, although entropy gradients have been ignored here, one generally has $\bv \cdot \nabla  s = 0$ in the single fluid case, meaning that $s = s(\psi)$. \citet{yosh12} argue that this forces a vanishing meridional flow else convectively unstable regions with $\nabla p \cdot \nabla s < 0$ exist in a realistic, stratified star. 
This is clearly in conflict with the BO conclusion, though the above considerations, which naturally carry over to GR, may remedy the situation.

\subsection{The stellar surface and other singularities}
\label{sec:alfvenpoints}

In the context of twisted-torus models where one considers the GS equation in isolation, the vanishing density at the stellar surface generally poses no problem to regularity. 
In fact, in the exterior of a static star where $n=0$ and there is a purely poloidal `test' field, Eq.~\eqref{eqGS1} decouples with respect to the angular coordinates and admits an 
analytical solution in terms of hypergeometric functions \cite{konno99}. Although rotating stars will not be surrounded by vacuum in reality \cite{gj69}, the vacuum GS equation 
necessarily describes a \emph{force-free} magnetosphere (see Ref.~\cite{pet16} for a discussion of `force-free' in the GR context).

When considering non-zero velocity fields, the limit $n \to 0$ needs special attention as a result of equations  \eqref{eqBmu} and \eqref{eqB2}.
The two subequations which are most relevant here are written again for convenience, viz.
\begin{align}
 B^i &= - \cC n \left( u_t + \Omega u_\varphi \right) u^i, \label{eqBi2}
 \\
 B^\varphi &= - \cC n \left [  \Omega +  ( u_t + \Omega u_\varphi ) u^\varphi   \right ].
 \label{eqBph2}
\end{align}
The toroidal field can be assumed to be confined in the stellar interior, rendering 
Eq.~\eqref{eqBph2} trivial at the surface and beyond, without placing any restriction on $\cC$.
The poloidal equations are clearly problematic though at $n=0$, unless $\cC \to \infty$ for every $\psi$
curve that crosses the surface or one carefully treats the multi-fluid dynamics, as in Sec.~\ref{sec:induction}. This and a related problem can be described in terms 
of the poloidal Alfv{\'e}n Mach number\footnote{It is somewhat unintuitive that $M_{\rm A}$ does not depend explicitly on the magnetic field, though this is because 
the poloidal flow scales with the poloidal magnetic field, as evident from expression \eqref{eqBi2}.},
\be 
\label{eq:malf}
M^2_{\rm A} = -\frac{4 \pi \mu} {\cC^2 n \left(g_{tt} +2 \Omega g_{t \phi} + \Omega^2 g_{\phi \phi} \right)}.
\ee
From Eqs.~\eqref{eqL1} and \eqref{eqB2}, it can be shown that $u_{\varphi}$ tends to diverge as $M_{\rm A} \to 1$ (i.e. at the Alfv{\'e}n surface) unless $L$ and $\Omega$ are 
proportional to each other there \citep{is04}. This problem is discussed in the Newtonian context by Mestel \cite{mest68} and Ogilvie \cite{og16} (see also section VI~C in Ref.~\cite{gour11}). 
Unfortunately, this does not really help reduce the pool of flow constants since proportionality is only required in a particular limit. The more obvious problem is that for a finite value 
of $M_{\rm A}$ to be maintained one requires $\cC^2 n > 0$ everywhere, which would seem to demand $\cC \sim \psi^{-1}$ with a field confined to the interior such that $\psi \to 0$ 
towards the surface (see also Sec.~\ref{sec:perturbative}). Although this choice was the one adopted in IS04, it is clearly unrealistic.

There are several possibilities worth considering in order to avoid this conundrum:
\begin{itemize}

\item[(i)]{The ideal MHD approximation (upon which  the BO relations are based) breaks down in the limit $n \to 0$. 
This is somewhat expected as, in reality, it is meaningless to discuss a current or velocity gradient without matter 
(e.g., manipulations of the ideal MHD assumption $E_{\mu} = F_{\mu \nu} u^{\nu} = 0$ become dubious; see Appendix \ref{sec:appa}).}

\item[(ii)]{A non-relaxed crust alters (or invalidates) some of the BO relations. That is, the term $T^{\mu \nu}_{\rm shear}$, defined through 
the Carter-Quintana equations \cite{cq72}, may also enter into the total stress-energy tensor. This term will naturally adjust the boundary conditions 
and flow constant structure (see Ref.~\cite{koj22} for some results in the Newtonian context). The presence of an ocean further complicates the dynamics.}

\item[(iii)]{The function $\cC(\psi)$ behaves in such a way that it tends to infinity everywhere except in an equatorial torus. Suppose that $\psi$ attains a 
maximum $\psi_{c}$ at the stellar surface, and we write $\cC = \tcC / \left[ \psi (\psi - \psi_{c}) \Theta(\psi - \psi_{c}) \right] $ for constant $\tcC$
with the meridional flow confined to the region $\psi > \psi_{c}$. Outside of this torus (e.g. at the surface) we have $\cC \to \infty$ but $u^{i} \to 0$ and so $B^{i}$ 
from Eq.~\eqref{eqBi2} can be non-zero even if $n \to 0$ there, while inside both $u^{i}$ and $\cC$ are finite with $n > 0$ and so $B^{i}$ is well behaved there also. 
Although there are mathematical issues with this approach related to the undefined limit of the Heaviside function as $\psi \to \psi_{c}$, one has to bear in mind
that the use of discontinuous functions is a convenient way of modelling sharp yet continuous gradients of the real physical system.}

\item[(iv)]{The neutron star is unlikely to be surrounded by vacuum in reality as mentioned previously, arguably rendering this issue moot. Care must be taken in 
this case to choose appropriate boundary conditions, for instance using the Goldreich-Julian number density $n_{\rm GJ} \sim B_{\mu} u^{\mu}$ \cite{gj69}. 
If $n \to 0$ anywhere in the exterior universe however, this problem reappears at that interface. Note this is related to point (i), as the ideal MHD approximation itself 
may not be valid in a tenuous magnetosphere.}

\end{itemize} 


\section{Perturbative equilibria}
\label{sec:perturbative}

This section considers perturbative MHD equilibria, where  `weak' (see below) magnetic fields are superimposed as perturbations on a non-magnetic background. 
For most of what follows, the background star is assumed to be static; the rotating case is discussed separately in Sec. \ref{sec:weakB2}. 
The hydrostatic metric we use takes the form 
\begin{align}
ds^2 & = g_{\mu\nu}^{\rm TOV} dx^\mu dx^\nu 
\nn
\\
&= - e^{2 \nu} dt^2 + e^{2 \lambda} dr^2 + r^2 \left(d \theta^2 + \sin^2\theta d \varphi^2 \right),
\end{align}
where we use the superscript `TOV' to indicate that, at a background level, we work with a static and spherically symmetric star whose 
geometry is described by the TOV equations.

\subsection{General remarks}

The following sections consider the leading-order behaviour of the flow constants, essentially through a power-counting analysis. 
A similar procedure was undertaken by IS04, however they assumed that rotation is always sub-leading and that $\psi \to 0$ at 
the stellar surface (cf. Sec. \ref{sec:alfvenpoints}). One goal of this section is to reexamine their results when these conditions are relaxed.

Firstly, it is instructive to define \emph{perturbative} when considering an expansion in terms of magnetic parameters. If, as is typical in numerical GR investigations, 
we further consider `units' of length such that the star has unit mass, $M_{\star} = 1$, the basic measure of magnetic flux becomes the dimensionless number 
$[\text{G cm}^{2}] = 1.39 \times 10^{-30} (1.4 M_{\odot}/\tilde{M})$ for physical-units mass $\tilde{M}$. Given that the magnetic flux is related to the local field 
strength through $\psi \approx B \Rs^2$, where $\Rs$ denotes the stellar radius, we see that expansions in powers of $\psi$ are well behaved ($\psi \ll 1$) provided that 
$(B/\text{G}) (\Rs/\text{cm})^2 \times 1.39 \times 10^{-30} (1.4 M_{\odot}/\tilde{M})  \ll 1$, which requires $B \ll 7.2 \times 10^{17}\,$G for  $R_{\star} = 10^{6}\,$cm 
and $\tilde{M} = 1.4 M_{\odot}$. This same limit can be independently derived by demanding that the magnetic pressure at the center of the star is much less than 
the minimum central hydrostatic pressure. This limit is physically equivalent to a magnetic-to-gravitational-binding energy ratio much below unity.

Although we consider expansions in $\psi$, it is not obvious that this is the most natural approach when considering a rotating star. In reality, most if not all 
neutron stars are both `slow' (rotating much slower than the break-up limit) and `weakly magnetised' (as defined above), and it may be more appropriate to 
consider a double expansion in both $u^{\varphi}$ and $\psi$ simultaneously in the form of a magnetic-Hartle-Thorne scheme (see Ref.~\cite{steil18} 
for such an approach). We ignore such complications here (though see Footnote 5 below).


\subsection{The magnetic field as a perturbation of a non-rotating star}
\label{sec:weakB1}

Here we treat the magnetic field as a perturbation on a static star, with the magnetic stream 
function $\psi$ used as a perturbative bookkeeping parameter. We assume the following scaling for the magnetic field components
\be \label{eq:bscalings}
B^i, B^\varphi = {\cal O} (\psi),  \quad B^t = {\cal O} (\psi^\gamma ),
\ee
for some $\gamma \geq 1$. To be more precise, the scaling $B^i =  {\cal O} (\psi) $ is a direct consequence of the definition of $\psi= A_\varphi$ (see Sec. \ref{sec:BO}) and we impose 
a comparable toroidal field to maintain linearity (cf. Sec. \ref{sec:previous}).
This `3+1' scaling accommodates the presence of a mixed poloidal-toroidal field to leading order,
while also taking into account the post-Newtonian character of $B^t$.

Metric perturbations appear first at ${\cal O} (\psi^2)$ since $T^{\alpha \beta}_{\rm EM} \sim B^2$ \cite{is03}, i.e. 
\be
g_{\mu\nu} = g_{\mu\nu}^{\rm TOV} + {\cal O} (\psi^2).
\ee
As discussed earlier, a magnetic field may induce fluid motion in an otherwise static configuration. We suppose that azimuthal flow appears at some sub-leading order,
\be \label{eq:omegascaling}
u^\varphi = {\cal O} (\psi^\eta), \qquad \Omega = {\cal O} (\psi^\eta),
\ee
for $\eta \geq 1$, together with  $u^t = {\cal O} (1)$ set by the wind equation $u^{t} u_{t} = -1$ at leading order. 
The meridional flow scaling is taken to be 
\be
u^i =  {\cal O} (\psi^\beta),
\ee
for some $\beta \geq \eta$. The flow constants $E, L$, and $D$ are hydrodynamical. That is, they are present even as $\psi \to 0$. 
We Taylor-expand these through
\begin{align} 
E & = E_0 + E_1 \psi + E_{2} \psi^2 + {\cal O} (\psi^3),  \label{eq:eeqn}
\\
L & = L_0 + L_1 \psi + L_{2} \psi^2 +  {\cal O} (\psi^3), 
\end{align}
where $E_0$ and $L_0$ are non-magnetic parameters, and the expansion for $D$ follows from Eq.~\eqref{eqD}. Last but not least, the flow constant  $\cC$ is assumed to scale as
\be
\cC = {\cal O} ( \psi^\alpha ),
\ee
where $\alpha \neq 0$, since induction-like constants should not enter at background order. The above scalings are not independent; inserting them in 
Eqs.~\eqref{eqBi1}, \eqref{eqBphi1}, \eqref{eqBt1} and~\eqref{eqB2}, we obtain
\be
\psi \sim n u_t \psi^{\alpha+\beta},
\ee
\be
\psi  \sim  n u_t u^\varphi \psi^\alpha \sim n \, \psi^{\alpha +\eta},
\ee
\be
\psi^\gamma  \sim n  ( \Omega u^t - u^\varphi ) \psi^{\alpha+\eta},
\label{eqBtscale1}
\ee
and
\be
\psi^2  \sim n ( B_t + \Omega B_\varphi ) \psi^\alpha \sim n  \psi^{\alpha + k}, \, k = \mbox{min} [\gamma, \eta+1],
\label{eqB2scale1}
\ee
respectively, where the symbol $\sim$ is used to relate parameters with a non-trivial $\psi$-scaling. The balance of $\psi$-powers in the first two expressions 
leads to
\be \label{eq:alphascal}
\alpha = 1 - \eta, \qquad \beta = \eta.
\ee
Subsequently, Eqs.~\eqref{eqBtscale1} and \eqref{eqB2scale1} lead to
\be
\gamma \geq 1+ \eta, \qquad \Omega u^t - u^\varphi  = {\cal O} ( \psi^{\gamma-1} )  \lesssim {\cal O} (\psi^\eta).
\ee

The resulting scalings $\cC \sim \psi^{-1}$ and $u^i, u^\varphi \sim \psi^2 $ are consistent with the analysis of IS04 if we set $\eta = 2$. 
More generally, however, any $\eta > 1$ with $\alpha = 1 - \eta <0$ is viable, though demanding integer solutions to permit (Laurent/Taylor) 
expansions forces $\eta = 2$ and $\alpha = -1$.
The remaining `hydrodynamical' Eqs.~\eqref{eqE2}, \eqref{eqL2} are compatible with them and offer no new information about
the scalings. It is also easy to verify that 
\be
g_{tt}^{\rm TOV} ( u^t )^2  = -1 + {\cal O} (\psi^2 ).
\ee
One last remark concerns the non-magnetic ($\psi \to 0$) limit of the MHD equations. The balance of $\psi$-powers implemented here
means that all of them should automatically behave smoothly in that limit (for example, the above purely magnetic equations reduce to a trivial $0=0$). 

For a static star, the above analysis thus determines the leading-order behaviour of the flow constants irrespective of the true system's initial conditions.


\subsection{The magnetic field as a perturbation of a rotating star}
\label{sec:weakB2}

Much of the  previous section's results depend on the crucial assumption of a non-rotating star in the limit $\psi \to 0$.
More astrophysically relevant is the scenario where the star is \emph{rigidly} rotating in the non-magnetic limit and $g_{t\varphi} \neq 0$ (though cf. Footnote 5 below). 
In such a case,
\be
u^\varphi = {\cal O} (1), \qquad \Omega = {\cal O} (1).
\ee
As rigid rotation may preclude comparable energy partitions, we instead allow for a more general scaling of the toroidal field through
\be
B^\varphi = {\cal O} (\psi^\lambda),
\ee
together with\footnote{Note that if $A_{t} \sim u^{i}$ and $u^{\varphi} = \mathcal{O}(1)$, the arguments outlined at the beginning of Sec. \ref{sec:BO} 
would appear to imply $B^i = {\cal O } (\psi^\chi)$ with $\chi = \text{min}[\beta,1]$. This, however, forces $\lambda = 0$ if $\beta \leq 1$ which leads to a contradiction. 
Physically speaking though, this simply echoes the sentiment that it does not really make sense to talk about a non-magnetic star with circulating charged particles 
(i.e. the `background' does not truly exist). On the other hand, if the fluid is composed only of uncharged matter, then the flow constant $\cC$ does not 
exist; cf. Sec. \ref{sec:induction}.} $B^i = {\cal O } (\psi)$.  As before, we assume
\be
u^i =  {\cal O} (\psi^\beta), \qquad \cC = {\cal O} ( \psi^\alpha).
\ee
Upon inserting these in Eqs.~\eqref{eqBi1}, \eqref{eqBphi1}, and \eqref{eqBt1} we find
\begin{align}
\psi  &\sim  n u_t \psi^{\alpha+\beta},
\\ \nn
\\
\psi^\lambda  &\sim  n u_t u^\varphi \psi^\alpha \sim n \psi^{\alpha},
\\ \nn
\\
\psi^\gamma  &\sim n( \Omega u^t - u^\varphi ) \psi^\alpha,
\end{align}
from which we easily deduce that
\be
\alpha = \lambda, \qquad   \alpha + \beta =1, \qquad \gamma = \alpha.
\ee
If we have comparable poloidal and toroidal field strengths, then we find that $\alpha = 1,~\beta = 0$. That is,
\be
u^i =  {\cal O} (1), \qquad \cC = {\cal O} ( \psi),
\ee
for $\lambda =1$. These scalings could be preempted from Eq.~\eqref{eq:alphascal} with $\eta = 0$, though are markedly different to those obtained for a 
non-rotating star in Sec.~\ref{sec:weakB1}. In particular, a leading-order meridional flow is an uncomfortable result as it implies a strong deviation from rigid
rotation in the non-magnetic limit.  This impasse prompts us to explore non-standard scalings for the toroidal field component, in combination with a sub-leading 
and non-integer-scaling meridional flow, $\beta > 0$. This requirement gives $ \lambda = 1 -\beta < 1$, which entails a toroidally-dominated MHD equilibrium unless 
$\psi$ itself is non-perturbative (i.e. $\psi^\lambda > \psi$ for $0 < \psi,\lambda < 1$). Some discussion on this point is given in Sec. \ref{sec:conclusions}.


\subsection{Linearised Grad-Shafranov equation}
\label{sec:lingssec}

Here we consider a perturbative and integer expansion of the GS equation \eqref{eqGS1} without $\mathcal{O}(1)$ rotation. 
Based on the previously obtained scalings and writing 
\be
\cC = {\tcC}/{\psi},
\ee
where $\tcC$ is a constant parameter, we have
\be \label{eq:lambdaapprox}
J^\varphi \sim {\cal O} (\psi), \,\, \Omega J^t  \lesssim  {\cal O} (\psi^2), \,\, \Lambda = \frac{1}{4\pi} u_t B_\varphi + {\cal O} (\psi^5 ),
\ee
\be
\label{eq:meridscaling}
\frac{1}{\cC \sqrt{-g}} \left [\,  ( \mu u_r )_{,\theta} - ( \mu u_\theta )_{,r} \, \right ] \sim \frac{\psi^{\beta-\alpha}}{\tilde{\cC}} \sim {\cal O} ( \psi^3),
\ee
\be
\ba
n u^\varphi \left  ( L^\p - \Lambda \cC^\p \right ) &= n u^\varphi \left ( L_1 + 2 L_2 \psi - \alpha \Lambda \tilde{\cC} \psi^{\alpha-1}  \right ) \\
&=  n \tilde{\cC} u^\varphi  \frac{\Lambda}{\psi^2} + {\cal O} (\psi^2 ),
 \ea
\ee
and
\be
\ba
 - n u^t \left [  E^\p - \Lambda ( \cC \Omega )^\p \right ] &= -n u^t \Big[ E_1 + 2 E_2 \psi  \\
 & - \frac{\Lambda  \tilde{\cC}}{\psi} \left ( \Omega^\p - \Omega \psi^{-1} \right )  \Big]  + {\cal O} (\psi^2 ).
 \ea
\ee
The above imply that \eqref{eqGS1} takes the form
\be
\mathcal{O}(\psi^2) = J^\varphi  +  n \tilde{\cC}  \frac{ u^\varphi  \Lambda}{\psi^2}  - n u^t \left [ E_1 + 2 E_2 \psi  - \Omega_{2} \Lambda  \tilde{\cC}  \right ],
\ee
with $\Omega(\psi) = \Omega_{2} \psi^2$.
Upon inserting the approximate $\Lambda$ from Eq.~\eqref{eq:lambdaapprox}, one finally obtains the $\mathcal{O}(\psi)$ GS equation
\be \label{eq:lings1}
\ba
0 =& J^\varphi  + \frac{n}{4\pi} \Big\{ \tilde{\cC}  \frac{ u^\varphi  u_t B_\varphi }{\psi^2} \\
&- u^t \left[  4\pi \left(  E_1 + 2 E_2 \psi  \right) - \Omega_{2} u_t B_\varphi \tilde{\cC}  \right]  \Big\}.
\ea
\ee
By using expression \eqref{eqL2}, one can also rewrite this equation using the flow constant $L$ (modulo some factors) in exchange for $u_{t} B_{\varphi}$; see Sec. \ref{sec:workedex}. 

The first interesting aspect of Eq.~\eqref{eq:lings1} we point out is that the term $\sim n u^t E_1 $ is ${\cal O} (1)$, and therefore we should expect $E_1 = 0$. 
That is to say, a strict power-counting argument appears to preclude the possibility of $E_{1} \neq 0$, else in the non-magnetic limit \eqref{eq:lings1} unphysically 
implies that $0 = n u^{t}$. The choice $E_{1} \neq 0$ is however made in the literature (where $E_{i+1}$ are usually rebranded as $c_{i}$), as discussed in the next section. 

A second point concerns the explicit appearance of $\Omega$, which is absent in the linearised equation presented by IS04 [see their Eq.~(64)]. We believe this is because these 
authors implicitly assume $\eta >2$ rather than $\eta=2$ as written in their Eq.~(58), meaning that all $\Omega$ terms are ignored as being higher order. [Note also the typo in their 
Eq.~(60) in the expansion of $D(\psi)$.]

\subsection{Some remarks on previous literature}
\label{sec:previous}

In this section, we discuss the perturbative GS equation and compare various approaches towards it found in the literature. Ignoring $\Omega$ corrections at all orders, 
writing $E'(\psi) \propto F(\psi)$ and $L_{0}/\tcC \propto \zeta$, one can eventually write the \emph{linear} GS equation in coordinate form as
\be 
\label{eq:c08gs}
0 = J_\varphi - {\zeta^2} e^{-2\nu} \psi - r^2 \sin^2 \theta ( \epsilon + p) F(\psi),
\ee
with $F''(\psi) = 0$. We write Eq.~\eqref{eq:c08gs} in this way so as to match the $\mathcal{O}(\psi)$ equation presented by C08 [i.e. their Eq.~(57)]. 
Sometimes the parameter $\zeta$ is upgraded to a general function of $\psi$, making \eqref{eq:c08gs} non-linear (see below).

Strict linearisation of the GS equation limits the possible set of equilibria matching to an exterior poloidal electrovacuum, with the only options 
being (i) the entire field is confined within the star, (ii) $B^{\varphi}$ is discontinuous and there is a non-zero surface current, or (iii) $\zeta = 0$ and $B^{\varphi} = 0$ identically. 
To see this, note that if one matches to an exterior where $\psi \neq 0$ then $\zeta^2 \psi \neq 0$ at the surface unless $\zeta = 0$ 
everywhere or $\zeta(\Rs) = 0$ imposed by hand discontinuously. Option (i) is that selected by IS04 and others (e.g. \cite{hask08}), while (ii) is chosen by C08.

While neither of these options characterise a realistic system, they are the only choices available for a mixed poloidal-toroidal field at \emph{strictly linear} order. 
An alternative is to consider \emph{non-linear} choices for $\zeta^2 \psi$ and/or $F(\psi)$, allowing a toroidal field that decays regularly towards the boundary ({\`a} la twisted torus), 
such as in Refs.~\cite{c09,c10,c13}. However, if one were to take $\zeta \propto \psi$ (for instance), quantities explicitly related to $u^{i}$ should appear in the GS equation since 
meridional flow enters at $\mathcal{O}(\psi^3)$. Omitting these terms thus implies that power counting is not handled self-consistently, presenting a conceptual problem.
At quadratic order one could avoid the appearance of meridional terms (i.e. taking $\zeta \propto \sqrt{\psi}$), though this again likely rules out a regular toroidal component. Moreover, metric perturbations should be accounted for as these appear at $\mathcal{O}(\psi^2)$, though this is often circumvented with the Cowling approximation. 

A related issue surrounds $F(\psi)$; C08 and others (e.g. \cite{c09,c10,g12,c13}) Taylor-expand this function as
\be
F (\psi) = c_0 + c_1 \psi + {\cal O} (\psi^2),
\ee
though declare $c_0 = {\cal O} (B)$. In C08, the authors study the simpler system~\eqref{eq:c08gs} with $c_1=0$, which admits
decoupled multipoles. In C09, by contrast, both $c_{0}$ and $c_{1}$ are kept and the multipole expansion becomes non-trivial (cf. Section~\ref{sec:workedex}). 
In either case, however, it seems inconsistent to keep $c_{0} \neq 0$ since this would imply one does not recover $0 = 0$ in the non-magnetic limit, $\psi \rightarrow 0$.

The above two points show that there are subtle issues involved in the construction of magnetic equilibria, aside from those related to induction explored previously. 
In the next section we resolve the second of these points, meaning that we numerically build equilibria where $c_{1} \neq 0$ but $c_{0} = 0$, doing so in the flow-constant 
language. Solving the non-linear system self-consistently with a twisted-torus configuration, meridional flow, and metric backreactions is left to future work.

\section{Worked examples of mixed, multipolar MHD equilibria}
\label{sec:workedex}

To start our numerical investigation of the impact of discarding the $E_{1}$ coefficient while keeping $E_{2}$, it is convenient to express \eqref{eq:lings1} in component form. 
While we ignore rotational corrections at background order, it turns out that introducing $\Omega = \Omega_{2} \psi^2$ simply leads to a rescaling of the $E_{2}$ constant due 
to Eq.~\eqref{eqD}, and thus setting $\Omega_{2} = 0$ leads to no loss of generality in our approach. We also opt to replace the constant $E_{2}$ by $D_{2}$ through Eq.~\eqref{eqD} 
to avoid confusion, as $E_{\ell}$ is often used to represent the energy stored in an $\ell$-pole. One eventually finds
\be
\ba
\label{eq:gscomp}
&0 = e^{- \nu} \left[ \partial_{r} \left( e^{\nu - \lambda} \partial_{r} \psi \right) + \frac{e^{\nu + \lambda}} {r^2} \sin \theta \partial_{\theta} \left( \frac{\partial_{\theta} \psi} {\sin \theta} \right) \right]  \\
& + \left( \frac{4 \pi L_{0}} {\tilde{\mathcal{C}}} \right)^2 e^{\lambda - 2\nu} \psi + 8 \pi D_{2} r^2 \sin^2\theta e^{\lambda} \left( p + \epsilon \right) \psi,
\ea
\ee
which agrees with previous literature once conventions have been accounted for (e.g., the $tt$-component of the TOV metric in C09 is $e^{\nu}$ rather than $e^{2 \nu}$). 

In order to ensure that the toroidal field is confined to the stellar interior, while maintaining a self-consistently linearised\footnote{As discussed in Sec.~\ref{sec:previous}, 
one cannot guarantee that $B^{\varphi}$ decays smoothly as $r \to R_{\star}$ in a strictly $\mathcal{O}(\psi)$ setup. It is necessary that nonlinearities enter into Eq.~\eqref{eq:gscomp} via $L$ or $\cC$ 
to avoid this; the choice $L_{0} \to L_{0} (\psi - \psi_{c}) \Theta(|\psi/\psi_{c}|-1)$, similar to that made by C09 and Ref.~\cite{g12}, is one possibility. 
Strict linearisation thus leads to 
some unrealistic features, whether it be $\psi \rightarrow 0$ towards the boundary, like IS04 impose, or a jump discontinuity.} equation that is not plagued by singularities at the stellar 
surface, we impose a linear variant on choice (iii) described in Sec.~\ref{sec:alfvenpoints} in the form
\be 
\label{eq:lchoice}
\tcC \to \tcC / \Theta(|\psi/\psi_{c}|-1).
\ee

Independently of the choice of $\cC$ or $L$, we note that $D_{2} \neq 0$ prevents a decoupling of the multipolar components in Eq.~\eqref{eq:gscomp}. 
Indeed, the penultimate term behaves like $\sim \psi$ while the last behaves as $\sim D_{2} \psi \sin^2 \theta$, breaking the angular symmetry. 
The obvious choice $D_{2} = 0$ is not permitted since the boundary conditions cannot be satisfied (see below). Therefore, even the linear GS problem requires 
a multipolar solution. To make progress, we follow the method of C09 by projecting the equation onto a Legendre basis. We introduce a multipolar expansion
\be
\psi(r ,\theta) = \sum_{\ell=1}^{\ell_{\rm max}} f_{\ell}(r) \sin \theta \frac {d P_{\ell} (\cos \theta)}{d \theta},
\ee
where a finite $\ell_{\rm max}$ is necessary because of numerical limitations. Making use of the orthogonality relations
\be
\int^{\pi}_{0} d \theta  \frac {d P_{\ell}(\cos \theta)} {d \theta} \frac { d P_{\ell'}(\cos \theta)} {d \theta} \sin \theta  \propto \delta_{\ell \ell'},
\ee
we project Eq.~\eqref{eq:gscomp} into harmonics by multiplying by Legendre polynomials and integrating. This produces a coupled set of $\ell_{\rm max}$ 
ordinary differential equations for the functions $f_{\ell}(r)$ (see Sec.~\ref{sec:numerics} for numerical details).

The system is closed by setting appropriate boundary conditions and choosing a hydrostatic equation of state (EOS). For the former, we make the standard 
choices that (i) the field behaves as regularly as possible towards the origin, $f_{\ell}(r \rightarrow 0) \rightarrow \alpha_{\ell} r^{\ell+1}$ for some constants 
$\alpha_{\ell}$ [see, e.g., Eq.~(24) of C09], and (ii) the field matches smoothly to electrovacuum. The latter condition can be straightforwardly expressed since 
Eq.~\eqref{eq:gscomp} can be solved analytically when $p = \epsilon = L_{0} = 0$ via hypergeometric functions that decay as $f_{\ell,\rm ext} \sim r^{-\ell}$ 
with amplitudes representing the electromagnetic multipole moments. 

We work with the analytic Tolman-VII solution for simplicity, the details of which are reproduced here for the  reader's convenience. The energy density takes the form (e.g. \cite{j19})
\be \label{eq:t71}
\epsilon(r) = \epsilon_{c} \left( 1 - x^2 \right),
\ee
for central value $\epsilon_{c} = 15 M_{\star}/ 8 \pi \Rs^3$ and dimensionless radius $x = r/\Rs$, which implies that $g_{rr}$ is found through
\be
e^{-2 \lambda} = 1 - \frac{8 \pi}{15}\Rs^2 \epsilon_{c} x^2 (5 - 3 x^2).
\ee
The pressure and $g_{tt}$ component of the metric are rather more involved, and are given by
\be
p(r) = \frac{1}{4 \pi \Rs^2} \left[ \sqrt{3 C e^{-2 \lambda}} \tan \phi_{\rm T} - \frac{C}{2} \left(5 - 3 x^2 \right) \right],
\ee
and
\be
e^{2 \nu} = \left(1 - 5 C/3 \right) \cos^2 \phi_{\rm T},
\ee
where $C = \Ms / \Rs$ is the compactness and
\be
\begin{aligned}
\label{eq:t74}
\phi_{\rm T} =& \arctan \sqrt{\frac{C}{3(1-2C)}} + \log \sqrt{ \frac{1}{6} + \sqrt{\frac{1-2C}{3C}} }\\
&- \frac{1}{2} \log \left(x^2 - \frac{5}{6} + \sqrt{ \frac {e^{-2 \lambda}} {3 C}} \right).
\end{aligned}
\ee
At the stellar surface $(r = \Rs)$ we have $p = \epsilon = 0$ and the metric functions match to the Schwarzschild exterior. We now have a well-defined boundary problem.

\subsection{Numerical methods}
\label{sec:numerics}

A few numerical remarks are in order before we can present a solution. Firstly, integrals involving the Heaviside function, i.e. over the toroidal terms that are 
radially-rescaled functions of the form
\be
\label{eq:iint}
I_{\ell}(r) = \int^{\pi}_{0} d \theta \Theta\left[ \left|\frac{\psi(r,\theta)}{\psi_{c}}\right|-1 \right] \psi(r,\theta) \frac {d P_{\ell}(\cos \theta)} {d \theta},
\ee
cannot be evaluated trivially like the poloidal components (though see Ref.~\cite{g12} for an analytic approach). One method is to use a smooth, sigmoid-like approximation 
for the Heaviside function, which allows $I_{\ell}$ to be evaluated analytically for low values of $\ell$. In this work, the integrals \eqref{eq:iint} are approximated with a sum via 
Simpson's method. That is, we approximate angular integrals over a small subregion $\theta \in (a,b)$ via 
\be
\frac{\int^{b}_{a} d \theta X(r,\theta)}{b-a} \approx \frac{X\left(r,a\right)}{6} + \frac{2 X\left(r, \frac{a+b}{2}\right)}{3} + \frac{X\left(r,b\right)}{6} ,
\ee
where the full range $0 \leq \theta < \pi$ is subdivided into a uniform grid of $n_{\phi}$ points. Experimentation reveals that a resolution of $n_{\phi} = 256$ produces solutions that are not visibly different from those with $n_{\phi} > 256$ for most cases. However, for fields with $L_{0}/\tcC \gtrsim 0.1$ the toroidal field occupies only a small volume (see Sec. \ref{sec:results}), and so higher resolution is needed to accurately evaluate $I_{\ell}$.

Solutions to the projected GS equations are achieved via a fourth-order Runge-Kutta method applied to the vector $(f_{1}, f_{2}, \ldots, f_{\ell_{\rm max}})$, with step size chosen such that the global error is controlled to be at most one part in $\gtrsim 10^6$ (see Sec. \ref{sec:tordom}). The surface boundary condition can be written as (see, e.g., C09)
\be
f_{\ell}'(\Rs) = \frac {f_{\ell, \rm ext}'(\Rs)} {f_{\ell, \rm ext}(\Rs)} f_{\ell}(\Rs),
\ee
which guarantees we match to some multipolar exterior.

Numerical implementation of the condition $f_{\ell}(r \rightarrow 0) \rightarrow \alpha_{\ell} r^{\ell+1}$ is trickier, and must be achieved by shooting for solutions with the most regularity (i.e. with the shallowest gradients around $r=0$). We perform a sequence of refining scans over values of $D_{2}$ and multipolar coefficients $f_{\ell}'(\Rs) / f_{\ell}(\Rs)$, for any given values of the toroidal amplitude $L_{0}$ and compactness $M_{\star}/R_{\star}$, until we converge to a solution that is regular within a specified tolerance. Such a procedure is computationally expensive for $\ell_{\rm max} > 3$ because we must scan over a set of dimension $\ell_{\rm max}$ (noting that we can fix the dipole moment without loss of generality, though cf. Sec. \ref{sec:tordom}), and typically we need a precision of order $\lesssim 10^{-4}$ in the eigenvalues to avoid divergences. Greater precision is required for strong toroidal fields, else the solutions display a numerical twist and makes it appear as if $\psi$ has additional nodes. Curiously, C08 found that disconnected solutions with multiple nodes are genuine features of dipolar, $c_{0} \neq 0$ equilibria (see their Fig. 2d), while in our case they only appear if the eigenvalues are approximated with insufficient precision. Regardless, keeping terms up to the octupole gives a representation of the true solution to \eqref{eq:gscomp} at the $\lesssim 10\%$ level, even for fields with strong toroidal fields, as found by C09 and us here. 

\subsection{Results}
\label{sec:results}

Implementing the approach described above, we present mixed poloidal-toroidal and multipolar solutions to \eqref{eq:gscomp} with a Tolman-VII background \eqref{eq:t71}--\eqref{eq:t74}. 
In general, we can plot the field line structure, as measured by a locally inertial observer, using the tetrad components of the field given by \cite{is04} (up to sign convention; cf. C09)
\be \label{eq:br}
B_{(r)} = -\frac {\psi_{,\theta}} {r^2 \sin \theta},
\ee
\be \label{eq:btheta}
B_{(\theta)} = \frac{ e^{- \lambda}} {r \sin \theta} \psi_{,r},
\ee
and
\be \label{eq:bphi}
B_{(\varphi)} = -\frac {4 \pi L_{0}}{\tcC r \sin \theta} \Theta(|\psi/\psi_{c}|-1) e^{-\nu} \psi.
\ee
For concreteness, we set the stellar compactness equal to $0.2$, i.e. $R_{\star} = 5 M_{\star}$ in geometric units. 

\begin{figure*}
\centering
\includegraphics[width=0.52\textwidth,height=0.35\textheight]{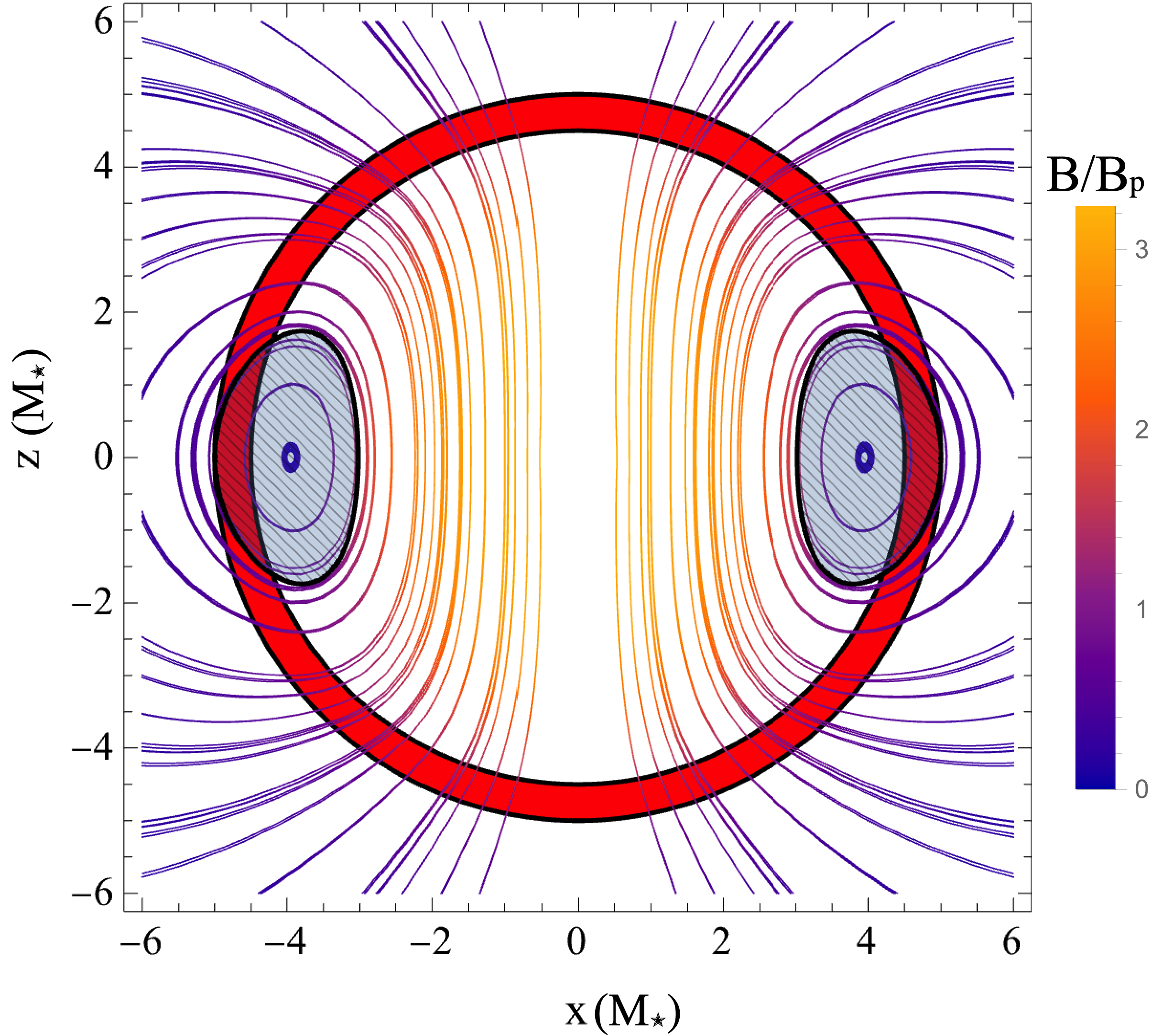}
\includegraphics[width=0.472\textwidth,height=0.35\textheight]{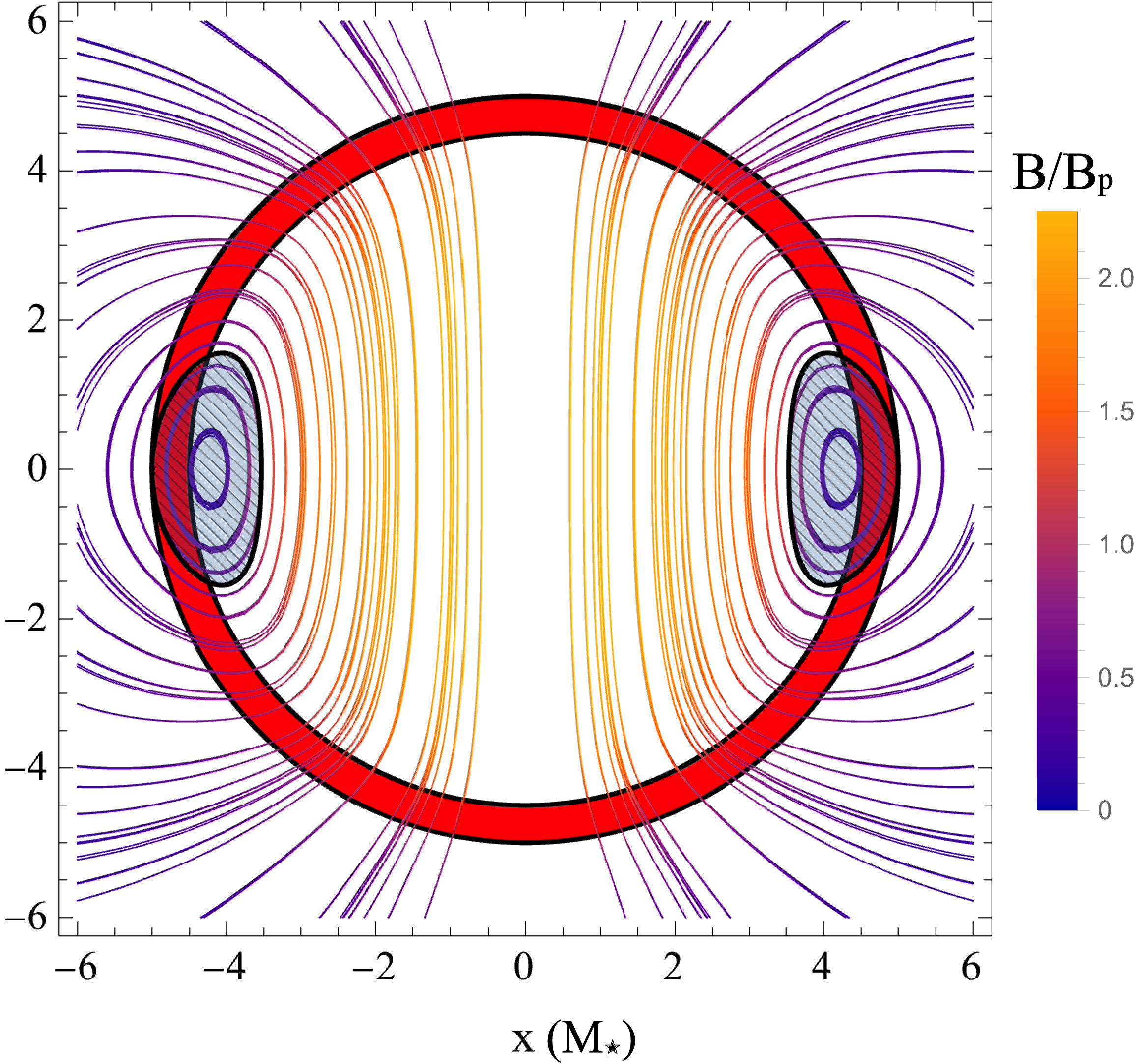}
\caption{Field line structure of the solution to the dipole projection of the linearised GS equation \eqref{eq:gscomp}, for a Tolman-VII background 
of compactness $R_{\star} = 5 M_{\star}$, where we set $L_{0}=0$ (left panel) and $L_{0} = 0.04 \tcC$ (right panel). The colour scale shows the relative 
magnitude of the poloidal field relative to the (arbitrary) polar field value, with redder shades indicating a stronger field. The shaded 
ellipsoids at the equator delimit the surface $|\psi| \geq \psi_{c}$, where the toroidal field resides, although $B_{(\varphi)}$ vanishes identically in the poloidal case (left). 
The red annulus depicts the `crustal' region $0.9 \Rs \leq r \leq \Rs$.}
\label{fig:dipole}
\end{figure*}

As a first (relatively crude) approximation, consider the pure dipole case, $f_{\ell\geq 2}(r)=0$. Fixing the dipole moment to some arbitrary (though sub-virial) value, we find the purely poloidal ($L_{0} = 0$) solution with $D_{2} \approx 0.69214$. The field line structure is shown in the left panel of Fig.~\ref{fig:dipole}, with the colour scale indicating the poloidal strength relative to the polar value, $B_{p} = |B_{(r)}(\Rs,0)|$. Although there is no toroidal field here by construction, we still show the region $|\psi| > \psi_{c}$ for comparison with mixed-field cases. The internal field attains a maximum of $\approx 3.2$ times the polar value $B_{p}$, as indicated by the colour scale. Instead setting $L_{0} = 0.04 \mathcal{\tilde{C}}$ to include a modest toroidal field, we find $D_{2} \approx 0.40386$. The associated field line structure is shown in the right panel of Fig. \ref{fig:dipole}.

The strength of the poloidal field is reduced for increasing $L_{0}$. For $L_{0} = 0.04 \tcC$, for instance, the field achieves a maximum $B_{\rm max} / B_{p} \approx 2.3$ (right panel). Moreover, as found in previous literature though in a different notation, increasing the ratio $L_{0} / \tcC$ shrinks the toroidal volume (as is evident from the size of the equatorial rings in Figure \ref{fig:dipole}). The eigenvalue $D_{2}$ decreases monotonically as we increase $L_{0}$, though the gradient softens after $L_{0} / \tcC \gtrsim 0.1$. Eventually, if one sets a very large $L_{0}$ then $B_{\varphi}$ occupies such a small volume that the solution in most of the star is unaffected, and the eigenvalue $D_{2}$ asymptotes towards a ($\ell_{\rm max}$-dependent) floor value of $D_{2, \rm min} \approx 0.236$. This is simply a consequence of the linearisation (cf. Refs.~\cite{g12,c13}); choosing more complicated flow constants allows one to control the toroidal volume and strength simultaneously (see Sec. \ref{sec:tordom}). 

We next construct a purely poloidal but multipolar solution with $\ell_{\rm max} = 3$. Although we could include a quadrupolar term, likely important for astrophysical systems as evidence from NICER and hotspot modelling suggests that hemispherically asymmetric magnetic geometries are prevalent in Nature (such as in  PSR J0030+0451 \cite{bil19}), the trivial solution $f_{2}(r) = 0$ is permitted because the even and odd parity sectors decouple, as also noted by C09. Although equilibria with $f_{2}(r) \neq 0$ exist\footnote{The situation is subtle. Mixed polarity solutions exist only for sufficiently large values of $L_{0}$ given a compactness. To see this, note that in the purely poloidal case the equations fully decouple and unless the eigenvalue $D_{2}$ happens to match in both equations there will be no solution. Although we do not show the field structure, an example of a nodeless, dipole-dominated solution is $L_{0} = 0.06 \tcC$, $D_{2} \approx 0.2936$, and $f_{1}(\Rs)/f_{2}(\Rs) \approx 90.11$.}, we focus on odd-parity multipoles for simplicity.

\begin{figure*}
\begin{center}
\includegraphics[width=0.52\textwidth,height=0.35\textheight]{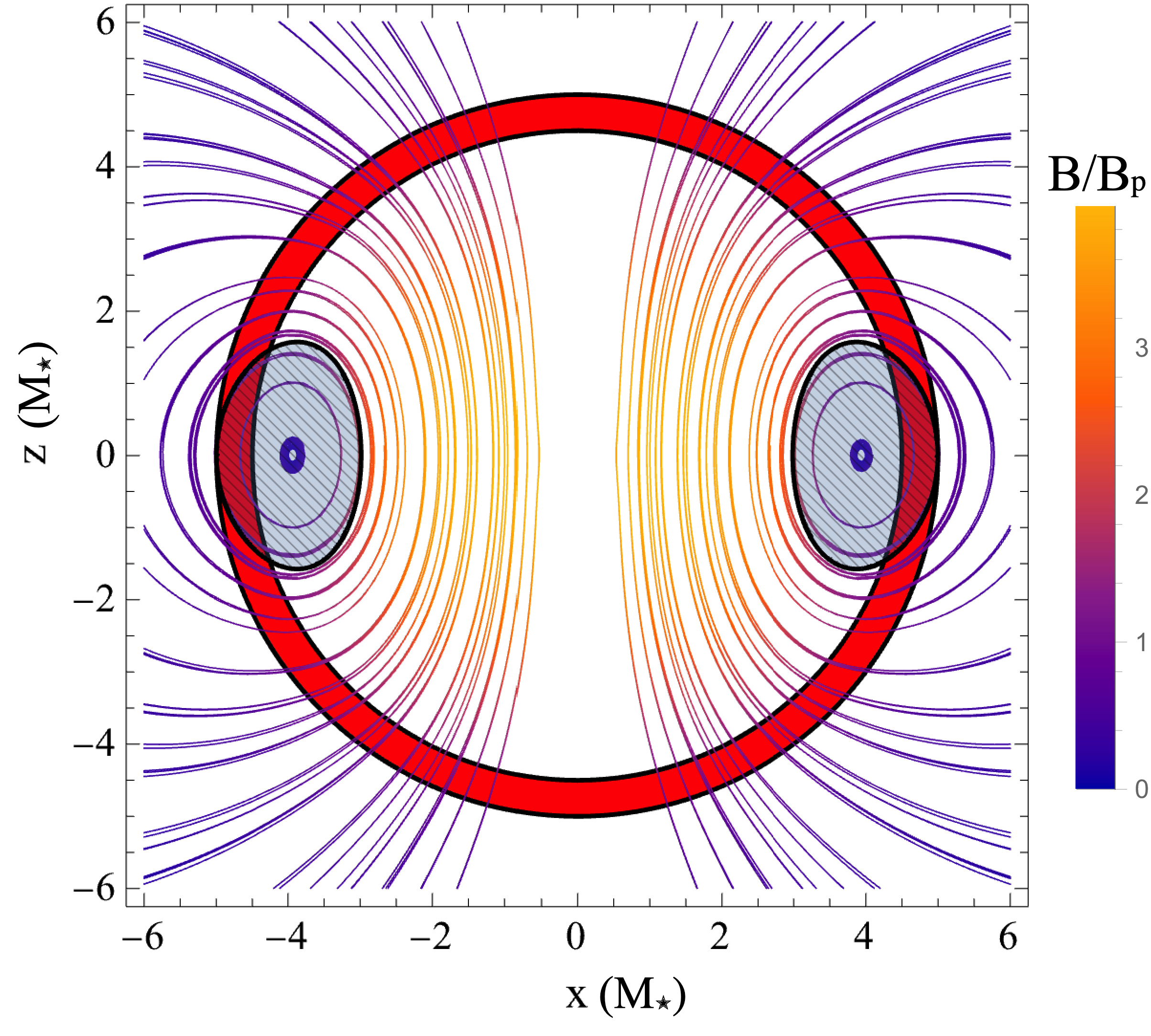}
\includegraphics[width=0.472\textwidth,height=0.35\textheight]{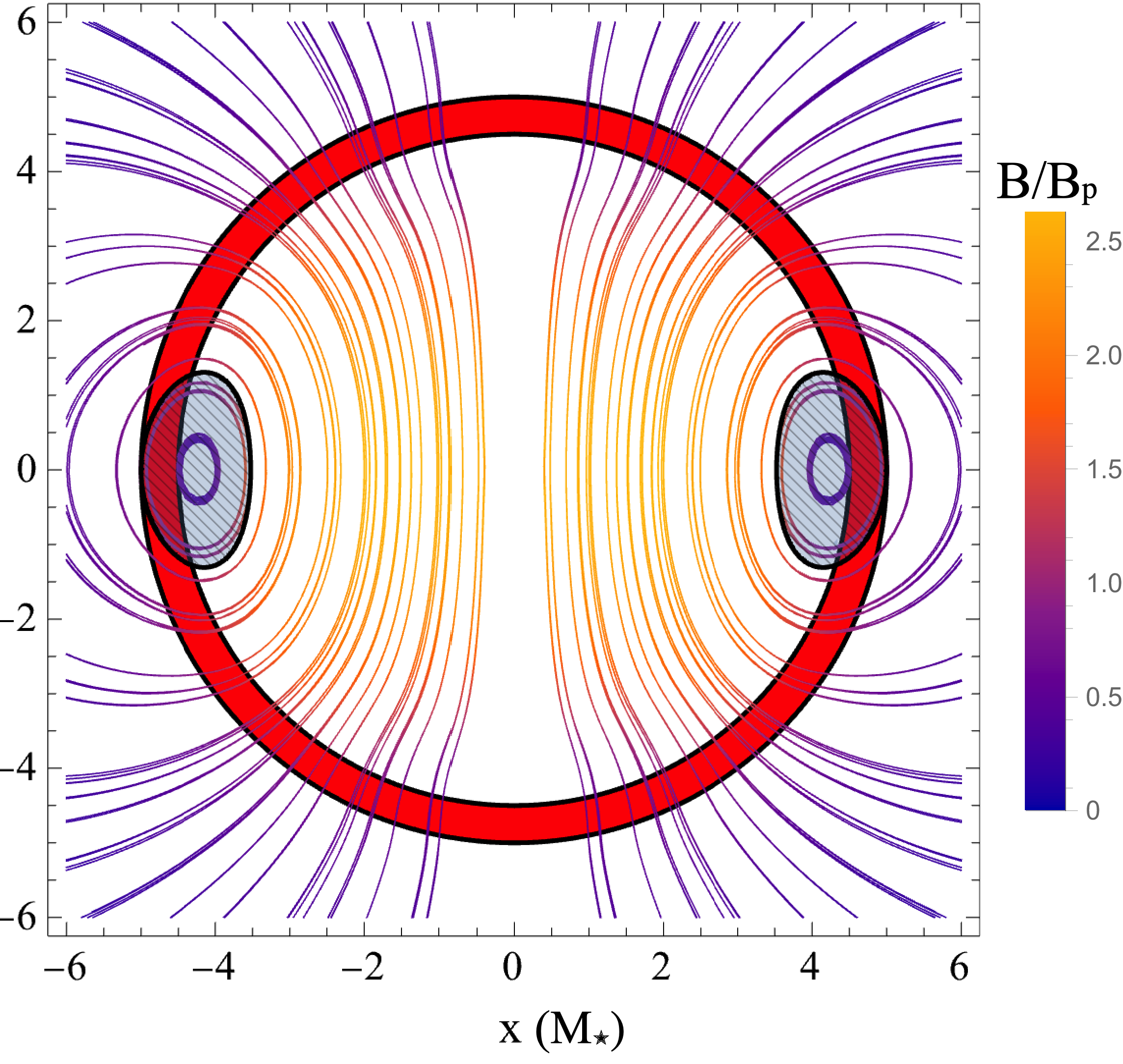}
\end{center}
\caption{Similar to Fig. \ref{fig:dipole}, though including octupole contributions, for $L_{0} = 0$ (left) and $L_{0} = 0.04 \tcC$ (right).}
\label{fig:octo}
\end{figure*}

Including an octupole moment, the numerical scan reveals $D_{2} \approx 0.6750$ and $f_{1}(\Rs)/f_{3}(\Rs) \approx 22.7795$ for $L_{0} = 0$. The field structure is shown in the left panel of Fig.~\ref{fig:octo}. The largeness of $f_{1}(\Rs)/f_{3}(\Rs)$ indicates that the octupole contribution is relatively weak, though its impact in the core is evident because the field maximum increases to $B / B_{p} \approx 3.7$ ($\approx 15\%$ larger than the dipole case; Fig. \ref{fig:dipole}). A mixed field solution with $L_{0} = 0.04 \tcC$, requiring $D_{2} \approx 0.3905$ and $f_{1}(\Rs)/f_{3}(\Rs) \approx 10.320$, is depicted in the right panel. As before, the volume of the toroidal region decreases as we increase $L_{0}$ and the poloidal field decreases in strength ($B_{\rm max} / B_{p} \approx 2.6$). Additionally, including a toroidal component also requires that the octupole contribution becomes stronger, with the `kink' in the field lines near polar latitudes around $r \sim 0.9 R_{\star}$ becoming more prominent. 

\begin{figure}[h]
\begin{center}
\includegraphics[width=0.487\textwidth]{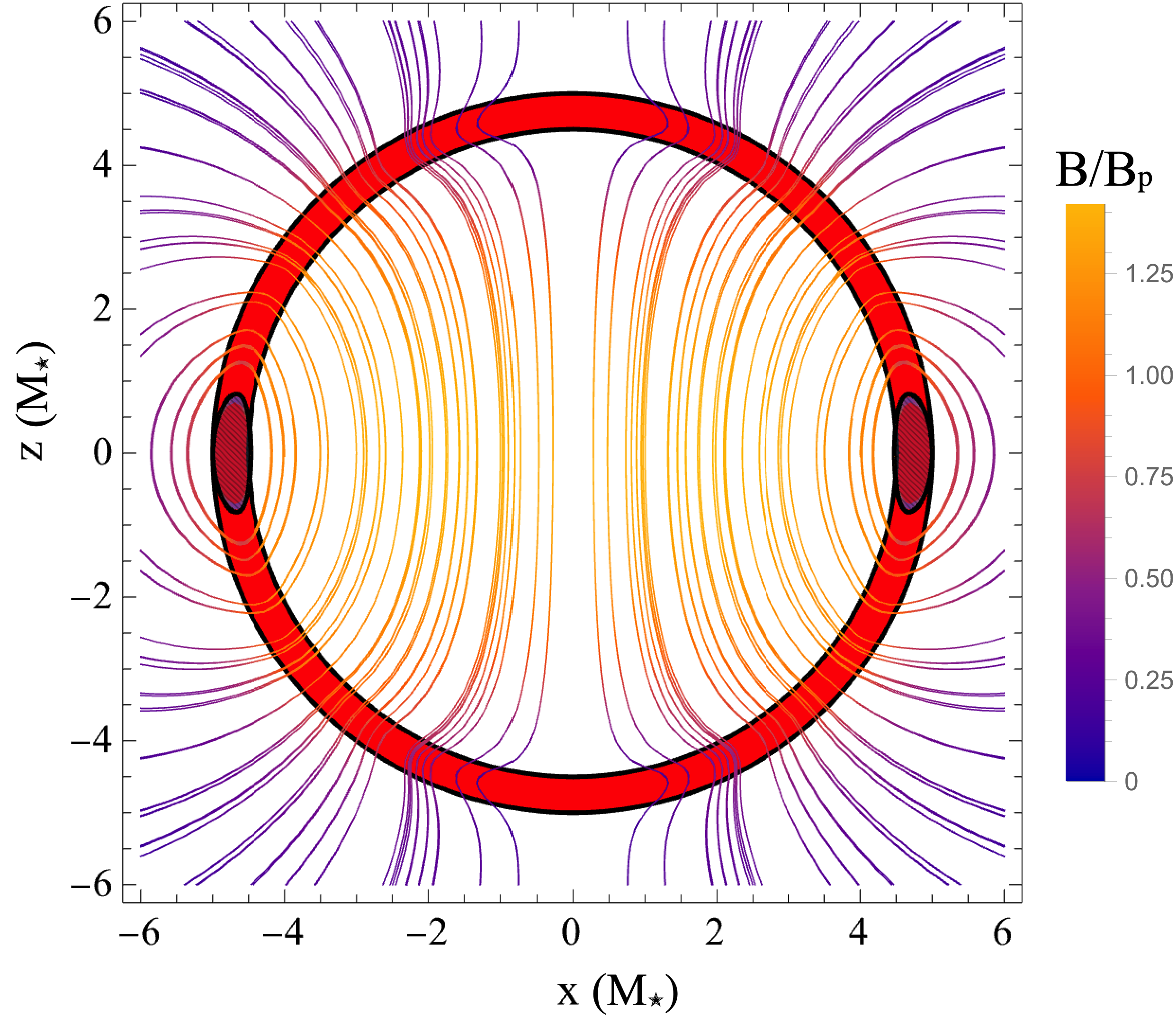}
\end{center}
\caption{Similar to Fig. \ref{fig:octo}, though with a stronger toroidal field, viz. $L_{0} = 0.1 \tilde{\mathcal{C}}$, that is confined to the `crust'.}
\label{fig:toroct2}
\end{figure}

Larger values of $L_{0}$ demand smaller ratios of $f_{1}(\Rs)/f_{3}(\Rs)$; for example, for $L_{0} = 0.1 \tilde{\mathcal{C}}$ we find $D_{2} \approx 0.234492$ and $f_{1}(\Rs)/f_{3}(\Rs) \approx 6.0101$. 
This solution, depicted in Fig.~\ref{fig:toroct2}, is interesting because the toroidal center migrates towards the surface and shrinks in volume enough that $B_{\varphi}$ becomes confined to the region $0.9 \lesssim x \leq 1$ representing a (hypothetical) crust. Such equilibria may be more realistic than ones with core-dominated fields owing to the torque instability described in Ref.~\cite{gl15}. Similar to Fig. \ref{fig:octo}, the solution we find is nodeless and the toroidal field is confined strictly by the neutral curves, unlike some of the peculiar solutions found by C08 and C09. Although not shown, a range of equilibria with varying $L_{0}$ were constructed and we found that the nodeless property persisted. In C08, by contrast, there are discrete ranges (of $\zeta$) such that magnetically-disconnected solutions exist (see their Fig.~2). In cases with mixed polarity, we find solutions with some minimum number of nodes only. 

We conclude by noting that the important distinction with cases where $D_{1} \neq 0$ (i.e. $c_{0} \neq 0$) is that solutions constructed for $D_{1} = 0$ are \emph{unique}, for a given EOS, because there are no free parameters. In C09, for instance, they were able to find solutions with an integer number of nodes in $\psi$ because of the $c_{0} \neq 0$ freedom (see their Fig.~6). One might argue that solutions where $\psi \rightarrow 0$ internally, at least for odd-multipoles, are physically undesirable as they indicate a singular limit of the flow constant formalism (cf. IS04 and Sec. \ref{sec:alfvenpoints}). In our case, it seems that only the minimum node solution can be found, indicating that that these extra solutions are artifacts of the additional freedom. Although this is difficult to prove rigorously, we have performed fine scans over the parameter space and can only consistently converge to the eigenvalues given above; no other choices give a regular solution. Overall though, the main quantitative features of our solutions are similar to those found in the aforementioned references.

Although we could continue this sequence of solutions by increasing $\ell_{\rm max}$, the field line geometry is unlikely to change dramatically because the ratio $f_{\ell}(\Rs)/f_{1}(\Rs)$ decreases as a function of $\ell$ (as found by C09). 

\subsubsection{Meridional flow}

Even though the meridional flow does not enter into Eq.~\eqref{eq:gscomp} at linear order, its presence is guaranteed by the existence of a toroidal field (see Sec. \ref{sec:notor}). The tetrad components of the velocity vector can be found through [see, e.g., Eqns.~(135--137) in IS04]
\be \label{eq:vr}
v_{(r)} = \frac{\Theta(|\psi/\psi_{c}|-1)}{\tilde{\mathcal{C}} \left( \epsilon + p \right) r^2} e^{-2 \nu} \psi \frac {\partial \psi} {\partial \theta},
\ee
\be \label{eq:vtheta}
v_{(\theta)} = - \frac{\Theta(|\psi/\psi_{c}|-1)}{ {\tilde{\mathcal{C}} \left(\epsilon + p\right) r \sin \theta}} e^{-\lambda -2 \nu} \psi \frac {\partial \psi} {\partial r},
\ee
and
\be \label{eq:vphi}
v_{(\varphi)} = e^{-\nu} \left[ \Omega  r \sin \theta - \frac {L_{0} \Theta(|\psi/\psi_{c}|-1) e^{-2 \nu}}{4 \pi \tilde{\mathcal{C}}^2 \left( \epsilon + p \right) r \sin \theta} \psi^2  \right].
\ee
It is interesting to remark that $B_{(\varphi)}$ depends on the combination $L_{0}/\tilde{\mathcal{C}}$, while the meridional components \eqref{eq:vr} and \eqref{eq:vtheta} depend only on $\tilde{\mathcal{C}}$. 
This is again an artifact of the linearisation procedure, and shows that a family of meridional flows of essentially arbitrary amplitude exist for a given $\psi$ (provided the perturbative criteria is satisfied). 

\begin{figure}[h]
\begin{center}
\includegraphics[width=0.487\textwidth]{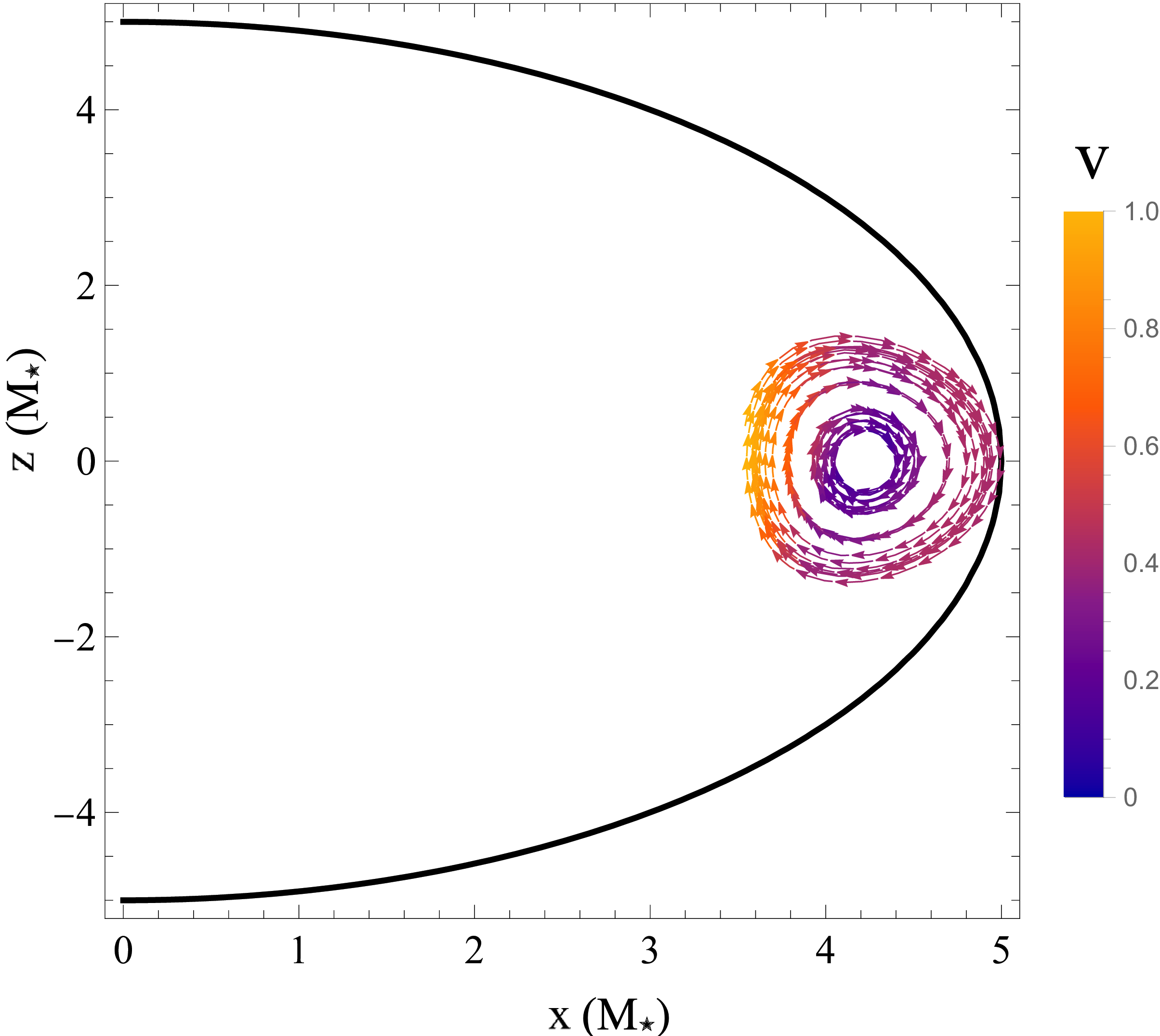}
\end{center}
\caption{Meridional flow, restricted to the toroidal region, for the solution shown in the right panel of Fig. \ref{fig:dipole}. For improved visibility, only the eastern 
hemisphere of the star is shown and the `crustal' region is not depicted. The colour scale shows the relative amplitude of the (clockwise) flow, normalised by an 
arbitrary constant as described in text.}
\label{fig:merid}
\end{figure}

Fig.~\ref{fig:merid} shows the meridional flow, in the case where $\psi$ corresponds to the dipole shown in the right panel of Fig.~\ref{fig:dipole}. 
Even though the denominator in the flow components \eqref{eq:vr} and \eqref{eq:vtheta} diverges at the stellar surface, the overall velocity vanishes there because of Eq.~\eqref{eq:lchoice}. 

\subsection{Regular toroidally-dominated configurations free of surface currents}
\label{sec:tordom}

\begin{figure}[h]
\begin{center}
\includegraphics[width=0.487\textwidth]{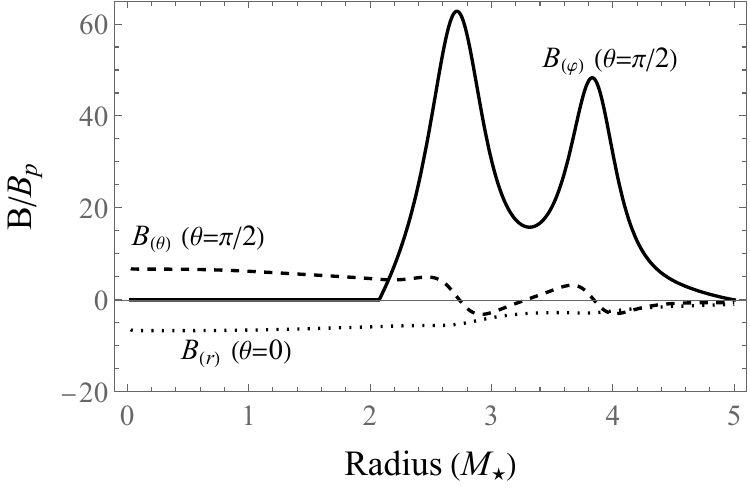}
\end{center}
\caption{Tetrad components of the magnetic field -- $B_{(r)}$ (dotted), $B_{(\theta)}$ (dashed), and $B_{(\varphi)}$ (solid) evaluated at either the equator or pole -- for a solution to the non-linear GS equation with the flow constant choices \eqref{eq:newe} and \eqref{eq:newl}, with $\kappa = 4$, $k_{1} = 1$, $k_{2} = 10$, $k_{3} = -0.1$, and $L_{0} = 1.1 \tcC$.}
\label{fig:domtor}
\end{figure}

As described in Sec. \ref{sec:results}, linearisation imposes a limit to the toroidal energy stored in the field: increasing $L_{0}$ in an 
attempt to boost $B_{(\varphi)}$ shrinks the toroidal volume. Moreover, the strict linearisation of the GS equation implies that it is 
not possible to have a regular toroidal field which is non-zero internally and vanishes externally if $\psi \neq 0$ at the boundary. These restrictions 
can be lifted simultaneously if we allow for a non-linear $E(\psi)$ and $L(\psi)$, though the price to pay is self-consistency with power-counting, 
as detailed in Sec. \ref{sec:previous}. Nevertheless, we can compute solutions with non-linear flow constants. Motivated by the choices made 
in Ref.~\cite{c13}, though again setting $E_{1} = 0$, we consider the functions
\be
\label{eq:newe}
\ba
E(\psi) =& E_{2} \psi \Big[ k_{1} \left(\left| { \psi}/{\psi_{c}}\right| - 1 \right)^{\kappa} \Theta\left(\left| { \psi}/{\psi_{c}}\right| -1 \right)  \\
&- k_{2} L(\psi) - k_{3} \Big],
\ea
\ee
\begin{equation} \label{eq:newl}
L(\psi) = L_{0} \psi \left(\left| { \psi}/{\psi_{c}}\right| - 1 \right) \Theta\left(\left| { \psi}/{\psi_{c}}\right| -1 \right),
\end{equation}
for parameters $L_{0}, \kappa, k_{1}, k_{2}$, and $k_{3}$. The choice $k_{1} = k_{2} = 0$ returns \eqref{eq:eeqn} with a rescaled value of $E_{2}$.  The motivation for the above inclusions are that one wishes to enhance the region of closed field lines -- achieved through $k_{1}$ and $\kappa$ -- while allowing the toroidal energy density to grow by minimising the azimuthal current $J_{\varphi}$ -- achieved through $k_{2}$ (see Ref.~\cite{c13} for a discussion). Furthermore, the non-linear choice \eqref{eq:newl} (also made in other works, e.g. \cite{g12}) ensures that the toroidal field decays \emph{smoothly} towards the stellar surface for suitable $\kappa$, thus preventing surface currents. A dipolar solution to Eq.~\eqref{eqGS1}, though again at the expense of ignoring meridional components and metric corrections, is found using the method described in Sec. \ref{sec:numerics} for the choices $\kappa = 4$, $k_{1} = 1$, $k_{2} = 10$, $k_{3} = -0.1$, and $L_{0} = 1.1 \tcC$. The eigenvalue we converge to is $D_{2} \approx 9.49276$. The field components, exhibiting a strongly dominant $B_{(\varphi)}$, are shown in Fig.~\ref{fig:domtor}. Note that because we consider a non-linear GS equation here, the exact normalisation of the dipole moment affects the solution\footnote{It is clear from these solutions that non-linear dynamics are important as concerns the magnetic field itself even for amplitudes well below the virial limit. The impact on the stellar structure is tiny however, even though metric corrections appear at $\mathcal{O}(\psi^2)$.}. The figures here correspond to a physical value of $B_{p} \gtrsim 5 \times 10^{14}\,$G.

The combined effect of these choices is that the toroidal field is both strong and vast. Furthermore, $B_{(\varphi)}$ decays smoothly to zero as $\psi \to \psi_{c}$, indicating an absence of surface currents and discontinuities. Direct integration reveals that the toroidal-to-total energy ratio is
\begin{equation}
\frac{E_{\rm tor}} {E_{\rm tot}} = \frac {\int dV B_{\varphi}^2 / 8 \pi} {\int dV B^2 / 8 \pi} = 0.956,
\end{equation}
meaning that the toroidal field houses $\approx 95.6\%$ of the total magnetic energy. Such a configuration may be relevant for magnetars. For instance, $\approx 36\,$ks phase-modulations in the X-ray pulses of 1E 1547.0--5408 have been consistently observed \cite{mak21}, possibly indicating a freely precessing source. Since the spin period of this object is $P = 2.087\,$s, the modulation periodicity suggests a quadrupolar ellipticity of order $\sim 6 \times 10^{-5}$, and thus a toroidal field of strength $B_{(\varphi)} \sim 10^{16}\,$G since the polar field strength is only $B_{p} \sim 10^{14}\,$G (see also Ref.~\cite{suv23}). 

\begin{figure}[h]
\begin{center}
\includegraphics[width=0.487\textwidth]{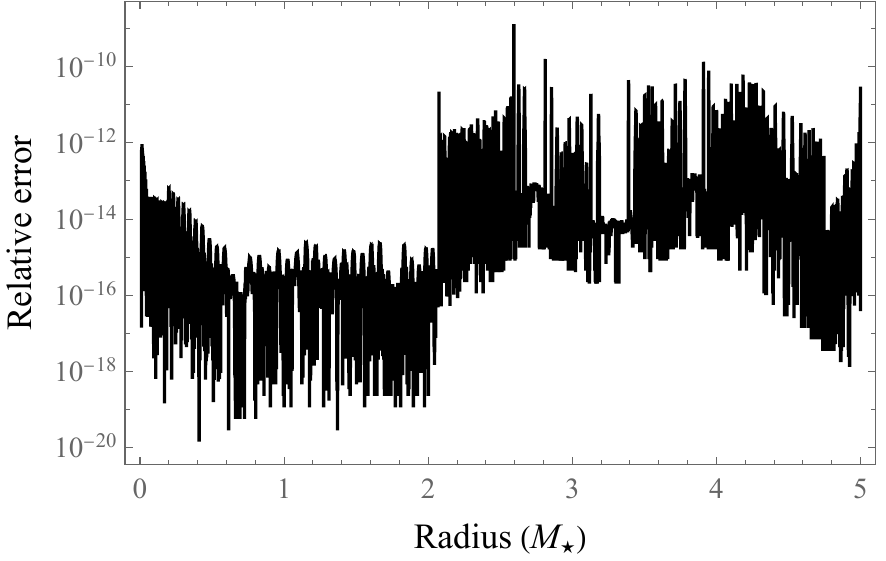}
\end{center}
\caption{{Residual of the} numerical solution shown in Fig. \ref{fig:domtor}. This plot is typical for solutions shown in this work.}
\label{fig:error}
\end{figure}

{Though more formal checks can be made with the GR virial relations (e.g.~\cite{kiu08}), the extent of numerical errors, quantified by the left-right mismatch in the radial projection(s) of Eq.~\eqref{eq:gscomp} in dimensionless units (i.e. the residual)}, is shown in Fig.~\ref{fig:error}. The local error reaches $\sim 10^{-9}$ at worst, indicating a reliable solution.

\section{Discussion}
\label{sec:conclusions}

In this paper we revisit the impact of `flow constants', introduced by BO in the GR context \cite{bek78,bek79}, on the construction of stationary and axisymmetric neutron star equilibria. The main results of this work are two-fold. The first part presents a modern review of the general formalism in both GR and Newtonian gravity (Sec. \ref{sec:flowconstants}), spelling out important corollaries at the general, non-linear level (Sec. \ref{sec:corollaries}). One key argument we present is that consideration of the GS equation alone gives the misleading impression that terms related to meridional flows can be discarded: such a choice is inconsistent with a mixed poloidal-toroidal field according to the BO theorem (Sec.~\ref{sec:notor}), required in canonical twisted-torus models and from a stability standpoint. The second part explores perturbative constructions, where an expansion in the sub-virial flux $\psi$ is taken (Sec. \ref{sec:perturbative}), with numerical equilibria built in Sec. \ref{sec:workedex}. We find that power-counting arguments suggest that some choices for flow constants made in previous literature are not, strictly speaking, self-consistent. Fortunately though, our numerical equilibria for static stars, where strict order-counting is enforced, are both qualitatively and quantitatively similar to existing models (see, e.g., Refs.~\cite{g12,col08,c09,c13}).

In considering the perturbative equilibria of stars that are rotating at background order, we find that power-balancing relations for the flow constants predict that the toroidal field and meridional flow scale as $B^{\varphi} = \mathcal{O}(\psi^{\lambda})$ and $u^{i} = \mathcal{O}(\psi^{1-\lambda})$, respectively, with $0< \lambda \leq 1$ (Sec. \ref{sec:weakB2}). Since $|\psi| \ll 1$ for the perturbative scheme to be valid, this implies that the toroidal field is larger than the poloidal one, though not necessarily in the global sense of total energies: the toroidal field may be locally stronger but confined to a small volume (cf. Sec. \ref{sec:results}). It is generally thought that strong toroidal fields are necessary to explain frequent magnetar outbursts \cite{pons11}, a conclusion which is supported by stability theory \cite{akg13} together with observations of large pulse fractions \cite{igo21} and $\gtrsim 10\,$ks phase modulations in X-ray lightcurves \cite{mak21}. Based on this, one may conjecture that magnetar birth conditions preferentially select values of $\lambda \ll 1$ while `ordinary' neutron stars have $\lambda \lesssim 1$. However, in modelling rotating stars one must be careful when taking expansions for the flow constants [such as in Eq.~\eqref{eq:eeqn}] as simple Taylor series are precluded by the necessity of non-integer powers (i.e. one should use Puiseux series instead). For future work, it would be worthwhile considering self-consistent GS equilibria at successive (fractional) orders in $\psi$, including meridional flow, metric corrections, and multi-fluids (see the discussion in Sec.~\ref{sec:induction}).

{Given that the equilibria constructed here are moulded around twisted tori, we expect that the stability results regarding such configurations carry over \cite{bs06}. Whether mixed poloidal-toroidal fields are long-term stable is not a fully settled matter though (especially in GR), as non-axisymmetric instabilities may be endemic to barotropic stars \cite{lj12} while convective instabilities can operate in non-barotropic systems with meridional flows \cite{yosh12}.}

In a full, time-dependent simulation, one should be able to specify initial data which eventually translate into flow constants when equilibrium is reached, if indeed it ever is. While in considering the time-independent GRMHD equations one is free to choose the flow constants arbitrarily, it is far from obvious whether a star set up with some initial data would actually reach such a state. This is highlighted by recent `long-term' GRMHD simulations, where late-time behaviour depends sensitively on the initial energy partition \cite{sur22}. Even if one ignores meridional flows and rotation, the toroidal function is totally free and rich families of GS-equilibria exist \cite{gl16}. Providing a definitive answer to how these constants behave given some initial data, {or which combinations of them are stable}, is an extremely difficult problem which we do not solve here. A future approach to this kind of `inverse problem' would be to try and fit the flow constants to (stable) numerical solutions. To this end, a crucial ingredient, not present in our investigation, is the magnetic helicity: it has been argued that magnetic fields evolve so as to minimise changes in the global helicity (i.e. to be as conserved as possible) \cite{spr08,c10}. 

The existence of meridional flow has important physical consequences. One effect was considered by IS04, who noted that non-circularity induces additional types of frame dragging into 
spacetime which violate reflection symmetry about the equator. Such effects could, in principle, be tied to equatorially-asymmetric hotspot formation \cite{bil19} or natal kicks \cite{is04}. 

The meridional flow could lead to an additional channel of magnetic diffusion via the viscous damping of the internal circulation.  
The associated dissipation timescale is identical to the standard viscous timescale, $t_{\rm visc} \sim R^2_\star \rho/\eta$, where $\eta$ is the shear  viscosity coefficient.
As an example, we consider  mature neutron stars with a neutron superfluid, in which case viscosity is dominated by electron-electron scattering. 
The corresponding viscosity coefficient is $\eta_{\rm ee} \approx 6 \times 10^{20} \rho_{15}^2 T_8^{-2}\, \mbox{g}/\mbox{cm s} $~\cite{cl87},
 where $T_8 = T/10^8\,\mbox{K}$ and $\rho_{15} = \rho/10^{15}\,\mbox{g}\, \mbox{cm}^{-3}$, and we find $t_{\rm visc} \sim 0.05 \, \rho_{15}^{-1} T_8^2\, \mbox{yr} $ 
 (we note that the corresponding timescale for non-superfluid matter is of the same order of magnitude with a modified density scaling $\rho^{-1} \to \rho^{-5/4}$). 
 This timescale is much shorter than typical neutron star ages. 
Since the poloidal field and meridional flow are proportional to each other [Eq. \eqref{eqBi1}], this na\"{\i}vely suggests that poloidal fields should be comparatively weak in mature stars, 
unless somehow regenerated by the toroidal field through MHD exchanges or if the proportionality factor $\cC$ adjusts in tandem. Either way, time-dependent effects of a meridional flow 
could have a profound influence on the magnetic evolution and observational appearance of neutron stars. Future observations of gravitational waves may elucidate the situation.


\acknowledgements{We thank Luciano Rezzolla, Riccardo Ciolfi, {and the anonymous referee} for comments. 
K. G. acknowledges support from research grant PID2020-1149GB-I00 of the Spanish Ministerio de Ciencia e Innovaci{\'o}n. A. G. S. was supported by the Alexander von Humboldt Foundation and the European Union's Horizon 2020 Programme under the AHEAD2020 project (Grant No. 871158) during the early stages of this work.} 


\appendix

\section{Flow constant derivation (GR)}
\label{sec:appa}

This Appendix provides a self-contained derivation of the flow constants originally discussed in BO78 \& BO79. 
Under the assumption of a stationary-axisymmetric system we have $Q_{,t} = Q_{,\varphi} =0$ where $Q$ stands for
any geometric or electromagnetic function. Note that \emph{nothing}  else is assumed about the metric or the fluid flow. 

The first group of flow constants does not depend on the Euler equation and therefore is a good starting point for our analysis.  
By definition, we have
\begin{align}
& F_{t\varphi} = 0,
\label{eq1}
\\
& F_{r\varphi,\theta} = F_{\theta\varphi,r},
\label{eq2}
\\ 
& F_{tr,\theta} = F_{t\theta,r}.
\label{eq3}
\end{align}
The $t,r,\varphi$ components of the ideal MHD condition $E_{\mu} = F_{\mu \nu} u^{\nu} = 0$ lead to
\begin{align}
& F_{t\theta} u^\theta + F_{tr} u^r = 0,
\label{eq4}
\\
& F_{r\varphi} u^r + F_{\theta\varphi} u^\theta = 0,
\label{eq5}
\\
& F_{rt} u^t + F_{r\theta} u^\theta + F_{r\varphi} u^\varphi = 0.
\label{eq6}
\end{align}
Ignoring the singular possibility $u^i = 0$ for now,
we may solve \eqref{eq4} and~\eqref{eq5} for $F_{t\theta}, F_{\rm \theta\varphi}$ and insert the results in~\eqref{eq3} and~\eqref{eq2}.
We find, respectively
\begin{align}
\frac{d}{d\tau} \log F_{tr} &= u^r \frac{ F_{tr,r} }{ F_{tr} } + u^\theta \frac{ F_{tr,\theta} }{ F_{tr} } =  - u^\theta \left ( \frac{u^r}{u^\theta} \right )_{,r}
\label{eq7}
\\
\frac{d}{d\tau} \log F_{r\varphi} &= u^r \frac{ F_{r\varphi,r} }{ F_{r\varphi} } + u^\theta \frac{F_{r\varphi,\theta} }{F_{r\varphi} } 
= - u^\theta \left ( \frac{u^r}{u^\theta} \right )_{,r}
\label{eq8}
\end{align}
It follows that 
\be
\frac{d}{d\tau} \log \left ( \frac{F_{r\varphi}}{F_{tr}} \right ) = 0 
~\Rightarrow~ F_{tr} / F_{r \varphi } = \Omega,
\label{eq9}
\ee
where $\Omega$ is a flow constant. Using this in \eqref{eq4} and \eqref{eq5}
\be
F_{t\theta} / F_{\theta \varphi} = \Omega.
\label{eq10}
\ee
Meanwhile, we can rewrite~\eqref{eq8} as
\be
\frac{d}{d\tau} \log F_{r\varphi} = - u^\mu_{~,\mu} + u^\mu \frac{u^\theta_{~,\mu}}{u^\theta}.
\ee
The baryon conservation equation $\nabla_\mu ( n u^\mu ) = 0$ can be written in the equivalent form
$ u^\mu_{~,\mu} = - d  [ \log ( \sqrt{-g} n )] /d\tau$ and the previous expression becomes
\be
\frac{d}{d\tau} \log \left ( \frac{F_{r\varphi}}{\sqrt{-g} n u^\theta} \right ) = 0~\Rightarrow~
 \frac{F_{\varphi r}}{\sqrt{-g} n u^\theta} = \cC,
 \label{eq11}
\ee
where $\cC$ is another flow constant. This result can be combined with~\eqref{eq5} to similarly show that
\be
 \frac{F_{\theta\varphi }}{\sqrt{-g} n u^r} = \cC.
 \label{eq12}
\ee
Finally,~\eqref{eq6} in combination with~\eqref{eq9}, \eqref{eq11} leads to
\be
  F_{r \theta} =  \sqrt{-g} \cC n ( u^\varphi - \Omega u^t ).
 \label{eq13}
\ee
Using the above expressions in the definition of the magnetic field we find,
\be
B^i = - \cC n ( u_t +  \Omega u_\varphi  )  u^i.
\label{eq21and22}
\ee
In fact, using the definition of $B^\mu$, we can produce the more general four-vectorial expression
\be
B^\alpha = - \cC n \left [\,  (  u_t + \Omega u_\varphi ) u^\alpha  + \delta_t^\alpha + \Omega \delta_\varphi^\alpha \, \right ].
\label{allB1}
\ee
This is a key relation between the magnetic field, the flow constants, and the fluid parameters.
With the help of this relation we can prove the following interesting property for any constant  
$dQ/d\tau = u^\alpha  Q_{,\alpha} = 0$:
\be
B^\alpha Q_{,\alpha} =  - \cC n \left [  ( u_t + \Omega u_\varphi  )  u^\alpha Q_{,\alpha}  +Q_{,t} + \Omega Q_{,\varphi} \right ] = 0,
\label{levelsurf1}
\ee
meaning that $B^\alpha$ and $Q$ share the same level surfaces.

The final conservation law of this first group relates to the  magnetic energy. It originates from the orthogonality between the Lorentz force 
and the four-velocity, i.e.
\be
u_\alpha \nabla_\beta T^{\alpha\beta}_{\rm EM}=  - u_\alpha F^{\alpha\beta} J_\beta   = 0.
\ee
Use of~\eqref{eq:emstress}  for $T_{\rm EM}^{\mu\nu}$ together with the orthogonality property  $u_\alpha B^\alpha =0$ allows us 
to rewrite this condition as
\be
\frac{d}{d\tau} B^2= - 2 B^2 \nabla_\alpha u^\alpha - 2 u_\alpha B^\beta \nabla_\beta B^\alpha.
\label{eq31}
\ee
With the help of  $ \nabla_\alpha u^\alpha = - u^\alpha n_{,\alpha} /n $ and $ u^\alpha B^\beta \nabla_\beta B_\alpha = - B^\alpha B^\beta \nabla_\beta u_\alpha$
this expression becomes
\begin{align}
\frac{d}{d\tau} B^2 &=  2 B^2 \frac{d}{d\tau}  \log n   + 2  B^\alpha B^\beta \nabla_\beta u_\alpha.
\label{eq32}
\end{align}
Expanding the covariant derivative in the first and third term,
\be
0 = B^2 \frac{d}{d\tau} \log n + B_\alpha B^\beta u^\alpha_{~,\beta} - B_\alpha B^\alpha_{~,\beta} u^\beta.
\label{eq36}
\ee
Moreover, we can exploit another equivalent expression for $dB^2 /d\tau$ in the form
\be
u^\beta B_\alpha B^\alpha_{~,\beta}  = \frac{d}{d\tau} B^2 -  u^\beta B^\alpha B_{\alpha,\beta},
\ee
together with $ B^\beta B_\alpha u^\alpha_{,~\beta} = - u^\alpha B^\beta B_{\alpha,\beta}$.  
Inserting these in~\eqref{eq36} and after some rearrangement we obtain
\be
\frac{d}{d\tau} \log \left (\frac{B^2}{n} \right ) 
=  \frac{1}{B^2} \left ( B^\alpha \frac{d B_\alpha}{d\tau} - u^\beta B^i B_{\beta,i} \right ).
\ee
After expanding the right-hand-side terms and using~\eqref{allB1} we can finally arrive 
to a `magnetic energy' conservation law
\be
\frac{d}{d\tau} \left [ \frac{B^2}{n} + \cC  ( B_t + \Omega B_\varphi ) \right ] = 0.
\ee
This implies the existence of a new flow constant $\cF$,
\be
 \frac{B^2}{n} + \cC  ( B_t + \Omega B_\varphi ) =  \cF.
 \label{eq39}
\ee
In the second group of conservation laws we find expressions derivable from the Euler equation. Projecting that
equation along $B^{\mu}$
\be
\left ( \epsilon + p + \frac{B^2}{4\pi} \right ) B_\alpha a^\alpha = - B^\alpha p_{,\alpha} + \frac{B^2}{4\pi} \nabla_\alpha B^\alpha.
\label{eq14}
\ee
For the divergence of the magnetic field we can write
\be
\nabla_\alpha B^\alpha = \frac{1}{2 \sqrt{-g}} \epsilon^{\alpha \beta \gamma \delta} \nabla_\alpha u_\beta F_{\gamma\delta}.
\label{divBfour}
\ee
This can be inverted
\be
 \frac{1}{2 \sqrt{-g}} \epsilon^{\alpha \beta \gamma \delta} F_{\gamma\delta}  = u^\alpha B^\beta - u^\beta B^\alpha ,
\ee
and we can rewrite~\eqref{divBfour} in the far more elegant form,
\be
\nabla_\alpha B^\alpha = B^\alpha u^\beta \nabla_\beta u_\alpha.
\label{eq16}
\ee
The same is true for Eq.~\eqref{eq14}, which becomes
\be
(\epsilon + p ) \nabla_\alpha B^\alpha = - B^\alpha p_{,\alpha}.
\label{eq17}
\ee
Introducing the chemical potential, which is just the specific enthalpy in the limit of zero temperature as considered here, 
\be
\mu = d\epsilon/dn =  ( \epsilon + p )/n,
\ee 
we can easily show that $ p_{,\alpha}=  n \mu_{,\alpha} =  ( \epsilon + p ) \mu_{,\alpha} / \mu $,
and as a result~\eqref{eq17} reduces to a total divergence
\be
\nabla_\alpha (\mu B^\alpha ) = 0.
\label{eq19}
\ee
An equivalent form for this expression is
\be
\mu \left( \sqrt{-g} B^{i} \right)_{,i} = - \sqrt{-g}   {\mu_{,i}} B^i.
\label{eq20}
\ee
Combining this with~\eqref{eq21and22} and the baryon conservation equation $(\sqrt{-g} n u^i )_{,i}=0$ leads to
to the conservation law
\be
\frac{d}{d\tau} \log \left [ \mu ( u_t + \Omega u_\varphi  ) \right ] =0.
\label{eq23dif}
\ee
The corresponding flow constant $D$ is defined as
\be
D =  \mu ( u_t + \Omega u_\varphi   ) =  \mu  u_\alpha (  \delta_t^\alpha +\Omega \delta_\varphi^\alpha  ).
\label{eq23}
\ee
The remaining conservation laws represent a generalisation of the hydrodynamical Bernoulli theorem (e.g. \cite{gour06}).
If $\xi^\alpha$ is a Killing vector, it is easy to show that $\nabla_\alpha ( T^{\alpha \beta} \xi_\beta ) = 0$.
For $T^{\alpha \beta} = T^{\alpha \beta}_{\rm EM} + T^{\alpha \beta}_{\rm fluid}$ this identity leads to
\begin{align}
&n u^\alpha \left (\chi u^\beta \xi_\beta \right )_{,\alpha}  - \frac{1}{4\pi} \left [  \xi_\beta B^\beta \nabla_\alpha B^\alpha 
+ B^\alpha \left ( \xi_\beta B^\beta \right )_{,\alpha} \right ] \nn \\
& + \xi_\beta g^{i\beta}  \left ( p + \frac{B^2}{8\pi} \right )_{,i}  = 0 ,
\label{eq25}
\end{align}
where we defined the generalised chemical potential $\chi = \mu + B^2/4\pi n$. Using~\eqref{eq19}
\begin{align}
&n u^\alpha \left (\chi u^\beta \xi_\beta \right )_{,\alpha}  + \frac{B^\alpha}{4\pi} \left [  \xi_\beta B^\beta  \frac{\mu_{,\alpha}}{\mu}
- \left ( \xi_\beta B^\beta \right )_{,\alpha} \right ] 
\nn \\
& + \xi_\beta g^{i\beta}  \left ( p + \frac{B^2}{8\pi} \right )_{,i}  = 0 .
\label{eq26}
\end{align}
Furthermore, using~\eqref{allB1} for $B^\alpha$ as well as~\eqref{eq23dif} allow us to rewrite this 
expression as
\be
\frac{d}{d\tau} \left [\, \chi u^\beta \xi_\beta   + \frac{\cC}{4\pi}   ( u_t + \Omega u_\varphi ) \xi_\beta B^\beta\,  \right ] 
= - \frac{ \xi^i}{n}  \left ( p + \frac{B^2}{8\pi} \right )_{,i}.
\label{eq28}
\ee
Insofar the spacetime is endowed with the two standard Killing vectors,
$\xi^\alpha = \{ \xi^\alpha_t, \xi^\alpha_\varphi \} = \{ \delta_t^\alpha, \delta_\varphi^\alpha \}$,
the right hand side term in the preceding expression vanishes and we arrive at the two conserved quantities
\begin{align}
E 
&=  - \left ( \chi u_\alpha  + \frac{\cC D}{4\pi \mu }  B_\alpha \right )  \xi_t^\alpha
=  - \chi u_t  - \frac{\cC D}{4\pi \mu }  B_t,
\label{eq29}
 \\
L 
&=  \left ( \chi u_\alpha + \frac{\cC D}{4\pi \mu}  B_\alpha \right ) \xi_\varphi^\alpha
=    \chi u_\varphi + \frac{\cC D}{4\pi \mu}  B_\varphi.
\label{eq30}
\end{align}
These expressions represent MHD generalisations of the conserved specific energy and angular momentum of a 
non-magnetic fluid. In particular, the bracketed terms can be identified as the system's canonical four-momentum.

A final non-trivial result follows from the combination of $E, L$ and Eq.~\eqref{eq39}. We have
\be \label{grf0}
E -\Omega L + D = {\cF D}/4\pi \mu,
\ee
and given that $\mu$ is not constant along flux surfaces, we should have $ \cF D =0$. Noticing that $D=0$ implies a 
vanishing poloidal field according to~\eqref{eq21and22}, we find that the only acceptable solution is $\cF = 0$. 
Then, $D = - (E -\Omega L) $ and the conservation law~\eqref{eq39} reduces to
\be
 B^2 = - \cC n  ( B_t + \Omega B_\varphi ).
 \label{eq39new}
\ee
It is clear now that the constants $E$ and $D$ are degenerate in the non-magnetic limit and thus one may say that $D$ has `no perfect fluid analogue' as commented by BO78.

For the remainder of this Appendix we assume, in addition to stationarity and axisymmetry, a strictly azimuthal fluid flow i.e.,
$u^\mu = ( u^t, 0,0,u^\varphi)$ [as discussed elsewhere in the paper, this type of flow entails a circular metric and as a consequence $u_\mu = ( u_t, 0,0,u_\varphi)$]. 
From the ideal MHD equation~\eqref{eq6} we have $ u^t F_{\alpha t} + u^\varphi F_{\alpha \varphi } = 0$
and therefore, in the absence of meridional flow (and a non-vanishing poloidal field),  the flow constant $\Omega$ can be identified with the angular frequency
\be
\Omega  = {u^\varphi}/{u^t}.
\ee 
In terms of the scalar magnetic potentials $ \Phi = A_t, \psi = A_\varphi$ we can write $F_{\alpha t} = \Phi_{,\alpha},F_{\alpha\varphi} = \psi_{,\alpha}$
and from the previous expressions we find $\Phi = \Phi (\psi)$ and $\Omega = - d\Phi / d\psi$.
The $\Omega=\Omega(\psi)$ result represents the relativistic version of Ferraro's theorem for differentially rotating MHD
systems (e.g. \cite{gour11}).

Finally, from the definition of $B^\alpha$ we find the following expressions for the poloidal field components
\be
B^r =  \frac{\psi_{,\theta}}{\sqrt{-g}\, u^t}, \qquad B^\theta =  - \frac{\psi_{,r}}{\sqrt{-g}\, u^t}.
\ee
These can be combined to give $ B^i \psi_{,i} = B^\alpha \psi_{,\alpha} = 0$.
This result  represents a well known geometric property, namely, that the poloidal field lines lie in $\psi$ level surfaces.


\section{Flow constant derivation (Newtonian)}
\label{sec:appNewton}

Here we present a self-contained derivation of the flow constants in the Newtonian context; unlike their GR counterparts these are well known, for example,
a concise introduction can be found in Mestel's textbook \cite{mestelbook}, though original derivations can be traced back to Prendergast \cite{p56}, Chandrasekhar \cite{c56}, 
and Woltjer \cite{w59}. Some important exceptions arise between the GR and Newtonian cases, which we point out here.

The ideal MHD equations for a `cold' Newtonian system are
\begin{align}
& \partial_t \rho + \bnabla \cdot ( \rho \bv )  =0,
 \\
\nonumber \\
& \rho ( \partial_t + \bv \cdot \bnabla ) \bv = - \rho \bnabla \Phi - \rho \bnabla h  + \frac{1}{4\pi} ( \bnabla \times \bB ) \times \bB
\nn \\
&=  - \rho \bnabla \left ( \Phi + h \right )  - \bnabla \left ( \frac{B^2}{8\pi} \right )  + \frac{1}{4\pi } ( \bB \cdot \bnabla ) \bB,
\label{EulerNewt}
\\
\nonumber \\
& \bnabla \cdot \bB = 0, \qquad \partial_t \bB = \bnabla \times (\bv \times \bB ),
\label{inductNewt}
\end{align}
together with an EOS and the Poisson equation $\nabla^2 \Phi = 4\pi G \rho$ for the gravitational potential. Note that
we have opted for working with \emph{enthalpy} instead of pressure,
\be
dp = \rho dh ~\Rightarrow ~ \bnabla p = \rho \bnabla h.
\ee 
A general strategy for deriving conservation laws for scalar quantities is to take the scalar product of the above equations with 
the two available vectors, $\{\bv, \bB \}$, and subsequently impose axisymmetry and stationarity. In the present context 
a conservation law means `conservation along flow lines', in other words
\be
\bv \cdot \bnabla Q = 0,
\label{consQN}
\ee
for any scalar function $Q$. Since by assumption $\partial_t Q =0$ this condition amounts to Lie transportation, ${\cal L}_v Q = 0 $. 

The ideal MHD condition is encapsulated in the induction equation and, as in the GR analysis, we start our discussion from there. 
The assumed symmetries of the system imply 
\be
\bv \times \bB = \bnabla \Phi_e,
\label{Phie0}
\ee
from which it follows that the electrostatic potential $\Phi_e$ is conserved, $\bv \cdot  \bnabla \Phi_e  = 0$.

The decomposition of the vectors $\bv, \bB$ into poloidal and toroidal components,
\be
\bv = \bv_{\rm p} + v_\varphi \bphi, \qquad \bB = \bB_{\rm p} + B_\varphi \bphi.
\ee
allows us to write~\eqref{Phie0} as
\be
\bv_{\rm p} \times \bB_{\rm p} + B_\varphi ( \bv_{\rm p} \times \bphi ) - v_\varphi ( \bB_{\rm p} \times \bphi)= \bnabla \Phi_e .
\label{Phie2}
\ee
The projection of this equation along $\bphi$ yields,
\be
\bv_{\rm p} = k (r,\theta)  \bB_{\rm p} .
\ee
This expression should represent the Newtonian analogue of Eq.~\eqref{eq21and22}, so we expect $k$ to be inversely proportional
to $\rho$. This is easy to show with the help of $ \bnabla \cdot \bB_{\rm p} = 0,  \bnabla \cdot (\rho \bv_{\rm p} ) = 0 $. 
From these we have $\bB_{\rm p} \cdot \bnabla ( \rho k ) = 0$ which is equivalent to the conservation law
\be
 \bv  \cdot  \bnabla  \cC_\rN   =0,
\ee
with $\cC_\rN = 1/\rho k$. 
Thus we have found
\be 
\label{polbnewt}
\bB_{\rm p} = \cC_\rN \rho \bv_{\rm p},
\ee
which represents the Newtonian analogue of~\eqref{eq21and22}. 

The poloidal field can be expressed in terms of the magnetic potential function  $\psi (r,\theta)$ in the usual way,
\be
\bB_{\rm p} = \bnabla \psi \times \bnabla \varphi.
\label{psi_def2}
\ee
This parametrisation guarantees that poloidal field lines as well as the meridional flow lines lie in constant $\psi$ surfaces,
$\bB_{\rm p} \cdot \bnabla \psi =0, \bv_{\rm p} \cdot \bnabla \psi = 0$. We can conclude that
\be
\bv  \cdot \bnabla \psi =0,
\ee
which means that conserved quantities are functions $Q=Q(\psi)$, see~\eqref{consQN}.
Using the above expressions in~\eqref{Phie2} gives
\begin{align}
\varpi^{-1} \left ( v_\varphi - k B_\varphi \right )\bnabla \psi = \bnabla \Phi_e,
\end{align}
where $\varpi = r\sin\theta$. The curl of this expression leads to the conserved quantity
\be 
 \varpi^{-1} \left ( v_\varphi - k B_\varphi \right ) = \Omega_\rN (\psi) 
\ee
This result allows us to write the Newtonian counterpart of~\eqref{allB1}
\be
\bB =  \cC_\rN \rho  \left( \bv - \varpi \Omega_\rN \bphi \right ),
\label{allB2}
\ee
and identify $\Omega_\rN$ as the Newtonian analogue of $\Omega$ (notice that in the absence of
a toroidal field this constant reduces to the angular frequency parameter, $v_\varphi = \varpi \Omega_\rN$).

The remaining `hydrodynamical' constants $D_\rN$, $E_\rN$, $L_\rN$ should emerge from projections of the Euler equation. 
The azimuthal component of that equation gives
\be
 \bv_{\rm p} \cdot \bnabla (\varpi v_\varphi ) = \frac{1}{4\pi \rho} \bB_{\rm p} \cdot \bnabla ( \varpi B_\varphi ),
\ee
Using~\eqref{allB2} we can write this expression as a conservation law, $\bv \cdot \bnabla L_\rN =0$, where
\be 
\label{lcstnewt}
L_\rN = \varpi  \left (v_\varphi - \frac{\cC_\rN}{4\pi} B_\varphi \right ) .
\ee
The last two constants $ E_\rN$ and $D_\rN$ are found by dotting the Euler equation with $\bv$ and $\bB$. 
We find,
\begin{align} \label{eulerdotv1}
& \frac{1}{2} \rho \partial_t v^2 + \rho \bv \cdot [ (\bv \cdot \bnabla ) \bv] + \rho \bv \cdot  \bnabla  \left ( \Phi + h \right ) 
\nn \\
& + \bv \cdot \bnabla  \left (  \frac{B^2}{8\pi} \right )   - \frac{1}{4\pi} \bv \cdot [ (\bB \cdot \bnabla) \bB]= 0  ,
\end{align}
\be
\rho \bB \cdot \partial_t \bv + \rho \bB \cdot [  (\bv \cdot \bnabla ) \bv] + \rho \bB \cdot \bnabla \left ( \Phi + h \right ) = 0.
\label{eulerdotB1}
\ee
We can manipulate some of the terms appearing in these expressions, viz.
\begin{align}
 \rho \bv \cdot [ (\bv \cdot \bnabla ) \bv] 
 & =  \bnabla \cdot \left ( \frac{1}{2} \rho v^2 \bv \right ) +  \frac{1}{2} v^2 \partial_t \rho,
\nn \\
 \rho \bv \cdot \bnabla \left ( \Phi + h \right ) 
 & =   \bnabla \cdot \left [ \rho \bv \left ( \Phi + h \right ) \right ]  + \left ( \Phi + h \right ) \partial_t \rho.
 \end{align}
For our stationary system, Eq.~\eqref{eulerdotv1} becomes
\begin{align}
0 =& \bnabla \cdot \left [ \rho \bv  \left ( \frac{1}{2}  v^2  +  \Phi + h  \right  ) \right ] 
\\
&+ \bv \cdot \bnabla  \left ( \frac{B^2}{8\pi} \right )   
- \frac{1}{4\pi} \bv \cdot [ (\bB \cdot \bnabla) \bB] .
\label{eulerdotv3}
\end{align}
In the \emph{non-magnetic} case this equation would have led to the familiar Bernoulli flow-line constant ${\cal B} =  v^2/2  +  \Phi + h $ \cite{c56}.
The absence of $\rho$ in the magnetic terms in \eqref{eulerdotv3} does not allow the same kind of manipulation (unless $\rho$ is constant). 
However, the fact that $\bnabla \cdot \bB_{\rm p} = 0$ and $ \bv_{\rm p} = \bB_{\rm p}/\cC_\rN \rho$ may allow for a $\bB_{\rm p} \cdot \bnabla Q = 0$ conservation law.
Using index notation $(i,j=\{r,\theta,\varphi\})$ for clarity, the magnetic tension term can be written as
\be
 v_i B^j \nabla_j  B^i =  B^j \nabla_j  (  v_i B^i) - B^i B^j \nabla_j v_i.
\ee
At the same time, from the induction equation we have
\be
B^j \nabla_j v^i - \nabla_j ( v^j B^i ) = 0,
\ee
and the magnetic tension term becomes
\be
 v_i B^j \nabla_j  B^i = \bnabla \cdot [ ( \bv \times \bB) \times \bB ] + \frac{1}{2} \bv \cdot \bnabla B^2.
\ee
The resulting Euler equation is
\be
\hspace{-0.1cm}0= \bnabla \cdot \left [ \rho \bv_{\rm p}  \left ( \frac{1}{2}  v^2  +  \Phi + h  \right  )  - \frac{1}{4\pi}  (\bv \times \bB) \times \bB\right ],
\ee 
where we have exploited axisymmetry to set $\bv \to \bv_{\rm p}$ in the first term. The last term can be written as
\be
 (\bv \times \bB) \times \bB  = - \varpi \Omega_\rN ( B^2_{\rm p} \bphi - \bB_{\rm p} B_\varphi ).
\ee
In axisymmetry the first right-hand-side term is divergence free and the Euler equation reduces to
\be
\bv \cdot \bnabla \left [    \left ( \frac{1}{2}  v^2  +  \Phi + h  \right  )  -  \frac{1}{4\pi} \varpi \Omega_\rN \cC_\rN B_\varphi  \right ] = 0.
\ee
We have thus arrived to the MHD generalisation of the Bernoulli constant 
\be 
\label{eq:newtE}
E_\rN =  \frac{1}{2}  v^2  +  \Phi + h  -  \frac{\Omega_\rN \cC_\rN}{4\pi} \varpi  B_\varphi .
\ee
Unlike the analogous GR result \eqref{eq29}, \eqref{eq:newtE} does not depend on the chemical potential.

Our last order of business for this section is the manipulation of Eq.~\eqref{eulerdotB1}. Removing the time-derivative
\be
 \bB \cdot \left[  (\bv \cdot \bnabla ) \bv \right] +  \bB \cdot \bnabla \left ( \Phi + h \right ) = 0 .
\label{eulerdotB2}
\ee
The first term can be manipulated with the help of the induction equation to give
\be
B_i v^j \nabla_j v^i =  \nabla_j ( v^j v^i B_i ) - \frac{1}{2} B^j \nabla_j v^2  .
\ee
The axisymmetric Euler equation is equivalent to
\be
\bv \cdot \bnabla  \left [   \frac{ ( \bB \cdot \bv )}{\cC_\rN \rho}   + \Phi + h  - \frac{1}{2} v^2\right ] = 0, 
\ee
and the corresponding flow constant is
\be
D_\rN = \frac{\bB \cdot \bv}{\cC_\rN \rho}   + \Phi + h  - \frac{1}{2} v^2 .
\ee
With the help of~\eqref{allB2} we can rewrite this as
\be
D_\rN =\frac{1}{2} v^2 + \Phi + h -\varpi \Omega_\rN v_\varphi  .
\ee
As in GR, it can be noticed that the constants $E_\rN$ and $D_\rN$ are degenerate in the non-magnetic limit (both being
equal to the Bernoulli constant ${\cal B}$ in different frames).

We remark that we have been unable to obtain an analogous $\mathcal{F}$ flow constant in the Newtonian case, though the relation 
$D_\rN = E_\rN -\Omega_\rN L_\rN$ essentially corresponds to setting $\mathcal{F} = 0$ in GRMHD, see Eq.~\eqref{grf0}.

The `no toroidal field theorem' discussed in Sec.~\ref{sec:notor} persists in the present Newtonian context.  Indeed, considering the limit
of vanishing meridional flow,  $\bv_{\rm p} \to 0$, we find that~\eqref{polbnewt} predicts $\cC_\rN \to \infty$ if we require a poloidal magnetic field to be present.
However, we can see that the divergence of $\cC_\rN$ makes both the hydrodynamic constants $E_\rN$ and $L_\rN$ divergent unless $B_\varphi \to 0$ 
since there are no other terms available for counterbalancing. As in GR, we are thus forced to conclude that the field must be purely poloidal.


%



\begin{thebibliography}{75}%
\makeatletter
\providecommand \@ifxundefined [1]{%
 \@ifx{#1\undefined}
}%
\providecommand \@ifnum [1]{%
 \ifnum #1\expandafter \@firstoftwo
 \else \expandafter \@secondoftwo
 \fi
}%
\providecommand \@ifx [1]{%
 \ifx #1\expandafter \@firstoftwo
 \else \expandafter \@secondoftwo
 \fi
}%
\providecommand \natexlab [1]{#1}%
\providecommand \enquote  [1]{``#1''}%
\providecommand \bibnamefont  [1]{#1}%
\providecommand \bibfnamefont [1]{#1}%
\providecommand \citenamefont [1]{#1}%
\providecommand \href@noop [0]{\@secondoftwo}%
\providecommand \href [0]{\begingroup \@sanitize@url \@href}%
\providecommand \@href[1]{\@@startlink{#1}\@@href}%
\providecommand \@@href[1]{\endgroup#1\@@endlink}%
\providecommand \@sanitize@url [0]{\catcode `\\12\catcode `\$12\catcode
  `\&12\catcode `\#12\catcode `\^12\catcode `\_12\catcode `\%12\relax}%
\providecommand \@@startlink[1]{}%
\providecommand \@@endlink[0]{}%
\providecommand \url  [0]{\begingroup\@sanitize@url \@url }%
\providecommand \@url [1]{\endgroup\@href {#1}{\urlprefix }}%
\providecommand \urlprefix  [0]{URL }%
\providecommand \Eprint [0]{\href }%
\providecommand \doibase [0]{http://dx.doi.org/}%
\providecommand \selectlanguage [0]{\@gobble}%
\providecommand \bibinfo  [0]{\@secondoftwo}%
\providecommand \bibfield  [0]{\@secondoftwo}%
\providecommand \translation [1]{[#1]}%
\providecommand \BibitemOpen [0]{}%
\providecommand \bibitemStop [0]{}%
\providecommand \bibitemNoStop [0]{.\EOS\space}%
\providecommand \EOS [0]{\spacefactor3000\relax}%
\providecommand \BibitemShut  [1]{\csname bibitem#1\endcsname}%
\let\auto@bib@innerbib\@empty
\bibitem [{\citenamefont {{Duncan}}\ and\ \citenamefont
  {{Thompson}}(1992)}]{dt92}%
  \BibitemOpen
  \bibfield  {author} {\bibinfo {author} {\bibfnamefont {R.~C.}\ \bibnamefont
  {{Duncan}}}\ and\ \bibinfo {author} {\bibfnamefont {C.}~\bibnamefont
  {{Thompson}}},\ }\href {\doibase 10.1086/186413} {\bibfield  {journal}
  {\bibinfo  {journal} {\apjl}\ }\textbf {\bibinfo {volume} {392}},\ \bibinfo
  {pages} {L9} (\bibinfo {year} {1992})}\BibitemShut {NoStop}%
\bibitem [{\citenamefont {{Thompson}}\ and\ \citenamefont
  {{Duncan}}(1993)}]{td93}%
  \BibitemOpen
  \bibfield  {author} {\bibinfo {author} {\bibfnamefont {C.}~\bibnamefont
  {{Thompson}}}\ and\ \bibinfo {author} {\bibfnamefont {R.~C.}\ \bibnamefont
  {{Duncan}}},\ }\href {\doibase 10.1086/172580} {\bibfield  {journal}
  {\bibinfo  {journal} {\apj}\ }\textbf {\bibinfo {volume} {408}},\ \bibinfo
  {pages} {194} (\bibinfo {year} {1993})}\BibitemShut {NoStop}%
\bibitem [{\citenamefont {{Beskin}}\ \emph {et~al.}(1993)\citenamefont
  {{Beskin}}, \citenamefont {{Gurevich}},\ and\ \citenamefont
  {{Istomin}}}]{besk93}%
  \BibitemOpen
  \bibfield  {author} {\bibinfo {author} {\bibfnamefont {V.~S.}\ \bibnamefont
  {{Beskin}}}, \bibinfo {author} {\bibfnamefont {A.~V.}\ \bibnamefont
  {{Gurevich}}}, \ and\ \bibinfo {author} {\bibfnamefont {Y.~N.}\ \bibnamefont
  {{Istomin}}},\ }\href@noop {} {\emph {\bibinfo {title} {{Physics of the
  pulsar magnetosphere}}}}\ (\bibinfo  {publisher} {Cambridge University
  Press},\ \bibinfo {address} {Cambridge, England},\ \bibinfo {year}
  {1993})\BibitemShut {NoStop}%
\bibitem [{\citenamefont {{Melrose}}\ and\ \citenamefont
  {{Yuen}}(2016)}]{mel16}%
  \BibitemOpen
  \bibfield  {author} {\bibinfo {author} {\bibfnamefont {D.~B.}\ \bibnamefont
  {{Melrose}}}\ and\ \bibinfo {author} {\bibfnamefont {R.}~\bibnamefont
  {{Yuen}}},\ }\href {\doibase 10.1017/S0022377816000398} {\bibfield  {journal}
  {\bibinfo  {journal} {Journal of Plasma Physics}\ }\textbf {\bibinfo {volume}
  {82}},\ \bibinfo {eid} {635820202} (\bibinfo {year} {2016})}\BibitemShut
  {NoStop}%
\bibitem [{\citenamefont {{Patruno}}\ and\ \citenamefont
  {{Watts}}(2021)}]{pw21}%
  \BibitemOpen
  \bibfield  {author} {\bibinfo {author} {\bibfnamefont {A.}~\bibnamefont
  {{Patruno}}}\ and\ \bibinfo {author} {\bibfnamefont {A.~L.}\ \bibnamefont
  {{Watts}}},\ }\href {\doibase 10.1007/978-3-662-62110-3_4} {\bibfield
  {journal} {\bibinfo  {journal} {Astrophys. Space Sci. Lib.}\ }\textbf
  {\bibinfo {volume} {461}},\ \bibinfo {pages} {143} (\bibinfo {year}
  {2021})}\BibitemShut {NoStop}%
\bibitem [{\citenamefont {{Glampedakis}}\ and\ \citenamefont
  {{Suvorov}}(2021)}]{gs21}%
  \BibitemOpen
  \bibfield  {author} {\bibinfo {author} {\bibfnamefont {K.}~\bibnamefont
  {{Glampedakis}}}\ and\ \bibinfo {author} {\bibfnamefont {A.~G.}\ \bibnamefont
  {{Suvorov}}},\ }\href {\doibase 10.1093/mnras/stab2689} {\bibfield  {journal}
  {\bibinfo  {journal} {\mnras}\ }\textbf {\bibinfo {volume} {508}},\ \bibinfo
  {pages} {2399} (\bibinfo {year} {2021})}\BibitemShut {NoStop}%
\bibitem [{\citenamefont {{Prendergast}}(1956)}]{p56}%
  \BibitemOpen
  \bibfield  {author} {\bibinfo {author} {\bibfnamefont {K.~H.}\ \bibnamefont
  {{Prendergast}}},\ }\href {\doibase 10.1086/146186} {\bibfield  {journal}
  {\bibinfo  {journal} {\apj}\ }\textbf {\bibinfo {volume} {123}},\ \bibinfo
  {pages} {498} (\bibinfo {year} {1956})}\BibitemShut {NoStop}%
\bibitem [{\citenamefont {{Wright}}(1973)}]{wri73}%
  \BibitemOpen
  \bibfield  {author} {\bibinfo {author} {\bibfnamefont {G.~A.~E.}\
  \bibnamefont {{Wright}}},\ }\href {\doibase 10.1093/mnras/162.4.339}
  {\bibfield  {journal} {\bibinfo  {journal} {\mnras}\ }\textbf {\bibinfo
  {volume} {162}},\ \bibinfo {pages} {339} (\bibinfo {year}
  {1973})}\BibitemShut {NoStop}%
\bibitem [{\citenamefont {{Tayler}}(1973)}]{tay73}%
  \BibitemOpen
  \bibfield  {author} {\bibinfo {author} {\bibfnamefont {R.~J.}\ \bibnamefont
  {{Tayler}}},\ }\href {\doibase 10.1093/mnras/161.4.365} {\bibfield  {journal}
  {\bibinfo  {journal} {\mnras}\ }\textbf {\bibinfo {volume} {161}},\ \bibinfo
  {pages} {365} (\bibinfo {year} {1973})}\BibitemShut {NoStop}%
\bibitem [{\citenamefont {{Braithwaite}}\ and\ \citenamefont
  {{Spruit}}(2004)}]{bs04}%
  \BibitemOpen
  \bibfield  {author} {\bibinfo {author} {\bibfnamefont {J.}~\bibnamefont
  {{Braithwaite}}}\ and\ \bibinfo {author} {\bibfnamefont {H.~C.}\ \bibnamefont
  {{Spruit}}},\ }\href {\doibase 10.1038/nature02934} {\bibfield  {journal}
  {\bibinfo  {journal} {\nat}\ }\textbf {\bibinfo {volume} {431}},\ \bibinfo
  {pages} {819} (\bibinfo {year} {2004})}\BibitemShut {NoStop}%
\bibitem [{\citenamefont {{Braithwaite}}\ and\ \citenamefont
  {{Nordlund}}(2006)}]{bn06}%
  \BibitemOpen
  \bibfield  {author} {\bibinfo {author} {\bibfnamefont {J.}~\bibnamefont
  {{Braithwaite}}}\ and\ \bibinfo {author} {\bibfnamefont
  {{\r{A}}.}~\bibnamefont {{Nordlund}}},\ }\href {\doibase
  10.1051/0004-6361:20041980} {\bibfield  {journal} {\bibinfo  {journal}
  {\aap}\ }\textbf {\bibinfo {volume} {450}},\ \bibinfo {pages} {1077}
  (\bibinfo {year} {2006})}\BibitemShut {NoStop}%
\bibitem [{\citenamefont {{Braithwaite}}\ and\ \citenamefont
  {{Spruit}}(2006)}]{bs06}%
  \BibitemOpen
  \bibfield  {author} {\bibinfo {author} {\bibfnamefont {J.}~\bibnamefont
  {{Braithwaite}}}\ and\ \bibinfo {author} {\bibfnamefont {H.~C.}\ \bibnamefont
  {{Spruit}}},\ }\href {\doibase 10.1051/0004-6361:20041981} {\bibfield
  {journal} {\bibinfo  {journal} {\aap}\ }\textbf {\bibinfo {volume} {450}},\
  \bibinfo {pages} {1097} (\bibinfo {year} {2006})}\BibitemShut {NoStop}%
\bibitem [{\citenamefont {{Braithwaite}}(2009)}]{bra09}%
  \BibitemOpen
  \bibfield  {author} {\bibinfo {author} {\bibfnamefont {J.}~\bibnamefont
  {{Braithwaite}}},\ }\href {\doibase 10.1111/j.1365-2966.2008.14034.x}
  {\bibfield  {journal} {\bibinfo  {journal} {\mnras}\ }\textbf {\bibinfo
  {volume} {397}},\ \bibinfo {pages} {763} (\bibinfo {year}
  {2009})}\BibitemShut {NoStop}%
\bibitem [{\citenamefont {{Akg{\"u}n}}\ \emph {et~al.}(2013)\citenamefont
  {{Akg{\"u}n}}, \citenamefont {{Reisenegger}}, \citenamefont {{Mastrano}},\
  and\ \citenamefont {{Marchant}}}]{akg13}%
  \BibitemOpen
  \bibfield  {author} {\bibinfo {author} {\bibfnamefont {T.}~\bibnamefont
  {{Akg{\"u}n}}}, \bibinfo {author} {\bibfnamefont {A.}~\bibnamefont
  {{Reisenegger}}}, \bibinfo {author} {\bibfnamefont {A.}~\bibnamefont
  {{Mastrano}}}, \ and\ \bibinfo {author} {\bibfnamefont {P.}~\bibnamefont
  {{Marchant}}},\ }\href {\doibase 10.1093/mnras/stt913} {\bibfield  {journal}
  {\bibinfo  {journal} {\mnras}\ }\textbf {\bibinfo {volume} {433}},\ \bibinfo
  {pages} {2445} (\bibinfo {year} {2013})}\BibitemShut {NoStop}%
\bibitem [{\citenamefont {{Mitchell}}\ \emph {et~al.}(2015)\citenamefont
  {{Mitchell}}, \citenamefont {{Braithwaite}}, \citenamefont {{Reisenegger}}
  \emph {et~al.}}]{mit15}%
  \BibitemOpen
  \bibfield  {author} {\bibinfo {author} {\bibfnamefont {J.~P.}\ \bibnamefont
  {{Mitchell}}}, \bibinfo {author} {\bibfnamefont {J.}~\bibnamefont
  {{Braithwaite}}}, \bibinfo {author} {\bibfnamefont {A.}~\bibnamefont
  {{Reisenegger}}},  \emph {et~al.},\ }\href {\doibase 10.1093/mnras/stu2514}
  {\bibfield  {journal} {\bibinfo  {journal} {\mnras}\ }\textbf {\bibinfo
  {volume} {447}},\ \bibinfo {pages} {1213} (\bibinfo {year}
  {2015})}\BibitemShut {NoStop}%
\bibitem [{\citenamefont {{Braithwaite}}(2008)}]{bra08}%
  \BibitemOpen
  \bibfield  {author} {\bibinfo {author} {\bibfnamefont {J.}~\bibnamefont
  {{Braithwaite}}},\ }\href {\doibase 10.1111/j.1365-2966.2008.13218.x}
  {\bibfield  {journal} {\bibinfo  {journal} {\mnras}\ }\textbf {\bibinfo
  {volume} {386}},\ \bibinfo {pages} {1947} (\bibinfo {year}
  {2008})}\BibitemShut {NoStop}%
\bibitem [{\citenamefont {{Becerra}}\ \emph {et~al.}(2022)\citenamefont
  {{Becerra}}, \citenamefont {{Reisenegger}}, \citenamefont {{Valdivia}},\ and\
  \citenamefont {{Gusakov}}}]{bec22}%
  \BibitemOpen
  \bibfield  {author} {\bibinfo {author} {\bibfnamefont {L.}~\bibnamefont
  {{Becerra}}}, \bibinfo {author} {\bibfnamefont {A.}~\bibnamefont
  {{Reisenegger}}}, \bibinfo {author} {\bibfnamefont {J.~A.}\ \bibnamefont
  {{Valdivia}}}, \ and\ \bibinfo {author} {\bibfnamefont {M.~E.}\ \bibnamefont
  {{Gusakov}}},\ }\href {\doibase 10.1093/mnras/stac102} {\bibfield  {journal}
  {\bibinfo  {journal} {\mnras}\ }\textbf {\bibinfo {volume} {511}},\ \bibinfo
  {pages} {732} (\bibinfo {year} {2022})}\BibitemShut {NoStop}%
\bibitem [{\citenamefont {{Philippov}}\ \emph {et~al.}(2014)\citenamefont
  {{Philippov}}, \citenamefont {{Tchekhovskoy}},\ and\ \citenamefont
  {{Li}}}]{phil14}%
  \BibitemOpen
  \bibfield  {author} {\bibinfo {author} {\bibfnamefont {A.}~\bibnamefont
  {{Philippov}}}, \bibinfo {author} {\bibfnamefont {A.}~\bibnamefont
  {{Tchekhovskoy}}}, \ and\ \bibinfo {author} {\bibfnamefont {J.~G.}\
  \bibnamefont {{Li}}},\ }\href {\doibase 10.1093/mnras/stu591} {\bibfield
  {journal} {\bibinfo  {journal} {\mnras}\ }\textbf {\bibinfo {volume} {441}},\
  \bibinfo {pages} {1879} (\bibinfo {year} {2014})}\BibitemShut {NoStop}%
\bibitem [{\citenamefont {{Pons}}\ \emph {et~al.}(2009)\citenamefont {{Pons}},
  \citenamefont {{Miralles}},\ and\ \citenamefont {{Geppert}}}]{pons09}%
  \BibitemOpen
  \bibfield  {author} {\bibinfo {author} {\bibfnamefont {J.~A.}\ \bibnamefont
  {{Pons}}}, \bibinfo {author} {\bibfnamefont {J.~A.}\ \bibnamefont
  {{Miralles}}}, \ and\ \bibinfo {author} {\bibfnamefont {U.}~\bibnamefont
  {{Geppert}}},\ }\href {\doibase 10.1051/0004-6361:200811229} {\bibfield
  {journal} {\bibinfo  {journal} {\aap}\ }\textbf {\bibinfo {volume} {496}},\
  \bibinfo {pages} {207} (\bibinfo {year} {2009})}\BibitemShut {NoStop}%
\bibitem [{\citenamefont {{Gourgouliatos}}\ \emph {et~al.}(2022)\citenamefont
  {{Gourgouliatos}}, \citenamefont {{De Grandis}},\ and\ \citenamefont
  {{Igoshev}}}]{gour22}%
  \BibitemOpen
  \bibfield  {author} {\bibinfo {author} {\bibfnamefont {K.~N.}\ \bibnamefont
  {{Gourgouliatos}}}, \bibinfo {author} {\bibfnamefont {D.}~\bibnamefont {{De
  Grandis}}}, \ and\ \bibinfo {author} {\bibfnamefont {A.}~\bibnamefont
  {{Igoshev}}},\ }\href {\doibase 10.3390/sym14010130} {\bibfield  {journal}
  {\bibinfo  {journal} {Symmetry}\ }\textbf {\bibinfo {volume} {14}},\ \bibinfo
  {pages} {130} (\bibinfo {year} {2022})}\BibitemShut {NoStop}%
\bibitem [{\citenamefont {{Markakis}}\ \emph {et~al.}(2017)\citenamefont
  {{Markakis}}, \citenamefont {{Ury{\={u}}}}, \citenamefont {{Gourgoulhon}}
  \emph {et~al.}}]{mark17}%
  \BibitemOpen
  \bibfield  {author} {\bibinfo {author} {\bibfnamefont {C.}~\bibnamefont
  {{Markakis}}}, \bibinfo {author} {\bibfnamefont {K.}~\bibnamefont
  {{Ury{\={u}}}}}, \bibinfo {author} {\bibfnamefont {E.}~\bibnamefont
  {{Gourgoulhon}}},  \emph {et~al.},\ }\href {\doibase
  10.1103/PhysRevD.96.064019} {\bibfield  {journal} {\bibinfo  {journal}
  {\prd}\ }\textbf {\bibinfo {volume} {96}},\ \bibinfo {eid} {064019} (\bibinfo
  {year} {2017})}\BibitemShut {NoStop}%
\bibitem [{\citenamefont {{Chandrasekhar}}(1956)}]{c56}%
  \BibitemOpen
  \bibfield  {author} {\bibinfo {author} {\bibfnamefont {S.}~\bibnamefont
  {{Chandrasekhar}}},\ }\href {\doibase 10.1086/146217} {\bibfield  {journal}
  {\bibinfo  {journal} {\apj}\ }\textbf {\bibinfo {volume} {124}},\ \bibinfo
  {pages} {232} (\bibinfo {year} {1956})}\BibitemShut {NoStop}%
\bibitem [{\citenamefont {{Woltjer}}(1959)}]{w59}%
  \BibitemOpen
  \bibfield  {author} {\bibinfo {author} {\bibfnamefont {L.}~\bibnamefont
  {{Woltjer}}},\ }\href {\doibase 10.1086/146732} {\bibfield  {journal}
  {\bibinfo  {journal} {\apj}\ }\textbf {\bibinfo {volume} {130}},\ \bibinfo
  {pages} {405} (\bibinfo {year} {1959})}\BibitemShut {NoStop}%
\bibitem [{\citenamefont {Mestel}(1999)}]{mestelbook}%
  \BibitemOpen
  \bibfield  {author} {\bibinfo {author} {\bibfnamefont {L.}~\bibnamefont
  {Mestel}},\ }\href@noop {} {\emph {\bibinfo {title} {Stellar magnetism}}}\
  (\bibinfo  {publisher} {Oxford University Press},\ \bibinfo {address}
  {Oxford},\ \bibinfo {year} {1999})\BibitemShut {NoStop}%
\bibitem [{\citenamefont {{Lovelace}}\ \emph {et~al.}(1986)\citenamefont
  {{Lovelace}}, \citenamefont {{Mehanian}}, \citenamefont {{Mobarry}},\ and\
  \citenamefont {{Sulkanen}}}]{love86}%
  \BibitemOpen
  \bibfield  {author} {\bibinfo {author} {\bibfnamefont {R.~V.~E.}\
  \bibnamefont {{Lovelace}}}, \bibinfo {author} {\bibfnamefont
  {C.}~\bibnamefont {{Mehanian}}}, \bibinfo {author} {\bibfnamefont {C.~M.}\
  \bibnamefont {{Mobarry}}}, \ and\ \bibinfo {author} {\bibfnamefont {M.~E.}\
  \bibnamefont {{Sulkanen}}},\ }\href {\doibase 10.1086/191132} {\bibfield
  {journal} {\bibinfo  {journal} {\apjs}\ }\textbf {\bibinfo {volume} {62}},\
  \bibinfo {pages} {1} (\bibinfo {year} {1986})}\BibitemShut {NoStop}%
\bibitem [{\citenamefont {{Ioka}}\ and\ \citenamefont {{Sasaki}}(2003)}]{is03}%
  \BibitemOpen
  \bibfield  {author} {\bibinfo {author} {\bibfnamefont {K.}~\bibnamefont
  {{Ioka}}}\ and\ \bibinfo {author} {\bibfnamefont {M.}~\bibnamefont
  {{Sasaki}}},\ }\href {\doibase 10.1103/PhysRevD.67.124026} {\bibfield
  {journal} {\bibinfo  {journal} {\prd}\ }\textbf {\bibinfo {volume} {67}},\
  \bibinfo {eid} {124026} (\bibinfo {year} {2003})}\BibitemShut {NoStop}%
\bibitem [{\citenamefont {{Bekenstein}}\ and\ \citenamefont
  {{Oron}}(1978)}]{bek78}%
  \BibitemOpen
  \bibfield  {author} {\bibinfo {author} {\bibfnamefont {J.~D.}\ \bibnamefont
  {{Bekenstein}}}\ and\ \bibinfo {author} {\bibfnamefont {E.}~\bibnamefont
  {{Oron}}},\ }\href {\doibase 10.1103/PhysRevD.18.1809} {\bibfield  {journal}
  {\bibinfo  {journal} {\prd}\ }\textbf {\bibinfo {volume} {18}},\ \bibinfo
  {pages} {1809} (\bibinfo {year} {1978})}\BibitemShut {NoStop}%
\bibitem [{\citenamefont {{Bekenstein}}\ and\ \citenamefont
  {{Oron}}(1979)}]{bek79}%
  \BibitemOpen
  \bibfield  {author} {\bibinfo {author} {\bibfnamefont {J.~D.}\ \bibnamefont
  {{Bekenstein}}}\ and\ \bibinfo {author} {\bibfnamefont {E.}~\bibnamefont
  {{Oron}}},\ }\href {\doibase 10.1103/PhysRevD.19.2827} {\bibfield  {journal}
  {\bibinfo  {journal} {\prd}\ }\textbf {\bibinfo {volume} {19}},\ \bibinfo
  {pages} {2827} (\bibinfo {year} {1979})}\BibitemShut {NoStop}%
\bibitem [{\citenamefont {{Ioka}}\ and\ \citenamefont {{Sasaki}}(2004)}]{is04}%
  \BibitemOpen
  \bibfield  {author} {\bibinfo {author} {\bibfnamefont {K.}~\bibnamefont
  {{Ioka}}}\ and\ \bibinfo {author} {\bibfnamefont {M.}~\bibnamefont
  {{Sasaki}}},\ }\href {\doibase 10.1086/379650} {\bibfield  {journal}
  {\bibinfo  {journal} {\apj}\ }\textbf {\bibinfo {volume} {600}},\ \bibinfo
  {pages} {296} (\bibinfo {year} {2004})}\BibitemShut {NoStop}%
\bibitem [{\citenamefont {{Gourgoulhon}}(2006)}]{gour06}%
  \BibitemOpen
  \bibfield  {author} {\bibinfo {author} {\bibfnamefont {E.}~\bibnamefont
  {{Gourgoulhon}}},\ }\href {\doibase 10.1051/eas:2006106} {\bibfield
  {journal} {\bibinfo  {journal} {EAS Publ. Ser.}\ }\textbf {\bibinfo {volume}
  {21}},\ \bibinfo {pages} {43} (\bibinfo {year} {2006})}\BibitemShut {NoStop}%
\bibitem [{\citenamefont {{Gourgoulhon}}\ \emph {et~al.}(2011)\citenamefont
  {{Gourgoulhon}}, \citenamefont {{Markakis}}, \citenamefont {{Ury{\={u}}}},\
  and\ \citenamefont {{Eriguchi}}}]{gour11}%
  \BibitemOpen
  \bibfield  {author} {\bibinfo {author} {\bibfnamefont {E.}~\bibnamefont
  {{Gourgoulhon}}}, \bibinfo {author} {\bibfnamefont {C.}~\bibnamefont
  {{Markakis}}}, \bibinfo {author} {\bibfnamefont {K.}~\bibnamefont
  {{Ury{\={u}}}}}, \ and\ \bibinfo {author} {\bibfnamefont {Y.}~\bibnamefont
  {{Eriguchi}}},\ }\href {\doibase 10.1103/PhysRevD.83.104007} {\bibfield
  {journal} {\bibinfo  {journal} {\prd}\ }\textbf {\bibinfo {volume} {83}},\
  \bibinfo {eid} {104007} (\bibinfo {year} {2011})}\BibitemShut {NoStop}%
\bibitem [{\citenamefont {{Ury{\={u}}}}\ \emph {et~al.}(2019)\citenamefont
  {{Ury{\={u}}}}, \citenamefont {{Yoshida}}, \citenamefont {{Gourgoulhon}}
  \emph {et~al.}}]{uryu19}%
  \BibitemOpen
  \bibfield  {author} {\bibinfo {author} {\bibfnamefont {K.}~\bibnamefont
  {{Ury{\={u}}}}}, \bibinfo {author} {\bibfnamefont {S.}~\bibnamefont
  {{Yoshida}}}, \bibinfo {author} {\bibfnamefont {E.}~\bibnamefont
  {{Gourgoulhon}}},  \emph {et~al.},\ }\href {\doibase
  10.1103/PhysRevD.100.123019} {\bibfield  {journal} {\bibinfo  {journal}
  {\prd}\ }\textbf {\bibinfo {volume} {100}},\ \bibinfo {eid} {123019}
  (\bibinfo {year} {2019})}\BibitemShut {NoStop}%
\bibitem [{\citenamefont {{Tsokaros}}\ \emph {et~al.}(2022)\citenamefont
  {{Tsokaros}}, \citenamefont {{Ruiz}}, \citenamefont {{Shapiro}},\ and\
  \citenamefont {{Ury{\={u}}}}}]{uryu22}%
  \BibitemOpen
  \bibfield  {author} {\bibinfo {author} {\bibfnamefont {A.}~\bibnamefont
  {{Tsokaros}}}, \bibinfo {author} {\bibfnamefont {M.}~\bibnamefont {{Ruiz}}},
  \bibinfo {author} {\bibfnamefont {S.~L.}\ \bibnamefont {{Shapiro}}}, \ and\
  \bibinfo {author} {\bibfnamefont {K.}~\bibnamefont {{Ury{\={u}}}}},\ }\href
  {\doibase 10.1103/PhysRevLett.128.061101} {\bibfield  {journal} {\bibinfo
  {journal} {\prl}\ }\textbf {\bibinfo {volume} {128}},\ \bibinfo {eid}
  {061101} (\bibinfo {year} {2022})}\BibitemShut {NoStop}%
\bibitem [{\citenamefont {{Oron}}(2002)}]{oron02}%
  \BibitemOpen
  \bibfield  {author} {\bibinfo {author} {\bibfnamefont {A.}~\bibnamefont
  {{Oron}}},\ }\href {\doibase 10.1103/PhysRevD.66.023006} {\bibfield
  {journal} {\bibinfo  {journal} {\prd}\ }\textbf {\bibinfo {volume} {66}},\
  \bibinfo {eid} {023006} (\bibinfo {year} {2002})}\BibitemShut {NoStop}%
\bibitem [{\citenamefont {{Frieben}}\ and\ \citenamefont
  {{Rezzolla}}(2012)}]{fr12}%
  \BibitemOpen
  \bibfield  {author} {\bibinfo {author} {\bibfnamefont {J.}~\bibnamefont
  {{Frieben}}}\ and\ \bibinfo {author} {\bibfnamefont {L.}~\bibnamefont
  {{Rezzolla}}},\ }\href {\doibase 10.1111/j.1365-2966.2012.22027.x} {\bibfield
   {journal} {\bibinfo  {journal} {\mnras}\ }\textbf {\bibinfo {volume}
  {427}},\ \bibinfo {pages} {3406} (\bibinfo {year} {2012})}\BibitemShut
  {NoStop}%
\bibitem [{\citenamefont {{Lasky}}\ \emph {et~al.}(2011)\citenamefont
  {{Lasky}}, \citenamefont {{Zink}}, \citenamefont {{Kokkotas}},\ and\
  \citenamefont {{Glampedakis}}}]{lasky11}%
  \BibitemOpen
  \bibfield  {author} {\bibinfo {author} {\bibfnamefont {P.~D.}\ \bibnamefont
  {{Lasky}}}, \bibinfo {author} {\bibfnamefont {B.}~\bibnamefont {{Zink}}},
  \bibinfo {author} {\bibfnamefont {K.~D.}\ \bibnamefont {{Kokkotas}}}, \ and\
  \bibinfo {author} {\bibfnamefont {K.}~\bibnamefont {{Glampedakis}}},\ }\href
  {\doibase 10.1088/2041-8205/735/1/L20} {\bibfield  {journal} {\bibinfo
  {journal} {\apjl}\ }\textbf {\bibinfo {volume} {735}},\ \bibinfo {eid} {L20}
  (\bibinfo {year} {2011})}\BibitemShut {NoStop}%
\bibitem [{\citenamefont {{Ciolfi}}\ \emph {et~al.}(2011)\citenamefont
  {{Ciolfi}}, \citenamefont {{Lander}}, \citenamefont {{Manca}},\ and\
  \citenamefont {{Rezzolla}}}]{rezz11}%
  \BibitemOpen
  \bibfield  {author} {\bibinfo {author} {\bibfnamefont {R.}~\bibnamefont
  {{Ciolfi}}}, \bibinfo {author} {\bibfnamefont {S.~K.}\ \bibnamefont
  {{Lander}}}, \bibinfo {author} {\bibfnamefont {G.~M.}\ \bibnamefont
  {{Manca}}}, \ and\ \bibinfo {author} {\bibfnamefont {L.}~\bibnamefont
  {{Rezzolla}}},\ }\href {\doibase 10.1088/2041-8205/736/1/L6} {\bibfield
  {journal} {\bibinfo  {journal} {\apjl}\ }\textbf {\bibinfo {volume} {736}},\
  \bibinfo {eid} {L6} (\bibinfo {year} {2011})}\BibitemShut {NoStop}%
\bibitem [{\citenamefont {{Kiuchi}}\ \emph {et~al.}(2011)\citenamefont
  {{Kiuchi}}, \citenamefont {{Yoshida}},\ and\ \citenamefont
  {{Shibata}}}]{kiu12}%
  \BibitemOpen
  \bibfield  {author} {\bibinfo {author} {\bibfnamefont {K.}~\bibnamefont
  {{Kiuchi}}}, \bibinfo {author} {\bibfnamefont {S.}~\bibnamefont {{Yoshida}}},
  \ and\ \bibinfo {author} {\bibfnamefont {M.}~\bibnamefont {{Shibata}}},\
  }\href {\doibase 10.1051/0004-6361/201016242} {\bibfield  {journal} {\bibinfo
   {journal} {\aap}\ }\textbf {\bibinfo {volume} {532}},\ \bibinfo {eid} {A30}
  (\bibinfo {year} {2011})}\BibitemShut {NoStop}%
\bibitem [{\citenamefont {{Ciolfi}}\ and\ \citenamefont
  {{Rezzolla}}(2012)}]{rezz12}%
  \BibitemOpen
  \bibfield  {author} {\bibinfo {author} {\bibfnamefont {R.}~\bibnamefont
  {{Ciolfi}}}\ and\ \bibinfo {author} {\bibfnamefont {L.}~\bibnamefont
  {{Rezzolla}}},\ }\href {\doibase 10.1088/0004-637X/760/1/1} {\bibfield
  {journal} {\bibinfo  {journal} {\apj}\ }\textbf {\bibinfo {volume} {760}},\
  \bibinfo {eid} {1} (\bibinfo {year} {2012})}\BibitemShut {NoStop}%
\bibitem [{\citenamefont {{Ofengeim}}\ and\ \citenamefont
  {{Gusakov}}(2018)}]{gusakov18}%
  \BibitemOpen
  \bibfield  {author} {\bibinfo {author} {\bibfnamefont {D.~D.}\ \bibnamefont
  {{Ofengeim}}}\ and\ \bibinfo {author} {\bibfnamefont {M.~E.}\ \bibnamefont
  {{Gusakov}}},\ }\href {\doibase 10.1103/PhysRevD.98.043007} {\bibfield
  {journal} {\bibinfo  {journal} {\prd}\ }\textbf {\bibinfo {volume} {98}},\
  \bibinfo {eid} {043007} (\bibinfo {year} {2018})}\BibitemShut {NoStop}%
\bibitem [{\citenamefont {{Colaiuda}}\ \emph {et~al.}(2008)\citenamefont
  {{Colaiuda}}, \citenamefont {{Ferrari}}, \citenamefont {{Gualtieri}},\ and\
  \citenamefont {{Pons}}}]{col08}%
  \BibitemOpen
  \bibfield  {author} {\bibinfo {author} {\bibfnamefont {A.}~\bibnamefont
  {{Colaiuda}}}, \bibinfo {author} {\bibfnamefont {V.}~\bibnamefont
  {{Ferrari}}}, \bibinfo {author} {\bibfnamefont {L.}~\bibnamefont
  {{Gualtieri}}}, \ and\ \bibinfo {author} {\bibfnamefont {J.~A.}\ \bibnamefont
  {{Pons}}},\ }\href {\doibase 10.1111/j.1365-2966.2008.12966.x} {\bibfield
  {journal} {\bibinfo  {journal} {\mnras}\ }\textbf {\bibinfo {volume} {385}},\
  \bibinfo {pages} {2080} (\bibinfo {year} {2008})}\BibitemShut {NoStop}%
\bibitem [{\citenamefont {{Ciolfi}}\ \emph {et~al.}(2009)\citenamefont
  {{Ciolfi}}, \citenamefont {{Ferrari}}, \citenamefont {{Gualtieri}},\ and\
  \citenamefont {{Pons}}}]{c09}%
  \BibitemOpen
  \bibfield  {author} {\bibinfo {author} {\bibfnamefont {R.}~\bibnamefont
  {{Ciolfi}}}, \bibinfo {author} {\bibfnamefont {V.}~\bibnamefont {{Ferrari}}},
  \bibinfo {author} {\bibfnamefont {L.}~\bibnamefont {{Gualtieri}}}, \ and\
  \bibinfo {author} {\bibfnamefont {J.~A.}\ \bibnamefont {{Pons}}},\ }\href
  {\doibase 10.1111/j.1365-2966.2009.14990.x} {\bibfield  {journal} {\bibinfo
  {journal} {\mnras}\ }\textbf {\bibinfo {volume} {397}},\ \bibinfo {pages}
  {913} (\bibinfo {year} {2009})}\BibitemShut {NoStop}%
\bibitem [{\citenamefont {{Ciolfi}}\ \emph {et~al.}(2010)\citenamefont
  {{Ciolfi}}, \citenamefont {{Ferrari}},\ and\ \citenamefont
  {{Gualtieri}}}]{c10}%
  \BibitemOpen
  \bibfield  {author} {\bibinfo {author} {\bibfnamefont {R.}~\bibnamefont
  {{Ciolfi}}}, \bibinfo {author} {\bibfnamefont {V.}~\bibnamefont {{Ferrari}}},
  \ and\ \bibinfo {author} {\bibfnamefont {L.}~\bibnamefont {{Gualtieri}}},\
  }\href {\doibase 10.1111/j.1365-2966.2010.16847.x} {\bibfield  {journal}
  {\bibinfo  {journal} {\mnras}\ }\textbf {\bibinfo {volume} {406}},\ \bibinfo
  {pages} {2540} (\bibinfo {year} {2010})}\BibitemShut {NoStop}%
\bibitem [{\citenamefont {{Prix}}(2005)}]{prix05}%
  \BibitemOpen
  \bibfield  {author} {\bibinfo {author} {\bibfnamefont {R.}~\bibnamefont
  {{Prix}}},\ }\href {\doibase 10.1103/PhysRevD.71.083006} {\bibfield
  {journal} {\bibinfo  {journal} {\prd}\ }\textbf {\bibinfo {volume} {71}},\
  \bibinfo {eid} {083006} (\bibinfo {year} {2005})}\BibitemShut {NoStop}%
\bibitem [{\citenamefont {{Glampedakis}}\ \emph {et~al.}(2012)\citenamefont
  {{Glampedakis}}, \citenamefont {{Andersson}},\ and\ \citenamefont
  {{Lander}}}]{g12}%
  \BibitemOpen
  \bibfield  {author} {\bibinfo {author} {\bibfnamefont {K.}~\bibnamefont
  {{Glampedakis}}}, \bibinfo {author} {\bibfnamefont {N.}~\bibnamefont
  {{Andersson}}}, \ and\ \bibinfo {author} {\bibfnamefont {S.~K.}\ \bibnamefont
  {{Lander}}},\ }\href {\doibase 10.1111/j.1365-2966.2011.20112.x} {\bibfield
  {journal} {\bibinfo  {journal} {\mnras}\ }\textbf {\bibinfo {volume} {420}},\
  \bibinfo {pages} {1263} (\bibinfo {year} {2012})}\BibitemShut {NoStop}%
\bibitem [{\citenamefont {{Carter}}(1969)}]{cart69}%
  \BibitemOpen
  \bibfield  {author} {\bibinfo {author} {\bibfnamefont {B.}~\bibnamefont
  {{Carter}}},\ }\href {\doibase 10.1063/1.1664763} {\bibfield  {journal}
  {\bibinfo  {journal} {Journal of Mathematical Physics}\ }\textbf {\bibinfo
  {volume} {10}},\ \bibinfo {pages} {70} (\bibinfo {year} {1969})}\BibitemShut
  {NoStop}%
\bibitem [{\citenamefont {{Carter}}(1973)}]{cart73}%
  \BibitemOpen
  \bibfield  {author} {\bibinfo {author} {\bibfnamefont {B.}~\bibnamefont
  {{Carter}}},\ }in\ \href@noop {} {\emph {\bibinfo {booktitle} {Black
  Holes}}},\ \bibinfo {editor} {edited by\ \bibinfo {editor} {\bibfnamefont
  {B.~S.}\ \bibnamefont {{DeWitt}}}\ and\ \bibinfo {editor} {\bibfnamefont
  {C.~M.}\ \bibnamefont {{DeWitt}}}}\ (\bibinfo  {publisher} {Gordon and
  Breach, New York},\ \bibinfo {year} {1973})\BibitemShut {NoStop}%
\bibitem [{\citenamefont {{Papapetrou}}(1966)}]{pap66}%
  \BibitemOpen
  \bibfield  {author} {\bibinfo {author} {\bibfnamefont {A.}~\bibnamefont
  {{Papapetrou}}},\ }\href@noop {} {\bibfield  {journal} {\bibinfo  {journal}
  {Ann. Inst. H. Poincar{\'e} A}\ }\textbf {\bibinfo {volume} {4}},\ \bibinfo
  {pages} {83} (\bibinfo {year} {1966})}\BibitemShut {NoStop}%
\bibitem [{\citenamefont {{Bocquet}}\ \emph {et~al.}(1995)\citenamefont
  {{Bocquet}}, \citenamefont {{Bonazzola}}, \citenamefont {{Gourgoulhon}},\
  and\ \citenamefont {{Novak}}}]{boq95}%
  \BibitemOpen
  \bibfield  {author} {\bibinfo {author} {\bibfnamefont {M.}~\bibnamefont
  {{Bocquet}}}, \bibinfo {author} {\bibfnamefont {S.}~\bibnamefont
  {{Bonazzola}}}, \bibinfo {author} {\bibfnamefont {E.}~\bibnamefont
  {{Gourgoulhon}}}, \ and\ \bibinfo {author} {\bibfnamefont {J.}~\bibnamefont
  {{Novak}}},\ }\href {\doibase 10.48550/arXiv.gr-qc/9503044} {\bibfield
  {journal} {\bibinfo  {journal} {\aap}\ }\textbf {\bibinfo {volume} {301}},\
  \bibinfo {pages} {757} (\bibinfo {year} {1995})}\BibitemShut {NoStop}%
\bibitem [{\citenamefont {{Pili}}\ \emph {et~al.}(2017)\citenamefont {{Pili}},
  \citenamefont {{Bucciantini}},\ and\ \citenamefont {{Del Zanna}}}]{pili17}%
  \BibitemOpen
  \bibfield  {author} {\bibinfo {author} {\bibfnamefont {A.~G.}\ \bibnamefont
  {{Pili}}}, \bibinfo {author} {\bibfnamefont {N.}~\bibnamefont
  {{Bucciantini}}}, \ and\ \bibinfo {author} {\bibfnamefont {L.}~\bibnamefont
  {{Del Zanna}}},\ }\href {\doibase 10.1093/mnras/stx1176} {\bibfield
  {journal} {\bibinfo  {journal} {\mnras}\ }\textbf {\bibinfo {volume} {470}},\
  \bibinfo {pages} {2469} (\bibinfo {year} {2017})}\BibitemShut {NoStop}%
\bibitem [{\citenamefont {{Fujisawa}}\ \emph {et~al.}(2013)\citenamefont
  {{Fujisawa}}, \citenamefont {{Takahashi}}, \citenamefont {{Yoshida}},\ and\
  \citenamefont {{Eriguchi}}}]{fuji13}%
  \BibitemOpen
  \bibfield  {author} {\bibinfo {author} {\bibfnamefont {K.}~\bibnamefont
  {{Fujisawa}}}, \bibinfo {author} {\bibfnamefont {R.}~\bibnamefont
  {{Takahashi}}}, \bibinfo {author} {\bibfnamefont {S.}~\bibnamefont
  {{Yoshida}}}, \ and\ \bibinfo {author} {\bibfnamefont {Y.}~\bibnamefont
  {{Eriguchi}}},\ }\href {\doibase 10.1093/mnras/stt275} {\bibfield  {journal}
  {\bibinfo  {journal} {\mnras}\ }\textbf {\bibinfo {volume} {431}},\ \bibinfo
  {pages} {1453} (\bibinfo {year} {2013})}\BibitemShut {NoStop}%
\bibitem [{\citenamefont {{Kiuchi}}\ and\ \citenamefont
  {{Yoshida}}(2008)}]{kiu08}%
  \BibitemOpen
  \bibfield  {author} {\bibinfo {author} {\bibfnamefont {K.}~\bibnamefont
  {{Kiuchi}}}\ and\ \bibinfo {author} {\bibfnamefont {S.}~\bibnamefont
  {{Yoshida}}},\ }\href {\doibase 10.1103/PhysRevD.78.044045} {\bibfield
  {journal} {\bibinfo  {journal} {\prd}\ }\textbf {\bibinfo {volume} {78}},\
  \bibinfo {eid} {044045} (\bibinfo {year} {2008})}\BibitemShut {NoStop}%
\bibitem [{\citenamefont {{Goldreich}}\ and\ \citenamefont
  {{Julian}}(1969)}]{gj69}%
  \BibitemOpen
  \bibfield  {author} {\bibinfo {author} {\bibfnamefont {P.}~\bibnamefont
  {{Goldreich}}}\ and\ \bibinfo {author} {\bibfnamefont {W.~H.}\ \bibnamefont
  {{Julian}}},\ }\href {\doibase 10.1086/150119} {\bibfield  {journal}
  {\bibinfo  {journal} {\apj}\ }\textbf {\bibinfo {volume} {157}},\ \bibinfo
  {pages} {869} (\bibinfo {year} {1969})}\BibitemShut {NoStop}%
\bibitem [{\citenamefont {{Yoshida}}\ \emph {et~al.}(2012)\citenamefont
  {{Yoshida}}, \citenamefont {{Kiuchi}},\ and\ \citenamefont
  {{Shibata}}}]{yosh12}%
  \BibitemOpen
  \bibfield  {author} {\bibinfo {author} {\bibfnamefont {S.}~\bibnamefont
  {{Yoshida}}}, \bibinfo {author} {\bibfnamefont {K.}~\bibnamefont {{Kiuchi}}},
  \ and\ \bibinfo {author} {\bibfnamefont {M.}~\bibnamefont {{Shibata}}},\
  }\href {\doibase 10.1103/PhysRevD.86.044012} {\bibfield  {journal} {\bibinfo
  {journal} {\prd}\ }\textbf {\bibinfo {volume} {86}},\ \bibinfo {eid} {044012}
  (\bibinfo {year} {2012})}\BibitemShut {NoStop}%
\bibitem [{\citenamefont {{Konno}}\ \emph {et~al.}(1999)\citenamefont
  {{Konno}}, \citenamefont {{Obata}},\ and\ \citenamefont
  {{Kojima}}}]{konno99}%
  \BibitemOpen
  \bibfield  {author} {\bibinfo {author} {\bibfnamefont {K.}~\bibnamefont
  {{Konno}}}, \bibinfo {author} {\bibfnamefont {T.}~\bibnamefont {{Obata}}}, \
  and\ \bibinfo {author} {\bibfnamefont {Y.}~\bibnamefont {{Kojima}}},\ }\href
  {\doibase 10.48550/arXiv.gr-qc/9910038} {\bibfield  {journal} {\bibinfo
  {journal} {\aap}\ }\textbf {\bibinfo {volume} {352}},\ \bibinfo {pages} {211}
  (\bibinfo {year} {1999})}\BibitemShut {NoStop}%
\bibitem [{\citenamefont {{P{\'e}tri}}(2016)}]{pet16}%
  \BibitemOpen
  \bibfield  {author} {\bibinfo {author} {\bibfnamefont {J.}~\bibnamefont
  {{P{\'e}tri}}},\ }\href {\doibase 10.1093/mnras/stv2613} {\bibfield
  {journal} {\bibinfo  {journal} {\mnras}\ }\textbf {\bibinfo {volume} {455}},\
  \bibinfo {pages} {3779} (\bibinfo {year} {2016})}\BibitemShut {NoStop}%
\bibitem [{\citenamefont {{Mestel}}(1968)}]{mest68}%
  \BibitemOpen
  \bibfield  {author} {\bibinfo {author} {\bibfnamefont {L.}~\bibnamefont
  {{Mestel}}},\ }\href {\doibase 10.1093/mnras/138.3.359} {\bibfield  {journal}
  {\bibinfo  {journal} {\mnras}\ }\textbf {\bibinfo {volume} {138}},\ \bibinfo
  {pages} {359} (\bibinfo {year} {1968})}\BibitemShut {NoStop}%
\bibitem [{\citenamefont {{Ogilvie}}(2016)}]{og16}%
  \BibitemOpen
  \bibfield  {author} {\bibinfo {author} {\bibfnamefont {G.~I.}\ \bibnamefont
  {{Ogilvie}}},\ }\href {\doibase 10.1017/S0022377816000489} {\bibfield
  {journal} {\bibinfo  {journal} {Journal of Plasma Physics}\ }\textbf
  {\bibinfo {volume} {82}},\ \bibinfo {eid} {205820301} (\bibinfo {year}
  {2016})}\BibitemShut {NoStop}%
\bibitem [{\citenamefont {{Carter}}\ and\ \citenamefont
  {{Quintana}}(1972)}]{cq72}%
  \BibitemOpen
  \bibfield  {author} {\bibinfo {author} {\bibfnamefont {B.}~\bibnamefont
  {{Carter}}}\ and\ \bibinfo {author} {\bibfnamefont {H.}~\bibnamefont
  {{Quintana}}},\ }\href {\doibase 10.1098/rspa.1972.0164} {\bibfield
  {journal} {\bibinfo  {journal} {Proceedings of the Royal Society of London
  Series A}\ }\textbf {\bibinfo {volume} {331}},\ \bibinfo {pages} {57}
  (\bibinfo {year} {1972})}\BibitemShut {NoStop}%
\bibitem [{\citenamefont {{Kojima}}\ \emph {et~al.}(2022)\citenamefont
  {{Kojima}}, \citenamefont {{Kisaka}},\ and\ \citenamefont
  {{Fujisawa}}}]{koj22}%
  \BibitemOpen
  \bibfield  {author} {\bibinfo {author} {\bibfnamefont {Y.}~\bibnamefont
  {{Kojima}}}, \bibinfo {author} {\bibfnamefont {S.}~\bibnamefont {{Kisaka}}},
  \ and\ \bibinfo {author} {\bibfnamefont {K.}~\bibnamefont {{Fujisawa}}},\
  }\href {\doibase 10.1093/mnras/stac036} {\bibfield  {journal} {\bibinfo
  {journal} {\mnras}\ }\textbf {\bibinfo {volume} {511}},\ \bibinfo {pages}
  {480} (\bibinfo {year} {2022})}\BibitemShut {NoStop}%
\bibitem [{\citenamefont {Steil}(2018)}]{steil18}%
  \BibitemOpen
  \bibfield  {author} {\bibinfo {author} {\bibfnamefont {M.~J.}\ \bibnamefont
  {Steil}},\ }\emph {\bibinfo {title} {Structure of slowly rotating magnetized
  neutron stars in a perturbative approach}},\ \href@noop {} {Ph.D. thesis},\
  \bibinfo  {school} {Technische Universit{\"a}t Darmstadt} (\bibinfo {year}
  {2018})\BibitemShut {NoStop}%
\bibitem [{\citenamefont {{Haskell}}\ \emph {et~al.}(2008)\citenamefont
  {{Haskell}}, \citenamefont {{Samuelsson}}, \citenamefont {{Glampedakis}},\
  and\ \citenamefont {{Andersson}}}]{hask08}%
  \BibitemOpen
  \bibfield  {author} {\bibinfo {author} {\bibfnamefont {B.}~\bibnamefont
  {{Haskell}}}, \bibinfo {author} {\bibfnamefont {L.}~\bibnamefont
  {{Samuelsson}}}, \bibinfo {author} {\bibfnamefont {K.}~\bibnamefont
  {{Glampedakis}}}, \ and\ \bibinfo {author} {\bibfnamefont {N.}~\bibnamefont
  {{Andersson}}},\ }\href {\doibase 10.1111/j.1365-2966.2008.12861.x}
  {\bibfield  {journal} {\bibinfo  {journal} {\mnras}\ }\textbf {\bibinfo
  {volume} {385}},\ \bibinfo {pages} {531} (\bibinfo {year}
  {2008})}\BibitemShut {NoStop}%
\bibitem [{\citenamefont {{Ciolfi}}\ and\ \citenamefont
  {{Rezzolla}}(2013)}]{c13}%
  \BibitemOpen
  \bibfield  {author} {\bibinfo {author} {\bibfnamefont {R.}~\bibnamefont
  {{Ciolfi}}}\ and\ \bibinfo {author} {\bibfnamefont {L.}~\bibnamefont
  {{Rezzolla}}},\ }\href {\doibase 10.1093/mnrasl/slt092} {\bibfield  {journal}
  {\bibinfo  {journal} {\mnras}\ }\textbf {\bibinfo {volume} {435}},\ \bibinfo
  {pages} {L43} (\bibinfo {year} {2013})}\BibitemShut {NoStop}%
\bibitem [{\citenamefont {{Jiang}}\ and\ \citenamefont {{Yagi}}(2019)}]{j19}%
  \BibitemOpen
  \bibfield  {author} {\bibinfo {author} {\bibfnamefont {N.}~\bibnamefont
  {{Jiang}}}\ and\ \bibinfo {author} {\bibfnamefont {K.}~\bibnamefont
  {{Yagi}}},\ }\href {\doibase 10.1103/PhysRevD.99.124029} {\bibfield
  {journal} {\bibinfo  {journal} {\prd}\ }\textbf {\bibinfo {volume} {99}},\
  \bibinfo {eid} {124029} (\bibinfo {year} {2019})}\BibitemShut {NoStop}%
\bibitem [{\citenamefont {{Bilous}}\ \emph {et~al.}(2019)\citenamefont
  {{Bilous}}, \citenamefont {{Watts}}, \citenamefont {{Harding}} \emph
  {et~al.}}]{bil19}%
  \BibitemOpen
  \bibfield  {author} {\bibinfo {author} {\bibfnamefont {A.~V.}\ \bibnamefont
  {{Bilous}}}, \bibinfo {author} {\bibfnamefont {A.~L.}\ \bibnamefont
  {{Watts}}}, \bibinfo {author} {\bibfnamefont {A.~K.}\ \bibnamefont
  {{Harding}}},  \emph {et~al.},\ }\href {\doibase 10.3847/2041-8213/ab53e7}
  {\bibfield  {journal} {\bibinfo  {journal} {\apjl}\ }\textbf {\bibinfo
  {volume} {887}},\ \bibinfo {eid} {L23} (\bibinfo {year} {2019})}\BibitemShut
  {NoStop}%
\bibitem [{\citenamefont {{Glampedakis}}\ and\ \citenamefont
  {{Lasky}}(2015)}]{gl15}%
  \BibitemOpen
  \bibfield  {author} {\bibinfo {author} {\bibfnamefont {K.}~\bibnamefont
  {{Glampedakis}}}\ and\ \bibinfo {author} {\bibfnamefont {P.~D.}\ \bibnamefont
  {{Lasky}}},\ }\href {\doibase 10.1093/mnras/stv638} {\bibfield  {journal}
  {\bibinfo  {journal} {\mnras}\ }\textbf {\bibinfo {volume} {450}},\ \bibinfo
  {pages} {1638} (\bibinfo {year} {2015})}\BibitemShut {NoStop}%
\bibitem [{\citenamefont {{Makishima}}\ \emph {et~al.}(2021)\citenamefont
  {{Makishima}}, \citenamefont {{Enoto}}, \citenamefont {{Yoneda}},\ and\
  \citenamefont {{Odaka}}}]{mak21}%
  \BibitemOpen
  \bibfield  {author} {\bibinfo {author} {\bibfnamefont {K.}~\bibnamefont
  {{Makishima}}}, \bibinfo {author} {\bibfnamefont {T.}~\bibnamefont
  {{Enoto}}}, \bibinfo {author} {\bibfnamefont {H.}~\bibnamefont {{Yoneda}}}, \
  and\ \bibinfo {author} {\bibfnamefont {H.}~\bibnamefont {{Odaka}}},\ }\href
  {\doibase 10.1093/mnras/stab149} {\bibfield  {journal} {\bibinfo  {journal}
  {\mnras}\ }\textbf {\bibinfo {volume} {502}},\ \bibinfo {pages} {2266}
  (\bibinfo {year} {2021})}\BibitemShut {NoStop}%
\bibitem [{\citenamefont {{Suvorov}}(2023)}]{suv23}%
  \BibitemOpen
  \bibfield  {author} {\bibinfo {author} {\bibfnamefont {A.~G.}\ \bibnamefont
  {{Suvorov}}},\ }\href {\doibase 10.1093/mnras/stad1672} {\bibfield  {journal}
  {\bibinfo  {journal} {\mnras}\ }\textbf {\bibinfo {volume} {523}},\ \bibinfo
  {pages} {4089} (\bibinfo {year} {2023})}\BibitemShut {NoStop}%
\bibitem [{\citenamefont {{Pons}}\ and\ \citenamefont
  {{Perna}}(2011)}]{pons11}%
  \BibitemOpen
  \bibfield  {author} {\bibinfo {author} {\bibfnamefont {J.~A.}\ \bibnamefont
  {{Pons}}}\ and\ \bibinfo {author} {\bibfnamefont {R.}~\bibnamefont
  {{Perna}}},\ }\href {\doibase 10.1088/0004-637X/741/2/123} {\bibfield
  {journal} {\bibinfo  {journal} {\apj}\ }\textbf {\bibinfo {volume} {741}},\
  \bibinfo {eid} {123} (\bibinfo {year} {2011})}\BibitemShut {NoStop}%
\bibitem [{\citenamefont {{Igoshev}}\ \emph {et~al.}(2021)\citenamefont
  {{Igoshev}}, \citenamefont {{Hollerbach}}, \citenamefont {{Wood}},\ and\
  \citenamefont {{Gourgouliatos}}}]{igo21}%
  \BibitemOpen
  \bibfield  {author} {\bibinfo {author} {\bibfnamefont {A.~P.}\ \bibnamefont
  {{Igoshev}}}, \bibinfo {author} {\bibfnamefont {R.}~\bibnamefont
  {{Hollerbach}}}, \bibinfo {author} {\bibfnamefont {T.}~\bibnamefont
  {{Wood}}}, \ and\ \bibinfo {author} {\bibfnamefont {K.~N.}\ \bibnamefont
  {{Gourgouliatos}}},\ }\href {\doibase 10.1038/s41550-020-01220-z} {\bibfield
  {journal} {\bibinfo  {journal} {Nature Astronomy}\ }\textbf {\bibinfo
  {volume} {5}},\ \bibinfo {pages} {145} (\bibinfo {year} {2021})}\BibitemShut
  {NoStop}%
\bibitem [{\citenamefont {{Lander}}\ and\ \citenamefont
  {{Jones}}(2012)}]{lj12}%
  \BibitemOpen
  \bibfield  {author} {\bibinfo {author} {\bibfnamefont {S.~K.}\ \bibnamefont
  {{Lander}}}\ and\ \bibinfo {author} {\bibfnamefont {D.~I.}\ \bibnamefont
  {{Jones}}},\ }\href {\doibase 10.1111/j.1365-2966.2012.21213.x} {\bibfield
  {journal} {\bibinfo  {journal} {\mnras}\ }\textbf {\bibinfo {volume} {424}},\
  \bibinfo {pages} {482} (\bibinfo {year} {2012})}\BibitemShut {NoStop}%
\bibitem [{\citenamefont {{Sur}}\ \emph {et~al.}(2022)\citenamefont {{Sur}},
  \citenamefont {{Cook}}, \citenamefont {{Radice}}, \citenamefont {{Haskell}},\
  and\ \citenamefont {{Bernuzzi}}}]{sur22}%
  \BibitemOpen
  \bibfield  {author} {\bibinfo {author} {\bibfnamefont {A.}~\bibnamefont
  {{Sur}}}, \bibinfo {author} {\bibfnamefont {W.}~\bibnamefont {{Cook}}},
  \bibinfo {author} {\bibfnamefont {D.}~\bibnamefont {{Radice}}}, \bibinfo
  {author} {\bibfnamefont {B.}~\bibnamefont {{Haskell}}}, \ and\ \bibinfo
  {author} {\bibfnamefont {S.}~\bibnamefont {{Bernuzzi}}},\ }\href {\doibase
  10.1093/mnras/stac353} {\bibfield  {journal} {\bibinfo  {journal} {\mnras}\
  }\textbf {\bibinfo {volume} {511}},\ \bibinfo {pages} {3983} (\bibinfo {year}
  {2022})}\BibitemShut {NoStop}%
\bibitem [{\citenamefont {{Glampedakis}}\ and\ \citenamefont
  {{Lasky}}(2016)}]{gl16}%
  \BibitemOpen
  \bibfield  {author} {\bibinfo {author} {\bibfnamefont {K.}~\bibnamefont
  {{Glampedakis}}}\ and\ \bibinfo {author} {\bibfnamefont {P.~D.}\ \bibnamefont
  {{Lasky}}},\ }\href {\doibase 10.1093/mnras/stw2115} {\bibfield  {journal}
  {\bibinfo  {journal} {\mnras}\ }\textbf {\bibinfo {volume} {463}},\ \bibinfo
  {pages} {2542} (\bibinfo {year} {2016})}\BibitemShut {NoStop}%
\bibitem [{\citenamefont {{Spruit}}(2008)}]{spr08}%
  \BibitemOpen
  \bibfield  {author} {\bibinfo {author} {\bibfnamefont {H.~C.}\ \bibnamefont
  {{Spruit}}},\ }\href {\doibase 10.1063/1.2900262} {\bibfield  {journal}
  {\bibinfo  {journal} {AIP Conf. Proc.}\ }\textbf {\bibinfo {volume} {983}},\
  \bibinfo {pages} {391} (\bibinfo {year} {2008})}\BibitemShut {NoStop}%
\bibitem [{\citenamefont {Cutler}\ and\ \citenamefont {Lindblom}(1987)}]{cl87}%
  \BibitemOpen
  \bibfield  {author} {\bibinfo {author} {\bibfnamefont {C.}~\bibnamefont
  {Cutler}}\ and\ \bibinfo {author} {\bibfnamefont {L.}~\bibnamefont
  {Lindblom}},\ }\href {\doibase 10.1086/165052} {\bibfield  {journal}
  {\bibinfo  {journal} {ApJ}\ }\textbf {\bibinfo {volume} {314}},\ \bibinfo
  {pages} {234} (\bibinfo {year} {1987})}\BibitemShut {NoStop}%
\end{thebibliography}
\end{document}